\documentclass[journal,onecolumn,11pt]{IEEEtran}
\usepackage[margin=1in]{geometry}
\usepackage{setspace}
\usepackage{etoolbox}
\AtBeginEnvironment{figure}{\setstretch{1.0}}
\AtBeginEnvironment{table}{\setstretch{1.0}}
\AtBeginEnvironment{algorithm}{\setstretch{1.0}}

\usepackage{amsmath,amssymb,amsfonts,amsthm}
\usepackage{graphicx}
\usepackage{booktabs}
\usepackage{enumitem}
\usepackage{algorithm}
\usepackage[noend]{algpseudocode}
\usepackage{cite}
\usepackage{xcolor}
\usepackage[hidelinks]{hyperref}
\usepackage{subfig}
\usepackage{tikz}
\usetikzlibrary{decorations.pathreplacing,shapes,positioning,patterns}

\newtheorem{theorem}{Theorem}
\newtheorem{lemma}{Lemma}
\newtheorem{proposition}{Proposition}
\newtheorem{corollary}{Corollary}
\theoremstyle{definition}
\newtheorem{definition}{Definition}
\newtheorem{remark}{Remark}
\newtheorem{example}{Example}

\begin{document}
\title{Rateless Lattices: Constructions,\\ Coding Theorems, and Capacity}
\author{Pegah~Sharifi, Hassan~Khodaiemehr, Khadijeh~Bagheri, and Chen~Feng%
\thanks{P. Sharifi is with the Department of Mathematics and Computer Science, Amirkabir University of Technology (Tehran Polytechnic), Tehran, Iran (e-mail: p.sharifi96@aut.ac.ir). H. Khodaiemehr, K. Bagheri and C. Feng are with the School of Engineering, University of British Columbia, Okanagan Campus, Kelowna, BC, Canada (e-mails: hassan.khodaiemehr@ubc.ca, khadijeh.bagheri@ubc.ca, chen.feng@ubc.ca).}}
\maketitle

\begin{abstract}
A rateless code sends a fixed message as an unbounded stream of symbols such that every prefix of the stream is a codeword. We introduce \emph{rateless lattices}: sequences of integral lattices of increasing dimension in which each lattice is the projection of the next one onto its first coordinates and the induced finite codebooks have a constant size. Our first contribution is a family of such lattices obtained from Construction~D$'$ with two nested Raptor codes. We determine its generator matrices, volume and encoder. For every noise or fading sequence that is unknown to the transmitter and independent of the code randomness, we prove an ensemble random-coding bound and a converse for multistage decoding, which quantify the loss caused by a fixed split of the message between the two levels; the bound remains valid for sparse LT stages of logarithmic degree. These results concern the recovery of the message, i.e., of the finite codeword or residue; we show that lattice-point recovery with a fixed alphabet has an additional coarse-lattice error term. We then show that the split loss comes from stage-by-stage decoding: for a dense ensemble with ring coefficients we determine the rateless capacity region under joint decoding within its admissible level sizes, and with equal level sizes the ensemble achieves the accumulated capacity of uniform pulse-amplitude modulation for two levels, and for more levels under a level-ordering condition that we verify numerically. Single-level $p$-ary rateless lattices achieve the same; they satisfy a finite-parameter rateless Poltyrev inequality, and with a geometrically shaped one-dimensional constellation they form lattice codes that come within any $\epsilon>0$ of the Gaussian incremental-redundancy capacity for bounded SNR. Repeating a codeword of any fixed-dimension lattice, in contrast, loses a factor that is unbounded in the number of retransmissions. For the practical code we analyze a joint belief-propagation decoder over $\mathbb Z_4$ by density evolution on a tree-like surrogate ensemble, which gives fixed-target operating points and guides the choice of the LT degree profile. Simulations of one undithered mother code on AWGN and block-fading channels, with exact sequential stopping, show that joint decoding removes most of the loss of multistage decoding.
\end{abstract}


\section{Introduction}\label{sec:intro}

\IEEEPARstart{L}{attice} codes are the Euclidean counterpart of linear codes. Poltyrev showed that there exist lattices of growing dimension that are reliable for unconstrained additive white Gaussian noise (AWGN) channels whenever the volume-to-noise ratio (VNR) exceeds one~\cite{3kah}. Nested lattice codes achieve $\frac12\log(1+\mathrm{SNR})$ on the power-constrained AWGN channel~\cite{6kah,EZ2004,23silva,24silva}. Lattices also underpin compute-and-forward, integer forcing and lattice-based physical-layer security~\cite{21kah,21silva,19silva}. Practical constructions with iterative decoders include low-density lattice codes (LDLC)~\cite{11kah,29silva}, Construction-A and Construction-D$'$ lattices built from LDPC codes~\cite{silva,fan,13fan,a}, low-density Construction-A (LDA) lattices~\cite{Pietro2018} and polar lattices~\cite{Polarm}. In all of them the dimension~$n$, and hence the rate, is fixed at the transmitter.

Modern wireless links deliver data through \emph{incremental redundancy} (IR). The transmitter keeps sending new coded symbols until the receiver, whose channel state is unknown to the transmitter, signals successful decoding~\cite{Mandelbaum1974,CT2001}. For binary codes, the cleanest model of this operation is the \emph{rateless} (fountain) code: one encoder emits a potentially infinite symbol stream, and every prefix is a codeword of a good code of the corresponding length~\cite{Luby2002,drshok,ES2006}. Shamai, Telatar and Verd\'u showed that rateless operation entails no loss of capacity on memoryless channels~\cite{STV2007}. For Gaussian channels, Erez, Trott and Wornell built rateless codes from layered fixed-rate base codes that are repeated with time-varying weights~\cite{ETW2012}. Spinal codes~\cite{Spinal2012} and rateless polar codes~\cite{LTCS2016} are further approaches; see also~\cite{BCH2011}. To the best of our knowledge, however, no \emph{lattice} has been defined whose prefixes form a family of lattices of increasing dimension with a common encoder, and there are no coding theorems for such objects. This paper fills that gap.

\subsection{Contributions}
We call a sequence of integral lattices $q\mathbb Z^n\subseteq\Lambda_n\subseteq\mathbb{Z}^n$, $n\ge n_0$, a \emph{rateless lattice} if every $\Lambda_n$ is the projection of every $\Lambda_{n'}$ with $n'\ge n$ onto its first $n$ coordinates, and the induced finite codebooks $\Lambda_n\cap\{0,1,\dots,q-1\}^n$ all have the same size. One encoder then maps a message to an infinite $q$-ary sequence whose every prefix is a lattice codeword. We construct rateless lattices from Construction~D$'$ with two nested Raptor codes. Our main contributions are the following.

\begin{enumerate}
\item \emph{Structure.} We prove that the proposed family is a rateless lattice and give its generator matrices, its volume $2^{2n-K-N}$ and a $\mathbb Z_4$-linear encoder (Theorem~\ref{thm:structure}).
\item \emph{Coding theorems for multistage decoding.} For an arbitrary noise sequence that is unknown to the transmitter and independent of the code randomness, which includes block fading and erasures, we prove an ensemble random-coding bound and a matching converse for the multistage architecture at every prefix length (Theorems~\ref{thm:achiev} and~\ref{thm:converse}). These are guarantees for a code drawn from a shared random seed, not for one deterministic code and all sequences simultaneously (Remark~\ref{rem:quant}), and they concern the message (equivalently, the residue modulo~$4$); lattice-point recovery has an additional coarse-lattice term (Remark~\ref{rem:residue}). They yield the exact loss caused by a fixed split of the message between the two levels (Corollary~\ref{cor:price}). The theorems remain valid for sparse LT stages whose average degree grows logarithmically in $N$ (Theorem~\ref{thm:sparse}, Corollary~\ref{cor:sparse}).
\item \emph{Capacity results.} For a dense ensemble of multilevel rateless lattices with ring coefficients we determine the rateless capacity region under joint ML decoding within its admissible level sizes, and show that, with equal level sizes, it achieves the accumulated capacity of uniform PAM for every noise sequence; this is proved for two levels and holds for more levels under a level-ordering condition that we verify numerically (Theorem~\ref{thm:mlcap}, Corollary~\ref{cor:mlcap}). Single-level $p$-ary rateless lattices achieve the same, satisfy a finite-parameter rateless Poltyrev inequality, and come within $0.254$ bits per dimension of the Gaussian incremental-redundancy capacity (Theorem~\ref{thm:capacity}, Corollaries~\ref{cor:poltyrev} and~\ref{cor:shaping}). Mapping the $p$-ary symbols to a nearly Gaussian constellation, by a many-to-one (probabilistic) or an injective (geometric) labeling, comes within any $\epsilon>0$ of that capacity for every noise sequence (Theorems~\ref{thm:shaped} and~\ref{thm:geoshape}); with geometric labeling every transmitted prefix is a point of a translate of the rateless lattice. In contrast, repeating a codeword of any fixed-dimension lattice loses a factor that is unbounded in the number of retransmissions, linear in it at high SNR (Theorem~\ref{thm:advantage}). We also compare rateless lattices with the layered rateless codes of~\cite{ETW2012} (Section~\ref{sec:vsrep}).
\item \emph{Practical realization.} We give a sparse construction whose Tanner graphs have girth at least $6$ at every prefix length (Lemma~\ref{lem:girth}), with $O(n)$ encoding and $O(\mathfrak I n)$ decoding, and a joint BP decoder over $\mathbb Z_4$ that decodes both levels at once (Section~\ref{sec:joint}). Density evolution on a locally tree-like surrogate ensemble gives its operating points (Section~\ref{sec:de}). Simulations of one undithered mother code with exact sequential stopping agree with these predictions on AWGN and block-fading channels and show that joint decoding recovers most of the split loss of multistage decoding.
\end{enumerate}

Construction-D$'$ lattices built from Raptor codes were studied, for fixed dimension and by simulation only, in~\cite{CCECE}. The preprint~\cite{arxivPrev} used such lattices for vehicular broadcast with a link-level abstraction. Neither work defines rateless lattices or contains the theorems of the present paper. Rateless lattice wiretap coding, which builds on the results of this paper, is treated in the companion letter~\cite{LCOMcomp}.

Sections~\ref{sec:structure}--\ref{sec:capacity} give the structure, the multistage coding theorems and the capacity results; Section~\ref{sec:construction} describes the sparse realization and its decoders, Section~\ref{sec:numerics} reports numerical results, and Section~\ref{sec:conclusion} concludes.

Vectors are column vectors in boldface; $\mathbf{x}_{[a:b]}$ denotes coordinates $a,\dots,b$ and $\pi_n$ the projection onto the first $n$ coordinates. $\mathbb{F}_2=\{0,1\}$; for a binary vector $\mathbf{c}$, the same symbol also denotes its embedding in $\{0,1\}^n\subset\mathbb{Z}^n$. Products such as $\mathbf{P}^{\mathsf T}\mathbf{x}$ with $\mathbf{P}$ binary and $\mathbf{x}$ integer are computed over $\mathbb{Z}$. $\mathrm{vol}(\Lambda)$ denotes the volume of a fundamental region of $\Lambda$, and $\mathrm{VNR}(\Lambda,\sigma^2)=\mathrm{vol}(\Lambda)^{2/n}/(2\pi e\sigma^2)$. Logarithms are to base $2$ unless stated otherwise, $[x]^+=\max(x,0)$, $Q(\cdot)$ is the Gaussian tail function, and $\mathbf 1\{\cdot\}$ is the indicator function. Table~\ref{tab:notation} lists the main symbols; each symbol has a single meaning throughout the paper.

\begin{table}[!t]
\centering
\footnotesize
\caption{Main notation.}
\label{tab:notation}
\setlength{\tabcolsep}{3pt}
\begin{tabular}{@{}ll@{\quad}ll@{}}
\toprule
\multicolumn{2}{@{}l}{\emph{Codes and lattices}} & \multicolumn{2}{l@{}}{\emph{Channels, information measures and decoding}}\\
\midrule
$N$, $K$ & intermediate length, precode dimension & $\sigma_j$ & noise std.\ of coordinate $j$ ($\infty$: erasure)\\
$\mathcal C_{\mathrm{pre}}$, $\mathbf H_{\mathrm{pre}}$, $\mathbf A_{\mathrm{pre}}$ & precode, its parity-check and generator matrix & $W_0,W_1$ & level channels~\eqref{eq:W0}--\eqref{eq:W1}\\
$\mathbf s_t$, $\mathcal N_t$ & $t$-th LT vector and its support & $C_0,C_1,C_{\mathrm X}$ & level and $4$-PAM mutual informations\\
$\mathbf P_n$ & $[\mathbf s_1,\dots,\mathbf s_{n-N}]$ & $E_0(\rho,W)$ & Gallager function~\eqref{eq:E0}, $\rho\in[0,1]$\\
$\mathbf m_0,\mathbf m_1$ & level-$0$/level-$1$ intermediate words & $W_{\ge l}$, $C_{\ge l}$ & multilevel channels, eq.~\eqref{eq:E0ml}\\
$\boldsymbol\xi=\mathbf m_0+2\mathbf m_1$ & systematic $\mathbb Z_4$ part & $I_p$ & mutual information of uniform $p$-PAM\\
$\mathcal C_0^{(n)},\mathcal C_1^{(n)}$ & level codes~\eqref{eq:codes} & $\mathsf Z_i$ & Bhattacharyya parameter of $W_i$\\
$\mathbf c_0,\mathbf c_1$ & level codewords & $\mathrm{SNR}_j$, $\gamma_i$ & SNR of coordinate $j$, of round $i$\\
$\Lambda_n$, $\mathbf G_n$ & prefix lattice, generator matrix~\eqref{eq:G} & $L_r$, $p_{\mathrm{er}}$ & round length, round erasure probability\\
$\mathcal S_n$ & codebook $\Lambda_n\cap\{0,\dots,q-1\}^n$ & $\varepsilon$ & error probability target\\
$\mathbf B_{\mathrm{pre}}$ & generator of $\mathcal C_{\mathrm{pre}}+2\mathbb Z^N$ & $\mathfrak I$ & maximum number of BP iterations\\
$\chi=K/N$ & level split & $\Omega$, $\bar d$ & LT degree distribution, mean LT degree\\
$\mathbf d$, $\mathbf D$ & level-$1$ dither (Sec.~\ref{sec:theorems}), public dither & $\theta$ & LT overhead $(n-N)/N$\\
$a$, $q=2^a$ & number of levels, alphabet size & $g$, $\Delta$, $\epsilon_g$ & labeling, grid step, entropy deficit\\
$p$, $k_p$, $\mathbf A_n$ & prime, dimension, generator of $p$-ary family & $\mathrm{SNR}_{\max}$ & largest SNR of a received coordinate\\
$Z$, $w_r$, $\nu_j$, $\kappa$ & circulant size, row weight, shifts, rank defect & $\varpi$ & density of sparse LT vectors\\
$\mathcal Q$, $d_i(n)$ & owed list, LT degree of coordinate $i$ in $\Lambda_n$ & $\mathbf e_j$, $N_0(n)$ & unit vector, number of uncovered coordinates\\
\bottomrule
\end{tabular}
\end{table}

\section{Rateless Construction-D$'$ Lattices}\label{sec:structure}
We recall Construction~D$'$ and systematic Raptor codes and combine them into a rateless lattice with an explicit generator matrix, volume and encoder (Theorem~\ref{thm:structure}).

Let $\mathcal{C}_0\subseteq\mathcal{C}_1\subseteq\mathbb{F}_2^n$ be nested binary linear codes with $k_l=\dim\mathcal C_l$. Choose linearly independent parity checks $\mathbf{h}_1,\dots,\mathbf{h}_{J_0}$ of $\mathcal{C}_0$ ($J_0=n-k_0$) such that the first $J_1=n-k_1$ of them are parity checks of $\mathcal{C}_1$. The two-level Construction-D$'$ lattice~\cite{11fan,silva,fan} is
\begin{equation}\label{eq:Dprime}
\Lambda=\big\{\mathbf{x}\in\mathbb{Z}^n:\ \mathbf{h}_j^{\mathsf T}\mathbf{x}\equiv 0 \!\!\pmod 4\ \text{for } j\le J_1,\quad \mathbf{h}_j^{\mathsf T}\mathbf{x}\equiv 0\!\!\pmod 2\ \text{for } J_1<j\le J_0\big\}.
\end{equation}
Every $\mathbf x\in\Lambda$ satisfies $\mathbf{x}\bmod 2\in\mathcal{C}_0$, and $4\mathbb{Z}^n\subseteq\Lambda$. The checks of $\mathcal C_1$ are imposed modulo $4$, and the remaining checks of $\mathcal C_0$ modulo $2$.

A Raptor code~\cite{drshok} concatenates a high-rate \emph{precode} with an LT code~\cite{Luby2002}. We use the systematic form~\cite{IET,2019}. The precode $\mathcal{C}_{\mathrm{pre}}\subseteq\mathbb{F}_2^N$ has dimension $K$ and full-row-rank parity-check matrix $\mathbf{H}_{\mathrm{pre}}\in\mathbb{F}_2^{(N-K)\times N}$, and encodes a message $\mathbf u_0\in\mathbb F_2^K$ into an \emph{intermediate word} $\mathbf{m}_0\in\mathcal{C}_{\mathrm{pre}}$. The LT stage first emits the $N$ intermediate bits and then, for $t=1,2,\dots$, the parity bit $\mathbf{s}_t^{\mathsf T}\mathbf{m}_0=\bigoplus_{i\in\mathcal{N}_t}m_{0,i}$, where $\mathcal{N}_t\subseteq\{1,\dots,N\}$ is the \emph{neighbourhood} of the $t$-th LT check and $\mathbf{s}_t$ its incidence vector. We write $\mathbf{P}_n=[\mathbf{s}_1,\dots,\mathbf{s}_{n-N}]$.

\subsection{The Raptor Lattice Family}
For $n\ge N$, define the two nested length-$n$ codes
\begin{equation}\label{eq:codes}
\mathcal{C}_1^{(n)}=\big\{(\mathbf{m},\mathbf{P}_n^{\mathsf T}\mathbf{m}\bmod2):\ \mathbf{m}\in\mathbb{F}_2^N\big\},\qquad
\mathcal{C}_0^{(n)}=\big\{(\mathbf{m},\mathbf{P}_n^{\mathsf T}\mathbf{m}\bmod 2):\ \mathbf{m}\in\mathcal{C}_{\mathrm{pre}}\big\}.
\end{equation}
$\mathcal{C}_0^{(n)}$ is the length-$n$ prefix of the systematic Raptor code, and $\mathcal{C}_1^{(n)}$ is the prefix of the systematic LT code alone. Hence $\mathcal{C}_0^{(n)}\subseteq \mathcal{C}_1^{(n)}$, $k_0=K$ and $k_1=N$. The matrices
\begin{equation}\label{eq:H}
\mathbf{H}_1^{(n)}=\big[\,\mathbf{P}_n^{\mathsf T}\ \ \mathbf{I}_{n-N}\,\big],\qquad
\mathbf{H}_0^{(n)}=\begin{bmatrix}\mathbf{H}_{\mathrm{pre}}&\mathbf{0}\\ \mathbf{P}_n^{\mathsf T}&\mathbf{I}_{n-N}\end{bmatrix}
\end{equation}
are full-row-rank parity-check matrices of $\mathcal{C}_1^{(n)}$ and $\mathcal{C}_0^{(n)}$, and the rows of $\mathbf{H}_1^{(n)}$ are rows of $\mathbf{H}_0^{(n)}$. Applying~\eqref{eq:Dprime} with these checks gives the following definition.

\begin{definition}[Raptor lattice]\label{def:raptorlattice}
For $n\ge N$, the length-$n$ Raptor lattice is
\begin{equation}\label{eq:lattice}
\Lambda_n=\Big\{\mathbf{x}\in\mathbb{Z}^n:\ \mathbf{H}_{\mathrm{pre}}\mathbf{x}_{[1:N]}\equiv\mathbf{0}\!\!\pmod 2,\ \ \mathbf{P}_n^{\mathsf T}\mathbf{x}_{[1:N]}+\mathbf{x}_{[N+1:n]}\equiv\mathbf{0}\!\!\pmod 4\Big\}.
\end{equation}
\end{definition}

\begin{definition}[Rateless lattice]\label{def:rateless}
Let $q\ge2$ be an integer. A sequence $(\Lambda_n)_{n\ge n_0}$ of full-rank lattices $\Lambda_n\subset\mathbb{R}^n$ is a \emph{$q$-ary rateless lattice} if
\begin{enumerate}[label=(\roman*)]
\item $q\mathbb{Z}^n\subseteq\Lambda_n\subseteq\mathbb{Z}^n$ for every $n$;
\item $\pi_n(\Lambda_{n'})=\Lambda_n$ for all $n'\ge n\ge n_0$;
\item the codebook $\mathcal{S}_n=\Lambda_n\cap\{0,1,\dots,q-1\}^n$ has a size that does not depend on $n$.
\end{enumerate}
\end{definition}

The integrality $\Lambda_n\subseteq\mathbb Z^n$ in~(i) is essential. Because $q\mathbb Z^n\subseteq\Lambda_n\subseteq\mathbb Z^n$, reduction modulo $q$ maps $\Lambda_n$ onto $\Lambda_n/q\mathbb Z^n$, every coset of $q\mathbb Z^n$ in $\Lambda_n$ has exactly one representative in $\{0,\dots,q-1\}^n$, and therefore $|\mathcal S_n|=[\Lambda_n:q\mathbb Z^n]=q^n/\mathrm{vol}(\Lambda_n)$. Condition~(iii) is therefore equivalent to $\mathrm{vol}(\Lambda_n)$ growing exactly by the factor $q$ per added dimension. Moreover, $\pi_n(\mathcal S_{n'})\subseteq\mathcal S_n$ by~(ii), and every $\mathbf x\in\mathcal S_n$ is the image of a point of $\mathcal S_{n'}$: lift it to $\Lambda_{n'}$ by~(ii) and reduce the new coordinates modulo $q$, which is allowed by~(i). Since the two finite sets have the same size by~(iii), $\pi_n:\mathcal S_{n'}\to\mathcal S_n$ is a bijection. A message therefore selects one point of $\mathcal{S}_{n'}$ for every $n'$, and these points are prefixes of one another. Without integrality these conclusions can fail: $\Lambda_1=\mathbb Z$, $\Lambda_2=\{(a,\frac a2+b):a,b\in\mathbb Z\}$ and $\Lambda_n=\Lambda_2\times(2\mathbb Z)^{n-2}$ satisfy the remaining conditions with $q=2$, but $\pi_1(\mathcal S_2)=\{0\}\neq\mathcal S_1$ and $|\mathcal S_2|=2\neq2^2/\mathrm{vol}(\Lambda_2)$. The next theorem shows that the Raptor lattice is a rateless lattice. Let $\Lambda_{\mathrm A}(\mathcal{C}_{\mathrm{pre}})=\mathcal{C}_{\mathrm{pre}}+2\mathbb{Z}^N$ be the Construction-A lattice of the precode and let $\mathbf{B}_{\mathrm{pre}}\in\mathbb{Z}^{N\times N}$ be any of its generator matrices.

\begin{theorem}[Structure]\label{thm:structure}
For every $n\ge N$:
\begin{enumerate}[label=(\alph*)]
\item $\Lambda_n$ is the Construction-D$'$ lattice of $\mathcal{C}_0^{(n)}\subseteq\mathcal{C}_1^{(n)}$, $4\mathbb{Z}^n\subseteq\Lambda_n\subseteq\mathbb{Z}^n$, and
\begin{equation}\label{eq:G}
\mathbf{G}_n=\begin{bmatrix}\mathbf{B}_{\mathrm{pre}}&\mathbf{0}\\ -\mathbf{P}_n^{\mathsf T}\mathbf{B}_{\mathrm{pre}}&4\mathbf{I}_{n-N}\end{bmatrix}
\end{equation}
is a generator matrix of $\Lambda_n$, i.e., $\Lambda_n=\{\mathbf{G}_n\mathbf{b}:\mathbf{b}\in\mathbb{Z}^n\}$.
\item $\mathrm{vol}(\Lambda_n)=2^{2n-K-N}$.
\item $\pi_n(\Lambda_{n'})=\Lambda_n$ for all $n'\ge n$. Equivalently, $\mathbf{G}_{n'}=\left[\begin{smallmatrix}\mathbf{G}_n&\mathbf{0}\\ \ast&4\mathbf{I}_{n'-n}\end{smallmatrix}\right]$.
\item $|\mathcal{S}_n|=2^{K+N}$, where $\mathcal{S}_n=\Lambda_n\cap\{0,1,2,3\}^n$, and the restriction of $\pi_n$ to $\mathcal{S}_{n'}$ is a bijection onto $\mathcal{S}_n$ for all $n'\ge n$. The map
\begin{equation}\label{eq:enc}
\mathrm{Enc}:\ \mathcal{C}_{\mathrm{pre}}\times\mathbb{F}_2^N\to\mathbb{Z}_4^{\infty},\quad
(\mathbf{m}_0,\mathbf{m}_1)\mapsto\big(\boldsymbol{\xi},\ (-\mathbf{s}_1^{\mathsf T}\boldsymbol{\xi})\bmod 4,\ (-\mathbf{s}_2^{\mathsf T}\boldsymbol{\xi})\bmod 4,\dots\big),\quad \boldsymbol{\xi}=\mathbf{m}_0+2\mathbf{m}_1,
\end{equation}
is injective, and $\pi_n\circ\mathrm{Enc}$ is a bijection onto $\mathcal{S}_n$.
\item Let $(\mathbf{m}_0,\mathbf{m}_1)$ be uniformly distributed. If no coordinate vanishes identically on $\mathcal{C}_{\mathrm{pre}}$ and $\mathbf{s}_t\notin\mathcal{C}_{\mathrm{pre}}^{\perp}$ for every $t$, then every coordinate of $\mathrm{Enc}(\mathbf{m}_0,\mathbf{m}_1)$ is uniformly distributed on $\{0,1,2,3\}$.
\end{enumerate}
Consequently, $(\Lambda_n)_{n\ge N}$ is a $4$-ary rateless lattice whose codebooks have $2^{K+N}$ points.
\end{theorem}

\begin{IEEEproof}
(a) The rows of $\mathbf H_1^{(n)}$ are checks of both $\mathcal{C}_1^{(n)}$ and $\mathcal{C}_0^{(n)}$ and are imposed modulo $4$ in~\eqref{eq:lattice}, whereas the rows $[\mathbf{H}_{\mathrm{pre}}\ \mathbf{0}]$ are checks of $\mathcal{C}_0^{(n)}$ only and are imposed modulo $2$; this is exactly~\eqref{eq:Dprime}. Since the defining congruences are linear over $\mathbb Z$ and are satisfied by every vector of $4\mathbb Z^n$, $\Lambda_n$ is a subgroup of $\mathbb Z^n$ containing $4\mathbb Z^n$, and therefore a full-rank lattice. To identify a basis, note that $\mathbf{H}_{\mathrm{pre}}\mathbf{x}_{[1:N]}\equiv\mathbf{0}\pmod 2$ holds iff $\mathbf{x}_{[1:N]}\bmod 2\in\mathcal{C}_{\mathrm{pre}}$, i.e., iff $\mathbf{x}_{[1:N]}=\mathbf{B}_{\mathrm{pre}}\mathbf{b}_1$ for some $\mathbf b_1\in\mathbb Z^N$, and that the modulo-$4$ congruences hold iff $\mathbf{x}_{[N+1:n]}=-\mathbf{P}_n^{\mathsf T}\mathbf{B}_{\mathrm{pre}}\mathbf b_1+4\mathbf{b}_2$ for some $\mathbf b_2\in\mathbb Z^{n-N}$. Hence $\mathbf x\in\Lambda_n$ iff $\mathbf x=\mathbf G_n[\mathbf b_1;\mathbf b_2]$ for an integer vector $[\mathbf b_1;\mathbf b_2]$, and the columns of $\mathbf G_n$ form a basis because $\mathbf G_n$ is block lower-triangular with nonsingular diagonal blocks.

(b) By the block structure, $|\det\mathbf{G}_n|=|\det\mathbf{B}_{\mathrm{pre}}|\cdot 4^{n-N}$. The first factor is the index of $\Lambda_{\mathrm A}(\mathcal{C}_{\mathrm{pre}})$ in $\mathbb{Z}^N$, which equals $2^N/|\mathcal{C}_{\mathrm{pre}}|=2^{N-K}$ because $\mathbb{Z}^N/2\mathbb{Z}^N\cong\mathbb F_2^N$ and $\Lambda_{\mathrm A}(\mathcal{C}_{\mathrm{pre}})/2\mathbb{Z}^N\cong\mathcal C_{\mathrm{pre}}$. Therefore $\mathrm{vol}(\Lambda_n)=2^{N-K}\,4^{n-N}=2^{2n-K-N}$.

(c) The congruences defining $\Lambda_n$ (the precode congruences and the LT congruences with $t\le n-N$) involve only the first $n$ coordinates and are among those defining $\Lambda_{n'}$; hence $\pi_n(\Lambda_{n'})\subseteq\Lambda_n$. Conversely, every $\mathbf x\in\Lambda_n$ extends to a point of $\Lambda_{n'}$ by setting $x_{N+t}=-\mathbf{s}_t^{\mathsf T}\mathbf{x}_{[1:N]}$ for $n-N<t\le n'-N$. The block form of $\mathbf G_{n'}$ follows from~\eqref{eq:G} and $\mathbf{P}_{n'}=[\mathbf{P}_n\ \ \mathbf{s}_{n-N+1}\cdots\mathbf{s}_{n'-N}]$.

(d) Because $4\mathbb{Z}^n\subseteq\Lambda_n$, the points of $\mathcal S_n$ are the representatives in $\{0,1,2,3\}^n$ of the cosets of $4\mathbb Z^n$ in $\Lambda_n$, and by~(b) there are $4^n/\mathrm{vol}(\Lambda_n)=2^{K+N}$ of them. By~(c), $\pi_n(\mathcal{S}_{n'})\subseteq\mathcal{S}_n$; moreover, the extensions of $\mathbf x\in\mathcal S_n$ to $\Lambda_{n'}$ are the vectors with $x_{N+t}\equiv-\mathbf{s}_t^{\mathsf T}\mathbf{x}_{[1:N]}\pmod 4$ for $t>n-N$, and exactly one of them has its new coordinates in $\{0,1,2,3\}$. Thus $\pi_n:\mathcal S_{n'}\to\mathcal S_n$ is a bijection. For $n=N$ there are no LT congruences, and $\boldsymbol\xi\in\{0,1,2,3\}^N$ lies in $\mathcal S_N$ iff $\boldsymbol\xi\bmod 2\in\mathcal C_{\mathrm{pre}}$; each such $\boldsymbol\xi$ is uniquely written as $\mathbf{m}_0+2\mathbf{m}_1$ with $\mathbf{m}_0=\boldsymbol\xi\bmod2\in\mathcal{C}_{\mathrm{pre}}$ and $\mathbf{m}_1=\lfloor\boldsymbol\xi/2\rfloor\in\mathbb{F}_2^N$. Hence $(\mathbf m_0,\mathbf m_1)\mapsto\boldsymbol\xi$ is a bijection onto $\mathcal S_N$, and combining it with the bijections $\pi_N:\mathcal S_n\to\mathcal S_N$ shows that $\pi_n\circ\mathrm{Enc}$ is a bijection onto $\mathcal S_n$ for every $n$ and that $\mathrm{Enc}$ is injective.

(e) A systematic symbol is $\xi_i=m_{0,i}+2m_{1,i}$, where $m_{1,i}$ is uniform and independent of $\mathbf m_0$, and $m_{0,i}$ is uniform because a linear functional that does not vanish on $\mathcal C_{\mathrm{pre}}$ takes each value on exactly half of the code; hence $\xi_i$ is uniform on $\{0,1,2,3\}$. For a parity symbol, write $-\mathbf{s}_t^{\mathsf T}\boldsymbol\xi=-\mathbf s_t^{\mathsf T}\mathbf m_0-2\mathbf s_t^{\mathsf T}\mathbf m_1$. Modulo~$4$, the second term depends on $\mathbf m_1$ only through the bit $\beta=\mathbf{s}_t^{\mathsf T}\mathbf{m}_1\bmod 2$, which is uniform and independent of $\mathbf m_0$ since $\mathbf s_t\neq\mathbf0$. The symbol is therefore uniform on the pair $\{a,a+2\}$ with $a\equiv-\mathbf s_t^{\mathsf T}\mathbf m_0\pmod 4$, and the parity of $a$, namely $\mathbf{s}_t^{\mathsf T}\mathbf{m}_0\bmod2$, is uniform because $\mathbf{s}_t\notin\mathcal{C}_{\mathrm{pre}}^\perp$. Hence the parity symbol is uniform on $\{0,1,2,3\}$ as well.
\end{IEEEproof}

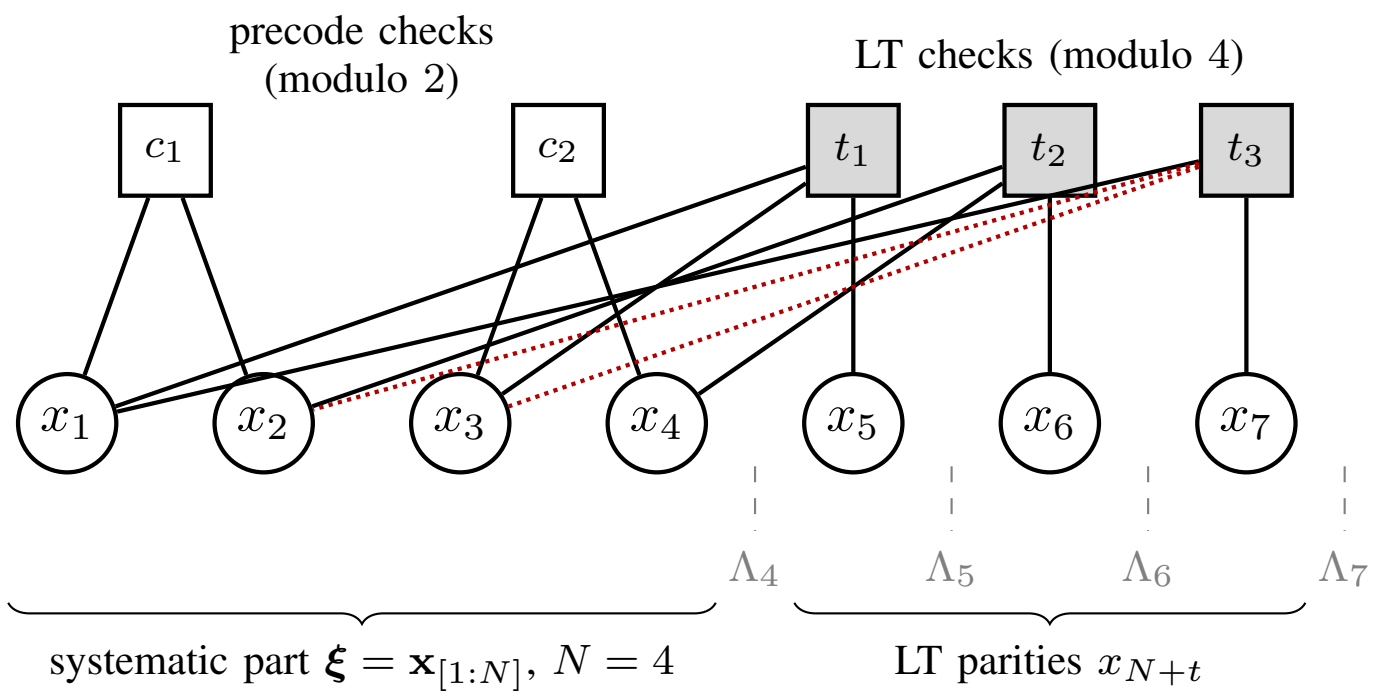
\begin{figure}[!t]
\centering
\begin{tikzpicture}[x=1.3cm,y=0.95cm,
  var/.style={circle,draw,thick,minimum size=6.5mm,inner sep=0pt,font=\small},
  pchk/.style={rectangle,draw,thick,minimum size=6mm,inner sep=0pt,font=\scriptsize,fill=white},
  lchk/.style={rectangle,draw,thick,minimum size=6mm,inner sep=0pt,font=\scriptsize,fill=black!15},
  edge/.style={thick}, cut/.style={densely dotted,thick,red!70!black}]
\foreach \i in {1,...,7}{\node[var] (x\i) at (\i,0) {$x_{\i}$};}
\node[pchk] (c1) at (1.5,1.9) {$c_1$};
\node[pchk] (c2) at (3.5,1.9) {$c_2$};
\node[lchk] (t1) at (5,1.9) {$t_1$};
\node[lchk] (t2) at (6,1.9) {$t_2$};
\node[lchk] (t3) at (7,1.9) {$t_3$};
\draw[edge] (c1)--(x1); \draw[edge] (c1)--(x2);
\draw[edge] (c2)--(x3); \draw[edge] (c2)--(x4);
\draw[edge] (t1)--(x1); \draw[edge] (t1)--(x3); \draw[edge] (t1)--(x5);
\draw[edge] (t2)--(x2); \draw[edge] (t2)--(x4); \draw[edge] (t2)--(x6);
\draw[edge] (t3)--(x1); \draw[edge] (t3)--(x7);
\draw[cut] (t3)--(x2); \draw[cut] (t3)--(x3);
\node[font=\scriptsize,align=center] at (2.5,2.55) {precode checks\\ (modulo $2$)};
\node[font=\scriptsize,align=center] at (6,2.55) {LT checks (modulo $4$)};
\foreach \xx/\lab in {4.5/4,5.5/5,6.5/6,7.5/7}{\draw[dashed,gray] (\xx,-0.3)--(\xx,-0.75); \node[font=\scriptsize,gray,anchor=north] at (\xx,-0.75) {$\Lambda_{\lab}$};}
\draw[decorate,decoration={brace,mirror,amplitude=4pt}] (0.7,-1.25)--(4.3,-1.25) node[midway,below=4pt,font=\scriptsize]{systematic part $\boldsymbol\xi=\mathbf x_{[1:N]}$, $N=4$};
\draw[decorate,decoration={brace,mirror,amplitude=4pt}] (4.7,-1.25)--(7.3,-1.25) node[midway,below=4pt,font=\scriptsize]{LT parities $x_{N+t}$};
\end{tikzpicture}
\caption{Tanner graph of Example~\ref{ex:toy} ($N=4$, $K=2$). White: precode checks (modulo~$2$); grey: LT checks (modulo~$4$), which alone form the level-$1$ graph. Each new dimension adds one variable node and one LT check; red dotted edges are rejected by Algorithm~\ref{alg:lt}.}
\label{fig:tanner}
\end{figure}

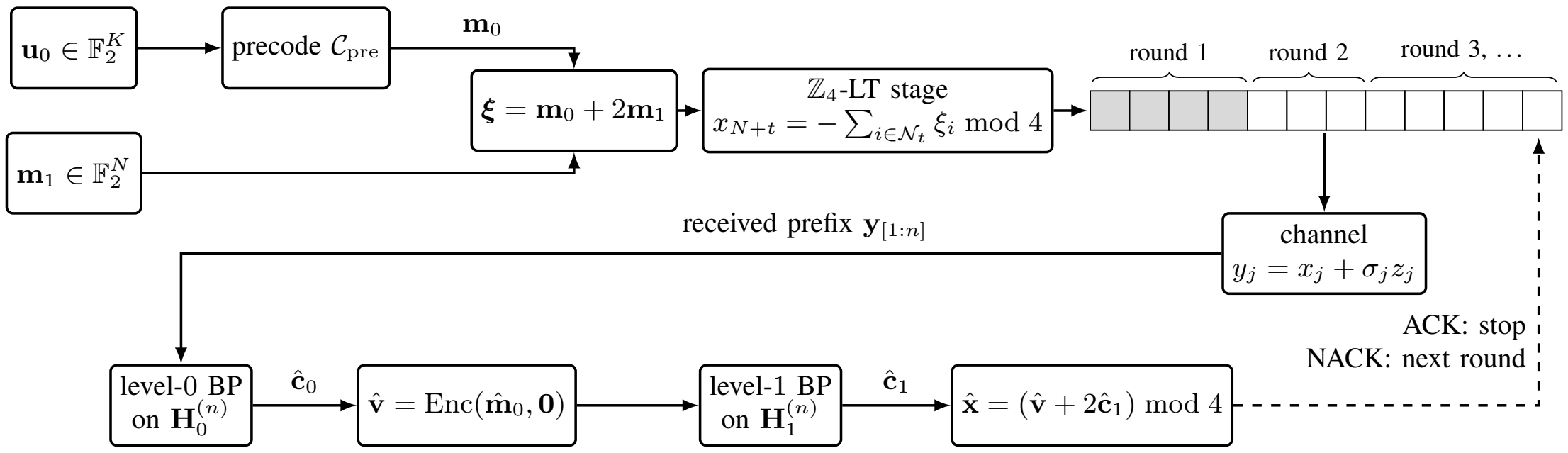
\begin{figure}[!t]
\centering
\resizebox{0.88\textwidth}{!}{%
\begin{tikzpicture}[font=\footnotesize,>=latex,
  blk/.style={draw,thick,rounded corners=2pt,align=center,inner sep=3pt,minimum height=9mm},
  sym/.style={draw,minimum width=4.4mm,minimum height=4.4mm,inner sep=0pt}]
\node[blk] (u0) at (0,0.7) {$\mathbf u_0\in\mathbb F_2^K$};
\node[blk] (u1) at (0,-0.7) {$\mathbf m_1\in\mathbb F_2^N$};
\node[blk] (pre) at (2.6,0.7) {precode $\mathcal C_{\mathrm{pre}}$};
\node[blk] (xi) at (5.6,0) {$\boldsymbol\xi=\mathbf m_0+2\mathbf m_1$};
\node[blk] (lt) at (9.0,0) {$\mathbb Z_4$-LT stage\\ $x_{N+t}=-\sum_{i\in\mathcal N_t}\xi_i\bmod 4$};
\draw[->,thick] (u0)--(pre);
\draw[->,thick] (pre.east)-|node[pos=0.25,above]{$\mathbf m_0$}(xi.north);
\draw[->,thick] (u1.east)-|(xi.south);
\draw[->,thick] (xi)--(lt);
\foreach \k in {0,...,11}{\ifnum\k<4 \node[sym,fill=black!15] at (11.6+0.44*\k,0) {};\else \node[sym] at (11.6+0.44*\k,0) {};\fi}
\draw[->,thick] (lt)--(11.35,0);
\draw[decorate,decoration={brace,amplitude=3pt}] (11.4,0.32)--(13.13,0.32) node[midway,above=3pt,font=\scriptsize]{round 1};
\draw[decorate,decoration={brace,amplitude=3pt}] (13.2,0.32)--(14.45,0.32) node[midway,above=3pt,font=\scriptsize]{round 2};
\draw[decorate,decoration={brace,amplitude=3pt}] (14.52,0.32)--(16.6,0.32) node[midway,above=3pt,font=\scriptsize]{round 3, \dots};
\node[blk] (ch) at (14.0,-1.6) {channel\\ $y_j=x_j+\sigma_jz_j$};
\draw[->,thick] (14.0,-0.25)--(ch.north);
\node[blk] (d0) at (1.2,-3.3) {level-$0$ BP\\ on $\mathbf H_0^{(n)}$};
\node[blk] (v) at (4.4,-3.3) {$\hat{\mathbf v}=\mathrm{Enc}(\hat{\mathbf m}_0,\mathbf 0)$};
\node[blk] (d1) at (7.8,-3.3) {level-$1$ BP\\ on $\mathbf H_1^{(n)}$};
\node[blk] (out) at (11.4,-3.3) {$\hat{\mathbf x}=(\hat{\mathbf v}+2\hat{\mathbf c}_1)\bmod 4$};
\draw[->,thick] (ch.west)-|node[pos=0.2,above]{received prefix $\mathbf y_{[1:n]}$}(d0.north);
\draw[->,thick] (d0)--node[above]{$\hat{\mathbf c}_0$}(v);
\draw[->,thick] (v)--(d1);
\draw[->,thick] (d1)--node[above]{$\hat{\mathbf c}_1$}(out);
\draw[->,thick,dashed] (out.east)-|node[pos=0.62,left,align=right]{ACK: stop\\ NACK: next round}(16.4,-0.3);
\end{tikzpicture}}
\caption{Rateless lattice transmission: the systematic part $\boldsymbol\xi$ is sent in round~1, followed by one $\mathbb Z_4$-LT symbol per new dimension; after each round the receiver decodes the received prefix and acknowledges or requests more. An erasure corresponds to $\sigma_j=\infty$.}
\label{fig:system}
\end{figure}

\begin{example}[A toy rateless lattice]\label{ex:toy}
Let $N=4$ and let $\mathcal C_{\mathrm{pre}}=\{0000,1100,0011,1111\}$ ($K=2$) with $\mathbf H_{\mathrm{pre}}=\left[\begin{smallmatrix}1&1&0&0\\0&0&1&1\end{smallmatrix}\right]$. Take the LT neighbourhoods $\mathcal N_1=\{1,3\}$, $\mathcal N_2=\{2,4\}$ and $\mathcal N_3=\{1\}$. These are the neighbourhoods that the pair-rejection rule of Algorithm~\ref{alg:lt} (Section~\ref{sec:construction}) produces from the proposed index sets $\{1,3\}$, $\{2,4\}$ and $\{1,2,3\}$: for $t_3$, index $1$ is accepted, and the indices $2$ and $3$ are rejected because the pairs $\{1,2\}$ and $\{1,3\}$ already lie in the checks $c_1$ and $t_1$; the owed list of Algorithm~\ref{alg:lt} proposes them again in a later row. The Tanner graph is shown in Fig.~\ref{fig:tanner}. A generator matrix of $\Lambda_{\mathrm A}(\mathcal C_{\mathrm{pre}})$ and the corresponding $\mathbf G_7$ from~\eqref{eq:G} are
\[
\mathbf B_{\mathrm{pre}}=\begin{bmatrix}1&0&0&0\\1&2&0&0\\0&0&1&0\\0&0&1&2\end{bmatrix},\qquad
\mathbf G_7=\left[\begin{array}{cccc|ccc}1&0&0&0&0&0&0\\1&2&0&0&0&0&0\\0&0&1&0&0&0&0\\0&0&1&2&0&0&0\\\hline-1&0&-1&0&4&0&0\\-1&-2&-1&-2&0&4&0\\-1&0&0&0&0&0&4\end{array}\right],
\]
since $\mathbf s_1^{\mathsf T}\mathbf B_{\mathrm{pre}}=(1,0,1,0)$, $\mathbf s_2^{\mathsf T}\mathbf B_{\mathrm{pre}}=(1,2,1,2)$ and $\mathbf s_3^{\mathsf T}\mathbf B_{\mathrm{pre}}=(1,0,0,0)$. For $n=4,5,6,7$, $\mathbf G_n$ is the leading $n\times n$ block of $\mathbf G_7$, and $\mathrm{vol}(\Lambda_n)=|\det\mathbf G_n|=4\cdot4^{n-4}=2^{2n-6}$, in agreement with Theorem~\ref{thm:structure}(b) for $K+N=6$. Every codebook $\mathcal S_n$ has $2^6=64$ points.

\emph{Encoding.} Take $\mathbf m_0=(1,1,0,0)\in\mathcal C_{\mathrm{pre}}$ and $\mathbf m_1=(0,1,1,0)$. Then $\boldsymbol\xi=\mathbf m_0+2\mathbf m_1=(1,3,2,0)$, and~\eqref{eq:enc} gives $x_5=-(\xi_1+\xi_3)=-3\equiv1$, $x_6=-(\xi_2+\xi_4)=-3\equiv1$ and $x_7=-\xi_1\equiv3$ (mod~$4$). Each prefix of length $n\ge4$ of the stream $(1,3,2,0\,|\,1,1,3,\dots)$ is a point of $\mathcal S_n$. Fig.~\ref{fig:system} shows how such a stream is sent in rounds and decoded after each round.

\emph{Levels.} Reducing modulo $2$ gives the level-$0$ word $\mathbf c_0=\mathbf x\bmod2=(1,1,0,0,1,1,1)\in\mathcal C_0^{(7)}$. Removing the point $\mathbf v=\mathrm{Enc}(\mathbf m_0,\mathbf 0)=(1,1,0,0,3,3,3)$, which depends on $\mathbf c_0$ only, leaves $\mathbf x-\mathbf v\equiv2\cdot(0,1,1,0,1,1,0)\pmod4$. The bracket is the level-$1$ word $\mathbf c_1=(\mathbf m_1,\mathbf P_7^{\mathsf T}\mathbf m_1\bmod2)\in\mathcal C_1^{(7)}$. The multistage decoder inverts this decomposition $\mathbf x=(\mathrm{Enc}(\mathbf m_0,\mathbf 0)+2\mathbf c_1)\bmod4$, decoding $\mathbf c_0$ first and then $\mathbf c_1$.
\end{example}

\begin{remark}[Which object is a rateless lattice?]\label{rem:nonsquare}
All information is carried by the $N\times N$ block $\mathbf{B}_{\mathrm{pre}}$, the generator of the Construction-A base lattice $\Lambda_{\mathrm A}(\mathcal{C}_{\mathrm{pre}})$. Each further dimension is a linear functional of the first $N$ coordinates, $x_{N+t}\equiv-\mathbf{s}_t^{\mathsf T}\mathbf{x}_{[1:N]}\pmod 4$. Equivalently, $\mathcal{S}_n$ is the image of $\mathcal{S}_N$ under a $\mathbb{Z}_4$-linear systematic LT code, and the encoder~\eqref{eq:enc} needs $|\mathcal{N}_t|$ additions modulo $4$ per transmitted symbol.

The second level is essential: with modulo-$2$ instead of modulo-$4$ conditions, the codebook in $\{0,1,2,3\}^n$ would have $2^{K+n}$ points, which violates condition~(iii) of Definition~\ref{def:rateless}.
\end{remark}

\subsection{VNR, Rate and SNR Along the Prefix Family}
Consider the unconstrained AWGN channel $\mathbf{y}=\mathbf{x}+\sigma\mathbf{z}$ with $\mathbf{z}\sim\mathcal{N}(\mathbf{0},\mathbf{I})$ and $\mathbf{x}\in\Lambda_n$. Consider also the $4$-PAM channel, in which $\mathbf{x}\in\mathcal{S}_n$ is transmitted after its mean $3/2$ is removed. The rate of $\mathcal{S}_n$ is $R_n=(K+N)/n$ bits per real dimension.

\begin{proposition}\label{prop:vnr}
For the Raptor lattice at noise variance $\sigma^2$,
\begin{equation}\label{eq:vnrn}
\mathrm{VNR}_n(\sigma^2)=\frac{2^{2(2-R_n)}}{2\pi e\,\sigma^2},
\end{equation}
which is strictly increasing in $n$ and tends to $16/(2\pi e\sigma^2)$ as $n\to\infty$. Under the conditions of Theorem~\ref{thm:structure}(e), the centred $4$-PAM codebook has average energy $P=5/4$ per dimension. With $\mathrm{SNR}=P/\sigma^2$,
\begin{equation}\label{eq:rate}
R_n=\tfrac12\log\mathrm{SNR}-\tfrac12\log\mathrm{VNR}_n(\sigma^2)-\tfrac12\log\tfrac{\pi e}{6}+\tfrac12\log\tfrac{16}{15}.
\end{equation}
\end{proposition}

\begin{IEEEproof}
By Theorem~\ref{thm:structure}(b), $\mathrm{vol}(\Lambda_n)^{2/n}=2^{2(2n-K-N)/n}=2^{2(2-R_n)}$, which gives~\eqref{eq:vnrn}. Since $R_n=(K+N)/n$ decreases strictly in $n$ and tends to $0$, $\mathrm{VNR}_n$ increases strictly to $2^4/(2\pi e\sigma^2)$. By Theorem~\ref{thm:structure}(e) each coordinate is uniform on $\{0,1,2,3\}$, whose variance is $\frac14\sum_{x=0}^3(x-\frac32)^2=\frac54$. Substituting $\sigma^2=5/(4\,\mathrm{SNR})$ into~\eqref{eq:vnrn} gives $\log\mathrm{VNR}_n=4-2R_n-\log(2\pi e)+\log\tfrac{4\,\mathrm{SNR}}{5}=-2R_n+\log\mathrm{SNR}-\log\tfrac{10\pi e}{64}$. Solving for $R_n$ and using $\frac{10\pi e}{64}=\frac{\pi e}{6}\cdot\frac{15}{16}$ gives~\eqref{eq:rate}.
\end{IEEEproof}

A receiver with fixed $\sigma^2$ sees $\mathrm{VNR}_n$ grow as dimensions arrive. If the threshold of $\Lambda_n$ had the same value $\nu^\star$ for every $n$, \eqref{eq:rate} would give a rate within $\frac12\log\nu^\star+\frac12\log\frac{\pi e}{6}-\frac12\log\frac{16}{15}$ bits of $\frac12\log\mathrm{SNR}$ at \emph{every} SNR ($1.53$~dB shaping loss, $0.28$~dB finite-constellation correction). This heuristic concerns the finite codebook $\mathcal S_n$, or equivalently the residue $\mathbf x\bmod4$; for unconstrained lattice-point decoding the sublattice $4\mathbb Z^n$ adds an error term that no fixed alphabet removes as $n\to\infty$ (Remark~\ref{rem:residue}). Section~\ref{sec:theorems} shows that, under MSD with a fixed split, $\nu^\star$ cannot be the same for all $n$.

\section{Coding Theorems for Rateless Lattices}\label{sec:theorems}
We now ask how long a prefix the receiver needs under multistage decoding, and quantify the price of a fixed level split.

\subsection{Channel Model and Level Channels}
The $j$-th transmitted coordinate $x_j\in\{0,1,2,3\}$ is received as
\begin{equation}\label{eq:channel}
y_j=x_j+\sigma_j z_j,\qquad z_j\sim\mathcal{N}(0,1)\ \text{i.i.d.},\qquad \sigma_j\in(0,\infty],
\end{equation}
where $\sigma_j=\infty$ denotes an erasure (the output carries no information about $x_j$). The sequence $(\sigma_j)_{j\ge1}$ is arbitrary and unknown to the transmitter, and the receiver knows it (CSIR). After coherent equalization, block fading with gain $h_r$ in round (block) $r$ takes this form with $\sigma_j=\sigma/|h_r|$, and packet erasures set $\sigma_j=\infty$. Constant shifts of the constellation (such as the removal of its mean) are known to the receiver and do not affect any result below.

Multistage decoding (MSD) of $\Lambda_n$ involves two binary-input channels. For a symbol $x\in\{0,1,2,3\}$ we call $x\bmod 2$ its \emph{level-$0$ bit}. Given the level-$0$ bit $b$, the symbol lies in the pair $\{b,b+2\}$, and its position in this pair is its \emph{level-$1$ bit}. Let $\varphi_\sigma$ be the $\mathcal{N}(0,\sigma^2)$ density. Define
\begin{align}
W_0(y\,|\,b;\sigma)&=\tfrac12\big[\varphi_\sigma(y-b)+\varphi_\sigma(y-b-2)\big], && b\in\{0,1\},\label{eq:W0}\\
W_1(y\,|\,b;\sigma)&=\varphi_\sigma(y-2b), && b\in\{0,1\}.\label{eq:W1}
\end{align}
$W_0$ is the channel seen by the level-$0$ bit when the level-$1$ bit is uniform and unknown. $W_1$ is the channel of the level-$1$ bit once the level-$0$ bit is known. Both are binary-input output-symmetric (BIOS) channels: the involution $y\mapsto3-y$ maps $W_0(\cdot|0)$ to $W_0(\cdot|1)$, and $y\mapsto2-y$ maps $W_1(\cdot|0)$ to $W_1(\cdot|1)$. Let $C_i(\sigma)$ be the capacity of $W_i(\cdot\,|\,\cdot;\sigma)$, which is attained by the uniform input. For a binary-input channel $W$ and $\rho\in[0,1]$, Gallager's function for the uniform input is~\cite{Gallager1968}
\begin{equation}\label{eq:E0}
E_0(\rho,W)=-\log\int\Big(\tfrac12W(y|0)^{\frac1{1+\rho}}+\tfrac12W(y|1)^{\frac1{1+\rho}}\Big)^{1+\rho}dy .
\end{equation}
All logarithms are to base~$2$. By the chain rule, $C_0(\sigma)+C_1(\sigma)=C_{\mathrm X}(\sigma)$, the mutual information of uniform $4$-PAM~\cite{WFH1999}. For an erasure, $C_i(\infty)=E_0(\rho,W_i(\cdot;\infty))=0$. For the unconstrained channel ($\mathbf x\in\Lambda_n$ instead of $\mathcal S_n$), the same statements hold with $W_0$ and $W_1$ replaced by the modulo-$2$ Gaussian channels at noise levels $\sigma$ and $\sigma/2$~\cite{4kah}, but only for the recovery of the \emph{residue} $\mathbf x\bmod4$ (Remark~\ref{rem:residue}).

\subsection{Random Ensemble, Multistage Decoding and Achievability}
\begin{definition}[Ensemble $\mathfrak{E}(K,N)$]\label{def:ens}
Let $1\le K\le N$. The precode $\mathcal{C}_{\mathrm{pre}}$ is the image of a generator matrix $\mathbf{A}_{\mathrm{pre}}\in\mathbb{F}_2^{N\times K}$ drawn uniformly from the matrices of rank $K$. The LT vectors $\mathbf{s}_1,\mathbf{s}_2,\dots$ are i.i.d.\ uniform on $\mathbb{F}_2^N$. A level-$1$ dither $\mathbf{d}=(d_1,d_2,\dots)$ with i.i.d.\ uniform bits is known to the receiver. All of these are independent. A message $(\mathbf{u}_0,\mathbf{u}_1)\in\mathbb{F}_2^K\times\mathbb{F}_2^N$ is mapped to $\mathbf{m}_0=\mathbf{A}_{\mathrm{pre}}\mathbf{u}_0$ and $\mathbf{m}_1=\mathbf{u}_1$. With the level codewords $\mathbf{c}_i=(\mathbf{m}_i,\mathbf{P}_n^{\mathsf T}\mathbf{m}_i\bmod 2)\in\mathcal{C}_i^{(n)}$, the transmitted prefix is
\begin{equation}\label{eq:ensenc}
\mathbf{x}=\big(\mathbf{v}+2(\mathbf{c}_1\oplus\mathbf{d}_{[1:n]})\big)\bmod 4,\qquad \mathbf{v}=\mathrm{Enc}(\mathbf{m}_0,\mathbf{0})_{[1:n]}.
\end{equation}
\end{definition}

Without the dither, \eqref{eq:ensenc} equals $\mathrm{Enc}(\mathbf{m}_0,\mathbf{m}_1)$ of~\eqref{eq:enc}. Indeed, for every $\mathbf s\in\mathbb F_2^N$, $-\mathbf{s}^{\mathsf T}(\mathbf{m}_0+2\mathbf{m}_1)\equiv-\mathbf{s}^{\mathsf T}\mathbf{m}_0+2\,(\mathbf{s}^{\mathsf T}\mathbf{m}_1\bmod 2)\pmod 4$, because $2a\equiv 2(a\bmod 2)\pmod4$ for every integer $a$. With the dither, the transmitted word is the codeword of $\mathcal{S}_n$ translated by $2\mathbf{d}$ modulo~$4$. The level-$0$ bits are not changed: $\mathbf{x}\bmod 2=\mathbf{v}\bmod2=\mathbf{c}_0$. The vector $\mathbf v$ is a function of $\mathbf c_0$, because $\mathbf m_0=(\mathbf c_0)_{[1:N]}$. The next lemma makes the level channels precise.

\begin{lemma}[Level channels]\label{lem:levels}
In $\mathfrak E(K,N)$, fix the codes, the message and $n$, and let $\mathbf Y=\mathbf x+\boldsymbol\sigma\odot\mathbf z$ be the channel output.
\begin{enumerate}[label=(\alph*)]
\item Averaged over $\mathbf d$, the output $\mathbf Y$ depends on the message only through $\mathbf c_0$, and its conditional density is $\prod_{j=1}^n W_0(y_j|c_{0,j};\sigma_j)$.
\item Given $\mathbf c_0$ and $\mathbf d$, the output $\mathbf Y$ depends on $\mathbf c_1$ through the memoryless channel $\prod_jW_1^{(j)}(y_j|c_{1,j})$. Each $W_1^{(j)}$ is obtained from $W_1(\cdot|\cdot;\sigma_j)$ by a relabelling of the input (when $d_j=1$) and an invertible affine map of the output, both known to the receiver. In particular, $W_1^{(j)}$ has the same capacity $C_1(\sigma_j)$ and the same function $E_0(\rho,\cdot)$ as $W_1(\cdot;\sigma_j)$.
\end{enumerate}
\end{lemma}

\begin{IEEEproof}
(a) The bits $b_j=c_{1,j}\oplus d_j$ are i.i.d.\ uniform and independent of $(\mathbf c_0,\mathbf c_1)$, because $\mathbf d$ is. Given $\mathbf c_0$, the coordinate $x_j=(v_j+2b_j)\bmod 4$ is therefore uniform on $\{v_j,(v_j+2)\bmod 4\}$, independently over $j$. This pair equals $\{0,2\}$ if $v_j$ is even and $\{1,3\}$ if $v_j$ is odd, i.e., it equals $\{c_{0,j},c_{0,j}+2\}$. Hence $Y_j$ has density $\frac12[\varphi_{\sigma_j}(y-c_{0,j})+\varphi_{\sigma_j}(y-c_{0,j}-2)]=W_0(y|c_{0,j};\sigma_j)$, independently over $j$.

(b) Given $(\mathbf c_0,\mathbf d)$, the coordinate $x_j$ is one of the two points $a_j<a_j+2$ of $\{c_{0,j},c_{0,j}+2\}$, and which one is determined by $c_{1,j}\oplus d_j$ and $v_j$. The map $y\mapsto y-a_j$ turns $Y_j$ into $W_1(\cdot|b;\sigma_j)$, where $b$ equals $c_{1,j}$ or $c_{1,j}\oplus1$, a relabelling known to the receiver. Relabelling the input and applying an invertible map to the output change neither the capacity nor~\eqref{eq:E0}.
\end{IEEEproof}

\emph{Multistage decoder.} Given $\mathbf{y}_{[1:n]}$, the level-$0$ decoder outputs the maximum-likelihood (ML) codeword $\hat{\mathbf{c}}_0\in\mathcal{C}_0^{(n)}$ for the channel of Lemma~\ref{lem:levels}(a). It then forms $\hat{\mathbf{v}}=\mathrm{Enc}(\hat{\mathbf m}_0,\mathbf 0)_{[1:n]}$ with $\hat{\mathbf m}_0=(\hat{\mathbf c}_0)_{[1:N]}$ and outputs the ML codeword $\hat{\mathbf{c}}_1\in\mathcal{C}_1^{(n)}$ for the channel of Lemma~\ref{lem:levels}(b) with $\mathbf v$ replaced by $\hat{\mathbf v}$. We assume ideal error detection, as is standard in the analysis of IR protocols~\cite{CT2001}. A rateless session then terminates at the first prefix length at which both levels are decoded correctly. Hence the probability that the session has not terminated by length $n$ is at most the MSD error probability $\bar P_{\mathrm e}(n)$ at length $n$.

\begin{theorem}[Rateless random-coding bound]\label{thm:achiev}
For every $n\ge N$ and every noise sequence $\sigma_1,\dots,\sigma_n\in(0,\infty]$, the ensemble-average MSD error probability of $\mathfrak{E}(K,N)$ at prefix length $n$ satisfies
\begin{equation}\label{eq:achiev}
\bar P_{\mathrm e}(n)\le\min_{\rho\in[0,1]}2^{-\left(\sum_{j=1}^nE_0(\rho,W_0(\cdot;\sigma_j))-\rho K\right)}
+\min_{\rho\in[0,1]}2^{-\left(\sum_{j=1}^nE_0(\rho,W_1(\cdot;\sigma_j))-\rho N\right)}.
\end{equation}
The bound holds for every transmitted message and does not depend on how the noise levels are ordered. It concerns the recovery of the message $(\mathbf u_0,\mathbf u_1)$, i.e., of the codeword of $\mathcal S_n$ (equivalently, of the residue $\mathbf x\bmod4$ in the unconstrained setting), and it is an average over the ensemble and the dither $\mathbf d$, for a noise sequence that does not depend on them.
\end{theorem}

\begin{IEEEproof}
MSD fails only if $\hat{\mathbf c}_0\neq\mathbf c_0$, or if $\hat{\mathbf c}_0=\mathbf c_0$ and $\hat{\mathbf c}_1\neq\mathbf c_1$. On the event $\hat{\mathbf c}_0=\mathbf c_0$, the level-$1$ decoder coincides with a \emph{genie-aided} decoder that is given the true $\mathbf c_0$. Hence $\bar P_{\mathrm e}(n)\le\bar P_{0}+\bar P_{1}$, where $\bar P_0$ is the level-$0$ ML error probability and $\bar P_1$ is the error probability of the genie-aided level-$1$ ML decoder. We first reduce to the all-zero message, \emph{before} averaging over the ensemble. For fixed codes, $\mathcal C_0^{(n)}$ and $\mathcal C_1^{(n)}$ are linear, the level-$0$ channel of Lemma~\ref{lem:levels}(a) is a memoryless BIOS channel, as is, given $(\mathbf c_0,\mathbf d)$, the level-$1$ channel of Lemma~\ref{lem:levels}(b). For ML decoding of a linear code over a memoryless BIOS channel the error probability does not depend on the transmitted codeword~\cite[Ch.~4]{RU2008}. Hence, for every realization of the codes, $P_0$ and $P_1$ are the same for every message, and we may take $(\mathbf u_0,\mathbf u_1)=(\mathbf 0,\mathbf 0)$, which gives $\mathbf c_0=\mathbf 0$ and $\mathbf c_1=\mathbf 0$ for every realization of the ensemble. We then bound both terms with Gallager's method: for an ML decoder for a channel with density $p(\mathbf y|\mathbf w)$, a transmitted word $\mathbf w$, a set $\mathcal W'$ of competing words, $\rho\in[0,1]$ and $\alpha=1/(1+\rho)$,~\cite[Sec.~5.6]{Gallager1968}
\begin{equation}\label{eq:gal}
\Pr\{\text{ML error}\}\le\int p(\mathbf{y}|\mathbf w)^{\alpha}\Big(\sum_{\mathbf{w}'\in\mathcal W'}p(\mathbf{y}|\mathbf{w}')^{\alpha}\Big)^{\rho}d\mathbf y.
\end{equation}
and, if $\mathcal W'$ is random and independent of the noise, Jensen's inequality for the concave map $t\mapsto t^\rho$ moves the expectation inside the $\rho$-th power:
\begin{equation}\label{eq:galavg}
\mathbb E\big[\Pr\{\text{ML error}\}\big]\le\int p(\mathbf{y}|\mathbf w)^{\alpha}\Big(\mathbb E\Big[\sum_{\mathbf{w}'\in\mathcal W'}p(\mathbf{y}|\mathbf{w}')^{\alpha}\Big]\Big)^{\rho}d\mathbf y.
\end{equation}

For level~$0$, Lemma~\ref{lem:levels}(a) gives $p(\mathbf y|\mathbf c_0)=\prod_jW_0(y_j|c_{0,j};\sigma_j)$, and the competitors are $\mathbf c_0'=\mathbf c_0\oplus\mathbf e^{(0)}$ with $\mathbf e^{(0)}=(\mathbf e,\mathbf P_n^{\mathsf T}\mathbf e\bmod2)$ and $\mathbf e=\mathbf A_{\mathrm{pre}}(\mathbf u_0'\oplus\mathbf u_0)$, $\mathbf u_0'\neq\mathbf u_0$ (with $\mathbf u_0=\mathbf 0$ and $\mathbf c_0=\mathbf 0$ by the reduction above; we keep the general notation because the computation is the same). For a fixed $\mathbf u_0'$, the vector $\mathbf e$ is nonzero because $\mathbf A_{\mathrm{pre}}$ is injective, and it is uniform on $\mathbb F_2^N\setminus\{\mathbf 0\}$ because the law of $\mathbf A_{\mathrm{pre}}$ is invariant under $\mathbf A_{\mathrm{pre}}\mapsto\mathbf T\mathbf A_{\mathrm{pre}}$ for every invertible $\mathbf T$ and the invertible matrices act transitively on the nonzero vectors. Given $\mathbf e$, the parity bits $\mathbf s_t^{\mathsf T}\mathbf e$, $t=1,\dots,n-N$, are i.i.d.\ uniform, since the $\mathbf s_t$ are i.i.d.\ uniform and independent of $\mathbf A_{\mathrm{pre}}$. Consequently, every nonzero word $\mathbf w\in\mathbb F_2^n$ satisfies
\begin{equation}\label{eq:ewlaw}
\Pr\{\mathbf e^{(0)}=\mathbf w\}\le\frac{1}{(2^N-1)\,2^{n-N}}.
\end{equation}
With $\mathsf g_j(y)=\frac12\sum_{b\in\{0,1\}}W_0(y|b;\sigma_j)^\alpha$, extending the sum to all $\mathbf w\in\mathbb F_2^n$ yields
\[
\mathbb E\big[p(\mathbf y|\mathbf c_0\oplus\mathbf e^{(0)})^\alpha\big]\le\frac{2^n}{(2^N-1)2^{n-N}}\sum_{\mathbf w\in\mathbb F_2^n}2^{-n}\prod_{j=1}^nW_0(y_j|c_{0,j}\oplus w_j;\sigma_j)^\alpha=\frac{2^N}{2^N-1}\prod_{j=1}^n\mathsf g_j(y_j).
\]
Since there are $2^K-1$ competitors and $(2^K-1)\frac{2^N}{2^N-1}\le2^K$ for $K\le N$, substituting into~\eqref{eq:galavg} and using the product form gives $\bar P_0\le2^{\rho K}\prod_{j=1}^n\int W_0(y|c_{0,j};\sigma_j)^{\alpha}\,\mathsf g_j(y)^{\rho}\,dy$. The output symmetry $y\mapsto3-y$ of $W_0$ leaves $\mathsf g_j$ invariant; hence each factor is the same for $c_{0,j}=0$ and $c_{0,j}=1$; averaging the two values shows that it equals $\int \mathsf g_j(y)^{1+\rho}dy=2^{-E_0(\rho,W_0(\cdot;\sigma_j))}$ by~\eqref{eq:E0}. Minimizing over $\rho$ gives the first term of~\eqref{eq:achiev}.

Level~$1$ is handled in the same way, with one difference: there is no precode, and hence the systematic part of a competitor is deterministic. By Lemma~\ref{lem:levels}(b), the genie-aided decoder sees the memoryless channel $\prod_jW_1^{(j)}$, and the competitors are $\mathbf c_1'=\mathbf c_1\oplus(\mathbf e,\mathbf P_n^{\mathsf T}\mathbf e\bmod 2)$ with $\mathbf e=\mathbf u_1'\oplus\mathbf u_1\neq\mathbf 0$, whose last $n-N$ coordinates are i.i.d.\ uniform. With $\mathsf g_j(y)=\frac12\sum_bW_1^{(j)}(y|b)^\alpha$,
\[
\mathbb{E}\Big[\sum_{\mathbf{u}_1'\neq\mathbf{u}_1}p(\mathbf{y}|\mathbf{c}_1')^{\alpha}\Big]
=\sum_{\mathbf{e}\neq\mathbf{0}}\prod_{j\le N}W_1^{(j)}(y_j|c_{1,j}\oplus e_j)^\alpha\prod_{j>N}\mathsf g_j(y_j)
\le 2^N\prod_{j=1}^n\mathsf g_j(y_j),
\]
where the inequality adds the term $\mathbf e=\mathbf 0$ and uses $\sum_{\mathbf e\in\mathbb F_2^N}\prod_{j\le N}W_1^{(j)}(y_j|c_{1,j}\oplus e_j)^\alpha=\prod_{j\le N}2\mathsf g_j(y_j)$. Substituting into~\eqref{eq:galavg} and using the output symmetry of $W_1^{(j)}$ as for level~$0$ gives $\bar P_1\le2^{\rho N}\prod_j2^{-E_0(\rho,W_1^{(j)})}$. By Lemma~\ref{lem:levels}(b), $E_0(\rho,W_1^{(j)})=E_0(\rho,W_1(\cdot;\sigma_j))$, which gives the second term of~\eqref{eq:achiev}. Finally, an erased coordinate has $W_i(y|0)=W_i(y|1)$, and its factor equals $1=2^{-0}$; and since both exponents are sums over $j$, the bound does not depend on the order of the noise levels.
\end{IEEEproof}

\begin{corollary}[Universality]\label{cor:universal}
Fix $0<\sigma_{\min}\le\sigma_{\max}<\infty$, $\beta\ge1$ and $\Gamma\ge1$. Assume that $\sigma_j\in[\sigma_{\min},\sigma_{\max}]\cup\{\infty\}$ for all $j$, that $K\le N\le\Gamma K$, and that $n\le\beta(K+N)$. For every $\delta>0$ there is a constant $c>0$, which depends only on $(\delta,\sigma_{\min},\sigma_{\max},\beta,\Gamma)$, such that
\begin{equation}\label{eq:univ}
\sum_{j=1}^nC_0(\sigma_j)\ge(1+\delta)K\quad\text{and}\quad\sum_{j=1}^nC_1(\sigma_j)\ge(1+\delta)N
\quad\Longrightarrow\quad \bar P_{\mathrm e}(n)\le 2\cdot2^{-cK}.
\end{equation}
\end{corollary}

\begin{IEEEproof}
For a BIOS channel $W$, $E_0(0,W)=0$ and $\partial_\rho E_0(\rho,W)|_{\rho=0}=C(W)$~\cite[Thm.~5.6.3]{Gallager1968}. For $i\in\{0,1\}$ and $\sigma\in[\sigma_{\min},\sigma_{\max}]$, the integrand in~\eqref{eq:E0} is a smooth function of $(\rho,\sigma)$ dominated by an integrable Gaussian envelope. By dominated convergence, $\partial_\rho^2E_0(\rho,W_i(\cdot;\sigma))$ is continuous on the compact set $[0,1]\times[\sigma_{\min},\sigma_{\max}]$. Hence there is $B\in(0,\infty)$ with $\partial_\rho^2E_0\ge-B$ on this set. Taylor's theorem with remainder then gives
\begin{equation}\label{eq:E0quad}
E_0(\rho,W_i(\cdot;\sigma))\ge\rho\,C_i(\sigma)-\tfrac{B}{2}\rho^2,\qquad\rho\in[0,1],
\end{equation}
which also holds for erased coordinates, where $E_0=C_i=0$. Summing~\eqref{eq:E0quad} over $j\le n$ and using $n\le\beta(K+N)\le\beta(1+\Gamma)K$, we find that the level-$0$ exponent in~\eqref{eq:achiev} satisfies
\[
\sum_jE_0(\rho,W_0(\cdot;\sigma_j))-\rho K\ \ge\ \rho(1+\delta)K-\rho K-\tfrac{B}{2}\rho^2\beta(1+\Gamma)K=\rho K\big(\delta-\tfrac{B}{2}\rho\beta(1+\Gamma)\big),
\]
and, because $N\ge K$, the level-$1$ exponent is at least $\rho\delta N-\frac B2\rho^2\beta(1+\Gamma)K$, which obeys the same lower bound. The choice $\rho=\min\{1,\delta/(B\beta(1+\Gamma))\}$ makes $\frac B2\rho\beta(1+\Gamma)\le\frac\delta2$; both exponents are then at least $\rho\delta K/2$, and~\eqref{eq:achiev} gives~\eqref{eq:univ} with $c=\rho\delta/2$.
\end{IEEEproof}

Corollary~\ref{cor:universal} is the lattice analogue of the fountain-capacity result of~\cite{STV2007}. Without channel knowledge at the transmitter, decoding succeeds once the \emph{accumulated} level capacities exceed the level sizes, for any admissible noise, fading or erasure sequence, with error probability exponentially small in $K$.

\begin{remark}[What ``universal'' means]\label{rem:quant}
Theorem~\ref{thm:achiev} and Corollary~\ref{cor:universal} bound $\sup_{\boldsymbol\sigma}\mathbb E_{\mathcal C,\mathbf d}\,P_{\mathrm e}(\mathcal C,\mathbf d;\boldsymbol\sigma)$, the ensemble-average error probability for each noise sequence $\boldsymbol\sigma$ that is fixed independently of the random code and dither. Operationally, this is a guarantee for a code drawn from a public random seed shared by transmitter and receiver, as in~\cite{STV2007}. It does \emph{not} imply that one deterministic code $\mathcal C$ satisfies $\sup_{\boldsymbol\sigma}\mathbb E_{\mathbf d}P_{\mathrm e}(\mathcal C,\mathbf d;\boldsymbol\sigma)\to0$, and the order of $\sup$ and $\mathbb E_{\mathcal C}$ cannot be exchanged in general. Indeed, for a fixed linear code an erasure pattern chosen after the code can defeat it: pick a nonzero message difference $\mathbf v$ and erase exactly the coordinates whose functional does not vanish on $\mathbf v$; the messages $\mathbf u,\mathbf u+\mathbf v$ then induce the same output law, even with a known dither, and for $p$-ary codes (Section~\ref{sec:capacity}) about a fraction $1/p$ of the coordinates survives; hence the accumulated-capacity condition can hold while the error probability is at least $\frac12$. By the union bound and Markov's inequality, \eqref{eq:univ} does give a single code that is good for any \emph{finite} family of $M$ sequences when $2M\cdot2^{-cK}<1$.
\end{remark}

\begin{remark}[Rayleigh fading and the compact noise range]\label{rem:rayleigh}
The constant $c$ in~\eqref{eq:univ} depends on $(\sigma_{\min},\sigma_{\max})$, whereas Rayleigh fading gives $\sigma_j=\sigma/|h_r|$ with neither a deterministic upper nor a positive deterministic lower bound. The non-asymptotic bounds~\eqref{eq:achiev}, \eqref{eq:sparse} and~\eqref{eq:capbound} hold conditionally on every fading realization. With a finite number $r$ of blocks whose lengths grow linearly with $K$, the realization lies almost surely in $(0,\infty)^r$, Corollary~\ref{cor:universal} applies conditionally, and dominated convergence extends the vanishing error to the average over the fading law on the event that the accumulated level capacities exceed $(1+\delta)K$ and $(1+\delta)N$. We do not claim a uniform exponent over the unbounded fading law, nor a statement for a number of blocks that grows with $K$.
\end{remark}

\begin{remark}[Residues versus lattice points]\label{rem:residue}
In the unconstrained setting, Theorem~\ref{thm:achiev} and Corollary~\ref{cor:universal} (with the modulo-$2$ level channels) bound the probability $P_{\mathrm{res}}$ of decoding the residue $\mathbf x\bmod4$ wrongly. They do not bound the lattice-point error $P_{\mathrm{lat}}$ of the same family. Because $4\mathbf e_j\in\Lambda_n$ for every $j$, Proposition~\ref{prop:short}(ii) gives $P_{\mathrm{lat}}\ge1-(1-2Q(2/\sigma))^n$ on the AWGN channel, which tends to one as $n\to\infty$ for every fixed $\sigma>0$, however small $P_{\mathrm{res}}$ is. For the two-stage decoder followed by rounding to the nearest point of the decoded coset of $4\mathbb Z^n$ (Section~\ref{sec:construction}), and without erasures, $P_{\mathrm{lat}}\le P_{\mathrm{res}}+\sum_{j=1}^n2Q(2/\sigma_j)$. An erased coordinate makes lattice-point recovery impossible, because $\mathbf x$ and $\mathbf x+4\mathbf e_j$ induce the same observations; hence this bound and every lattice-point statement require $\sigma_j<\infty$ for all $j\le n$. Hence a fixed-alphabet family cannot be Poltyrev-good as $n\to\infty$, which requires a coarse spacing growing with $n$ (Corollary~\ref{cor:poltyrev}); the unconstrained results of this section concern residues and finite $n$.
\end{remark}

\subsection{Sparse LT Stages: Logarithmic Degree Suffices}\label{sec:sparse}
The ensemble $\mathfrak E(K,N)$ uses dense LT vectors, with about $N/2$ ones each, whereas the realization of Section~\ref{sec:construction} uses a few ones per vector. We now show that the coding theorem survives sparsification down to a logarithmic number of ones.

\begin{definition}[Sparse ensemble $\mathfrak{E}_\varpi(K,N)$]\label{def:sparse}
For $\varpi\in(0,\frac12]$, $\mathfrak E_\varpi(K,N)$ is defined as $\mathfrak E(K,N)$ in Definition~\ref{def:ens}, except that the entries of the LT vectors $\mathbf s_1,\mathbf s_2,\dots$ are i.i.d.\ Bernoulli$(\varpi)$. The expected LT degree is $\varpi N$, and $\mathfrak E_{1/2}(K,N)=\mathfrak E(K,N)$.
\end{definition}

To state the result we use, for a binary-input channel $W$, the Bhattacharyya parameter $\mathsf Z(W)=\int\sqrt{W(y|0)W(y|1)}\,dy$; we write $\mathsf Z_i(\sigma)=\mathsf Z(W_i(\cdot;\sigma))$, and let $h_{\mathrm b}$ be the binary entropy function.

\begin{theorem}[Rateless coding theorem for sparse LT stages]\label{thm:sparse}
Let $\upsilon\in(0,\frac12]$, let $w_\upsilon=\lceil\upsilon N\rceil$, and put $b=(1-2\varpi)^{w_\upsilon}$ and $\Xi_1=\sum_{j=N+1}^{n}\big(1-\mathsf Z_1(\sigma_j)\big)$. For every $n\ge N$, every noise sequence and all $\rho_0,\rho_1\in[0,1]$, the ensemble-average MSD error probability of $\mathfrak E_\varpi(K,N)$ satisfies
\begin{multline}\label{eq:sparse}
\bar P_{\mathrm e}(n)\le(1+b)^{n-N}\Big[2^{-\left(\sum_jE_0(\rho_0,W_0(\cdot;\sigma_j))-\rho_0K\right)}+2^{-\left(\sum_jE_0(\rho_1,W_1(\cdot;\sigma_j))-\rho_1N\right)}\Big]\\
+2^{K+1-N(1-h_{\mathrm b}(\upsilon))}+\sum_{w=1}^{w_\upsilon-1}\binom Nw\exp\Big(-\tfrac12\big(1-e^{-2\varpi w}\big)\,\Xi_1\Big).
\end{multline}
\end{theorem}

\begin{IEEEproof}
We follow the proof of Theorem~\ref{thm:achiev}; the only new ingredient is the bias of the parity bits. For $\mathbf e\in\mathbb F_2^N$ of Hamming weight $w\ge1$, the bits $\mathbf s_t^{\mathsf T}\mathbf e$, $t=1,2,\dots$, are i.i.d., and $\mathbf s_t^{\mathsf T}\mathbf e$ is the parity of $w$ independent Bernoulli$(\varpi)$ variables; hence
\begin{equation}\label{eq:bias}
\Pr\{\mathbf s_t^{\mathsf T}\mathbf e=1\}=\tfrac12\big(1-(1-2\varpi)^w\big),\qquad \max_{\beta\in\{0,1\}}\Pr\{\mathbf s_t^{\mathsf T}\mathbf e=\beta\}=\tfrac12\big(1+(1-2\varpi)^w\big).
\end{equation}
In particular, if $w\ge w_\upsilon$, each value has probability at most $\frac12(1+b)$, and the parity bits are close to uniform. At level~$0$, the error event of the ML decoder is contained in the union of two events: some competitor $\mathbf u_0'\neq\mathbf u_0$ has a precode difference $\mathbf e=\mathbf A_{\mathrm{pre}}(\mathbf u_0'\oplus\mathbf u_0)$ of weight below $w_\upsilon$; or the ML decoder prefers a competitor whose difference has weight at least $w_\upsilon$. As shown before~\eqref{eq:ewlaw}, $\mathbf e$ is uniform on $\mathbb F_2^N\setminus\{\mathbf0\}$. Since $\upsilon\le\frac12$, $\sum_{w<w_\upsilon}\binom Nw\le2^{Nh_{\mathrm b}(\upsilon)}$, and the union bound over the $2^K-1$ competitors bounds the probability of the first event by $2^K\cdot2^{Nh_{\mathrm b}(\upsilon)}/(2^N-1)\le2^{K+1-N(1-h_{\mathrm b}(\upsilon))}$. For the second event we apply~\eqref{eq:galavg} to the competitors with difference weight at least $w_\upsilon$. By~\eqref{eq:bias}, on every parity coordinate
\[
\mathbb E\big[W_0(y_j|c_{0,j}\oplus\mathbf s_t^{\mathsf T}\mathbf e;\sigma_j)^\alpha\big]\le\tfrac12(1+b)\sum_{\beta}W_0(y_j|\beta;\sigma_j)^\alpha=(1+b)\,\mathsf g_j(y_j),
\]
with $\mathsf g_j$ as in Theorem~\ref{thm:achiev}, independently over $t$. The systematic coordinates are treated exactly as before. Hence the expected sum of the $\alpha$-th powers of the likelihoods of these competitors is at most $2^K(1+b)^{n-N}\prod_j\mathsf g_j(y_j)$, and~\eqref{eq:galavg} gives the level-$0$ term of~\eqref{eq:sparse} with the factor $(1+b)^{\rho_0(n-N)}\le(1+b)^{n-N}$.

At level~$1$ there is no precode to exclude low-weight differences, and we split the competitors $\mathbf c_1'=\mathbf c_1\oplus(\mathbf e,\mathbf P_n^{\mathsf T}\mathbf e)$ according to the weight $w$ of $\mathbf e=\mathbf u_1'\oplus\mathbf u_1$. For $w\ge w_\upsilon$, \eqref{eq:bias} and the level-$1$ computation in the proof of Theorem~\ref{thm:achiev} give the second Gallager term of~\eqref{eq:sparse}, again with the factor $(1+b)^{n-N}$. For $1\le w<w_\upsilon$ we use instead the union bound and the Bhattacharyya bound: for a BIOS channel, the probability that the ML decoder prefers a fixed codeword $\mathbf c'$ to the transmitted $\mathbf c$ is at most $\prod_{j:c_j\neq c'_j}\mathsf Z(W^{(j)})$~\cite{Gallager1968}. Bounding the systematic factors by $1$ and noting that the parity coordinate $N+t$ differs iff $\mathbf s_t^{\mathsf T}\mathbf e=1$, independently over $t$, with probability $\pi_w=\frac12(1-(1-2\varpi)^w)\ge\frac12(1-e^{-2\varpi w})$ by~\eqref{eq:bias}, we obtain after averaging over the LT vectors
\[
\mathbb E\Big[\prod_{t:\,\mathbf s_t^{\mathsf T}\mathbf e=1}\mathsf Z_1(\sigma_{N+t})\Big]=\prod_{j>N}\big(1-\pi_w(1-\mathsf Z_1(\sigma_j))\big)\le\exp(-\pi_w\Xi_1),
\]
where we used $1-x\le e^{-x}$ and $\mathsf Z(W_1^{(j)})=\mathsf Z_1(\sigma_j)$ (Lemma~\ref{lem:levels}(b)). Summing over the $\binom Nw$ vectors of weight $w$ gives the last term of~\eqref{eq:sparse}.
\end{IEEEproof}

\begin{corollary}[Logarithmic LT degree]\label{cor:sparse}
Assume the conditions of Corollary~\ref{cor:universal}, and in addition $K\le(1-h_{\mathrm b}(\upsilon)-\zeta_0)N$ for some $\upsilon\in(0,\frac12]$ and $\zeta_0>0$. Let $c_{\min}=C_1(\sigma_{\max})>0$ and $\xi=\frac12\delta c_{\min}$. If $h_{\mathrm b}(\upsilon)<0.45\,\xi$ and the expected LT degree satisfies
\begin{equation}\label{eq:logdeg}
\varpi N\ \ge\ \max\Big\{\frac{(1+\epsilon_0)\ln N}{(1-e^{-1})\,\xi},\ \frac{\ln n}{2\upsilon}\Big\}
\end{equation}
for some $\epsilon_0>0$, then $\bar P_{\mathrm e}(n)\le 2N^{-\epsilon_0}+2^{-c'N}$ for all sufficiently large $N$, where $c'>0$ does not depend on $N$. Hence an LT stage whose expected degree grows logarithmically in $N$ preserves the universality of Corollary~\ref{cor:universal}, with a polynomially rather than exponentially decaying error probability.
\end{corollary}

\begin{IEEEproof}
We bound the four terms of~\eqref{eq:sparse} in turn, starting with the quantity $\Xi_1$ that controls the low-weight term. For a BIOS channel, $C\le\sqrt{1-\mathsf Z^2}$~\cite{Arikan2009}, hence $1-\mathsf Z\ge1-\sqrt{1-C^2}\ge C^2/2$. Every non-erased coordinate has $C_1(\sigma_j)\ge c_{\min}$, and hence $1-\mathsf Z_1(\sigma_j)\ge\frac{c_{\min}}2C_1(\sigma_j)$, an inequality that also holds trivially for erased coordinates. Since the systematic coordinates contribute at most $N$ to $\sum_jC_1(\sigma_j)\ge(1+\delta)N$, the parity coordinates contribute at least $\delta N$, and therefore $\Xi_1\ge\frac{c_{\min}}2\delta N=\xi N$.

The second part of~\eqref{eq:logdeg} gives $(n-N)b\le n e^{-2\varpi\upsilon N}\le1$, hence $(1+b)^{n-N}\le e$, and the two Gallager terms are at most $e$ times their values in the proof of Corollary~\ref{cor:universal}, i.e., at most $2e\cdot2^{-cK}$. The precode term satisfies $2^{K+1-N(1-h_{\mathrm b}(\upsilon))}\le2^{1-\zeta_0N}$ by assumption. It remains to bound the low-weight sum, which we split at $w=1/(2\varpi)$. For $w\le1/(2\varpi)$, the inequality $1-e^{-x}\ge(1-e^{-1})x$ on $[0,1]$ and the first part of~\eqref{eq:logdeg} bound the summand by $N^w\exp(-(1-e^{-1})\varpi w\,\xi N)\le N^{-\epsilon_0w}$, and these summands add up to at most $2N^{-\epsilon_0}$ when $N^{\epsilon_0}\ge2$. For $1/(2\varpi)<w<w_\upsilon$, we have $\frac12(1-e^{-2\varpi w})\ge\frac12(1-e^{-1})>0.316$, and these summands add up to at most $2^{Nh_{\mathrm b}(\upsilon)}e^{-0.316\,\xi N}=2^{-N(0.456\,\xi-h_{\mathrm b}(\upsilon))}$, which is exponentially small because $h_{\mathrm b}(\upsilon)<0.45\,\xi$. Collecting the exponentially small terms into $2^{-c'N}$ completes the proof.
\end{IEEEproof}

Corollary~\ref{cor:sparse} is the rateless-lattice counterpart of the classical fact that LT codes need an average degree of order $\log N$~\cite{Luby2002}. The logarithmic degree is driven by low-weight errors at level~$1$, which has no precode; as in Raptor codes~\cite{drshok}, a high-rate precode on level~$1$ should reduce it to a constant. The realization of Section~\ref{sec:construction} (constant degree, cycle removal) is not covered by Theorem~\ref{thm:sparse}, which treats i.i.d.\ entries.

\subsection{Converse and the Price of a Fixed Level Split}
\begin{theorem}[Converse under multistage decoding]\label{thm:converse}
Consider any two-level rateless scheme of the form~\eqref{eq:ensenc}, with arbitrary nested codes, independent uniform messages $\mathbf{U}_0\in\mathbb{F}_2^K$ and $\mathbf{U}_1\in\mathbb{F}_2^N$, and a uniform level-$1$ dither $\mathbf d$ independent of the messages. Let the level-$0$ decoder $\phi_0(\mathbf{Y}_{[1:n]})$ use only the channel output, and let the level-$1$ decoder $\phi_1(\mathbf{Y}_{[1:n]},\mathbf{U}_0,\mathbf{d})$ use, in addition, a genie that reveals $\mathbf{U}_0$ and $\mathbf{d}$. If $\Pr\{\phi_i\neq\mathbf{U}_i\}\le\varepsilon_i$ for $i=0,1$, then
\begin{equation}\label{eq:converse}
(1-\varepsilon_0)K-1\le\sum_{j=1}^nC_0(\sigma_j),\qquad(1-\varepsilon_1)N-1\le\sum_{j=1}^nC_1(\sigma_j).
\end{equation}
\end{theorem}

\begin{IEEEproof}
Fano's inequality gives $H(\mathbf U_0|\mathbf Y)\le1+\varepsilon_0K$, and hence $(1-\varepsilon_0)K-1\le I(\mathbf{U}_0;\mathbf{Y})$. By Lemma~\ref{lem:levels}(a), which uses only the dither and not the random ensemble, $\mathbf{U}_0\to\mathbf{c}_0\to\mathbf{Y}$ is a Markov chain and, given $\mathbf c_0$, the outputs $Y_j$ are independent with laws $W_0(\cdot|c_{0,j};\sigma_j)$. Therefore
\[
I(\mathbf{U}_0;\mathbf{Y})\le I(\mathbf{c}_0;\mathbf{Y})=h(\mathbf Y)-\sum_jh(Y_j|c_{0,j})\le\sum_j\big[h(Y_j)-h(Y_j|c_{0,j})\big]=\sum_jI(c_{0,j};Y_j)\le\sum_jC_0(\sigma_j),
\]
where erased coordinates contribute zero and may be dropped from $\mathbf Y$. At level~$1$, Fano's inequality for $\phi_1$ gives $(1-\varepsilon_1)N-1\le I(\mathbf{U}_1;\mathbf{Y}|\mathbf{U}_0,\mathbf{d})$. Given $(\mathbf U_0,\mathbf d)=(\mathbf u_0,\mathbf d)$, $\mathbf U_1\to\mathbf c_1\to\mathbf Y$ is a Markov chain, and by Lemma~\ref{lem:levels}(b) the channel from $\mathbf c_1$ to $\mathbf Y$ is memoryless with components $W_1^{(j)}$. The same chain of inequalities gives $I(\mathbf{U}_1;\mathbf{Y}|\mathbf{U}_0=\mathbf u_0,\mathbf{d})\le\sum_jC_1(\sigma_j)$. Averaging over $(\mathbf u_0,\mathbf d)$ completes the proof.
\end{IEEEproof}

\begin{remark}[Scope of Theorem~\ref{thm:converse}]\label{rem:convscope}
The level-$0$ bound uses that $\phi_0$ sees $\mathbf Y$ only. Lemma~\ref{lem:levels}(a), an average over the hidden dither, then makes the level-$0$ channel memoryless with components $W_0$. It is therefore a converse for the MSD architecture analysed here, whose first stage uses the memoryless level-$0$ metric and ignores both the dither and the code constraints of level~$1$; it is not a converse for every decoder that merely decides level~$0$ first. A first stage that knows $\mathbf d$ sees a channel with memory: for $N=K=1$, $n=2$, $\mathbf s_1=1$, zero dither and level-$0$ bit $0$, the transmitted pair is $(0,0)$ or $(2,2)$, and the output density $\frac12\varphi_\sigma(y_1)\varphi_\sigma(y_2)+\frac12\varphi_\sigma(y_1-2)\varphi_\sigma(y_2-2)$ is not $W_0(y_1|0)W_0(y_2|0)$. A joint estimate from $(\mathbf Y,\mathbf d)$ is not restricted by~\eqref{eq:converse} either; the ``price of a fixed split'' below is the price of this first-stage restriction, not of the decision order alone.
\end{remark}

Theorems~\ref{thm:achiev} and~\ref{thm:converse} together determine the rateless performance of two-level lattices under MSD. Let $\chi=K/N\in(0,1]$ denote the \emph{level split}. On an AWGN channel with fixed $\sigma$, let $n^\star_{\mathrm{MSD}}(\sigma)$ be the smallest prefix length at which MSD succeeds with vanishing error probability.

\begin{corollary}[Price of a fixed level split]\label{cor:price}
Let $K,N\to\infty$ with $\chi=K/N$ fixed, and let $\sigma<\infty$ be fixed. Then
\begin{equation}\label{eq:nstar}
\lim_{N\to\infty}\frac{n^\star_{\mathrm{MSD}}(\sigma)}{N}=\max\Big\{\frac{\chi}{C_0(\sigma)},\frac{1}{C_1(\sigma)}\Big\}.
\end{equation}
Consequently, the asymptotic rateless efficiency under MSD, defined as the achieved rate $(K+N)/n^\star_{\mathrm{MSD}}$ divided by $C_{\mathrm X}(\sigma)$, is
\begin{equation}\label{eq:eta}
\eta_{\mathrm{MSD}}(\sigma;\chi)=\frac{(1+\chi)\min\{C_0(\sigma)/\chi,\ C_1(\sigma)\}}{C_0(\sigma)+C_1(\sigma)}\ \le 1,
\end{equation}
with equality iff $\chi=C_0(\sigma)/C_1(\sigma)$. For $\chi>C_0/C_1$ (level $0$ limits decoding) the loss is $1-\eta_{\mathrm{MSD}}=\frac{\chi C_1-C_0}{\chi(C_0+C_1)}$. For $\chi<C_0/C_1$ (level $1$ limits decoding) it is $\frac{C_0-\chi C_1}{C_0+C_1}$.
\end{corollary}

\begin{IEEEproof}
Write $n^\star_\infty=N\max\{\chi/C_0,1/C_1\}$. \emph{Achievability:} for $\delta>0$ and $n=\lceil(1+\delta)n^\star_\infty\rceil$ we have $nC_0\ge(1+\delta)K$ and $nC_1\ge(1+\delta)N$. Corollary~\ref{cor:universal} (with $\Gamma=1/\chi$, $\beta=(1+\delta)n^\star_\infty/(K+N)+1$) shows that the ensemble-average error probability tends to $0$, and therefore some code in the ensemble achieves this. \emph{Converse:} if $n<(1-\delta)n^\star_\infty$, then $nC_0<(1-\delta)K$ or $nC_1<(1-\delta)N$, and Theorem~\ref{thm:converse} forces $\varepsilon_0\ge\delta-1/K$ or $\varepsilon_1\ge\delta-1/N$. Since $\delta$ is arbitrary, \eqref{eq:nstar} follows. For~\eqref{eq:eta}, divide $K+N=N(1+\chi)$ by~\eqref{eq:nstar} and by $C_{\mathrm X}=C_0+C_1$, and use $1/\max\{a,b\}=\min\{1/a,1/b\}$. Equality in~\eqref{eq:eta} holds iff both terms of the minimum agree, i.e., iff $\chi=C_0/C_1$. The two loss expressions follow by direct substitution.
\end{IEEEproof}

The optimal split $C_0(\sigma)/C_1(\sigma)$ is not constant (Fig.~\ref{fig:theory}(a)). It tends to $1$ as $\mathrm{SNR}\to\infty$, and to $\frac14$ as $\mathrm{SNR}\to0$: since $X=B_0+2B_1$ with independent uniform bits of variances $\frac14$ and $1$, the low-SNR expansion of mutual information (or the I-MMSE relation~\cite{GSV2005}) gives $C_{\mathrm X}(\sigma)=\frac{5/4}{2\sigma^2\ln2}+o(\sigma^{-2})$ and $C_1(\sigma)=\frac{1}{2\sigma^2\ln2}+o(\sigma^{-2})$, hence $C_0/C_1\to\frac14$. The second-order terms give $C_0/C_1=\frac14-\frac18\mathrm{SNR}+o(\mathrm{SNR})$, and hence the ratio approaches $\frac14$ from below. It is not monotone: by numerical integration (Gauss--Hermite quadrature) it decreases to a minimum $r_{\min}\approx0.18624$ near $1.21$~dB and then increases to $1$ (it equals $0.249$, $0.189$, $0.211$ and $0.280$ at $-20$, $0$, $4$ and $6$~dB). Numerically, therefore, a split $\chi\in(r_{\min},\frac14)$ is lossless at two positive SNRs (e.g., $\chi=0.2$ at about $-2.0$ and $3.4$~dB), the split $r_{\min}$ at one, a split $\chi\in[\frac14,1)$ at one (e.g., $\chi=\frac14$ at about $5.3$~dB), and a split $\chi<r_{\min}$ at none. We verified this shape on a fine SNR grid but have not proved it analytically. In any case, \emph{no} fixed split is efficient at all SNRs. The next result shows that this loss comes from stage-by-stage decoding and not from the lattice itself.

\begin{proposition}[Outer bound for any decoder]\label{prop:outer}
Consider any scheme as in Theorem~\ref{thm:converse} whose symbols are uniform on $\{0,1,2,3\}$ given the dither (cf.\ Theorem~\ref{thm:structure}(e)), and any decoder $\phi(\mathbf{Y},\mathbf d)$ of $(\mathbf{U}_0,\mathbf{U}_1)$, which may use the public dither, with error probability at most $\varepsilon$. Then
\begin{equation}\label{eq:outer}
(1-\varepsilon)(K+N)-1\le\sum_{j=1}^n\big[C_0(\sigma_j)+C_1(\sigma_j)\big],\qquad (1-\varepsilon)N-1\le\sum_{j=1}^nC_1(\sigma_j).
\end{equation}
Consequently, the efficiency of any decoder satisfies $\eta(\sigma;\chi)\le\min\{1,(1+\chi)C_1/(C_0+C_1)\}$, which equals $1$ for every $\chi\ge C_0/C_1$.
\end{proposition}

\begin{IEEEproof}
Since $\mathbf d$ is independent of the messages, Fano's inequality gives $(1-\varepsilon)(K+N)-1\le I(\mathbf{U}_0,\mathbf{U}_1;\mathbf{Y}\,|\,\mathbf d)$. Given $\mathbf d$, the channel from $\mathbf X$ to $\mathbf Y$ is memoryless, and the chain of inequalities in the proof of Theorem~\ref{thm:converse} gives $I(\mathbf{U}_0,\mathbf{U}_1;\mathbf{Y}|\mathbf d=\mathbf d')\le\sum_jI(X_j;Y_j|\mathbf d=\mathbf d')$. Since $X_j$ is uniform given $\mathbf d=\mathbf d'$, each term equals $C_{\mathrm X}(\sigma_j)=C_0(\sigma_j)+C_1(\sigma_j)$, and averaging over $\mathbf d'$ gives the first inequality. For the second inequality, the decoder $\phi_1(\mathbf Y,\mathbf U_0,\mathbf d)=[\phi(\mathbf Y,\mathbf d)]_1$ (which ignores $\mathbf U_0$) errs with probability at most $\varepsilon$, and the level-$1$ part of Theorem~\ref{thm:converse}, which allows the decoder to know $(\mathbf U_0,\mathbf d)$, applies. For the efficiency bound, set $\varepsilon\to0$ and $n=(K+N)/(\eta C_{\mathrm X})$ in~\eqref{eq:outer} with $\sigma_j=\sigma$; the first inequality gives $\eta\le1$ and the second gives $\eta\le(1+\chi)C_1/C_{\mathrm X}$.
\end{IEEEproof}

\begin{remark}[Asymptotic benchmarks versus finite-session bounds]\label{rem:bench}
Theorem~\ref{thm:converse} and Proposition~\ref{prop:outer} are fixed-length Fano inequalities with an error allowance. Letting $\varepsilon\to0$ gives the limiting rate conditions $K\le\sum_jC_0$, $N\le\sum_jC_1$ and $K+N\le\sum_jC_{\mathrm X}$. In Section~\ref{sec:numerics} we turn them into \emph{asymptotic capacity-crossing benchmarks} by stopping at the first round at which the limiting conditions hold. They are not finite-session throughput upper bounds: a session with a decoding-dependent stopping time can exceed them, in particular at high SNR, where the uncoded level~$1$ at $n=N$ is often decoded correctly although $NC_1(\sigma)<N$. We do not derive a finite-session converse.
\end{remark}

In the level-$0$-limited regime $\chi>C_0/C_1$, Proposition~\ref{prop:outer} leaves room for efficiency $1$, whereas Corollary~\ref{cor:price} shows that MSD loses $\frac{\chi C_1-C_0}{\chi C_{\mathrm X}}$. A joint (or iterative) decoder can use the slack $\sum_jC_1(\sigma_j)-N$ of level~$1$ to help level~$0$. Section~\ref{sec:mlcap} shows that joint ML decoding achieves~\eqref{eq:outer} for a dense ensemble with uniform $\mathbb Z_4$ coefficients and $K\le N$ (Theorem~\ref{thm:mlcap}); for the binary-coefficient ensembles of this section, joint achievability is not proved.

\subsection{Throughput-Optimal Split Under Block Fading}\label{sec:split}
Consider IR transmission in rounds. Round $r$ carries $L_r$ new dimensions with gain $|h_r|^2\sim\mathrm{Exp}(1)$, i.i.d.\ over rounds, and each round is erased independently with probability~$p_{\mathrm{er}}$. A session stops at the first round after which decoding succeeds, or after $r_{\max}$ rounds (outage). By the renewal-reward theorem~\cite{CT2001}, the long-term throughput in bits per real dimension is
\begin{equation}\label{eq:thr}
\mathsf{T}=\frac{(K+N)\,(1-P_{\mathrm{out}})}{\mathbb{E}[\text{dimensions used per session}]}.
\end{equation}
By Theorems~\ref{thm:achiev}--\ref{thm:converse}, the asymptotic MSD benchmark of~\eqref{eq:thr} (Remark~\ref{rem:bench}; we also call it the MSD limit) is obtained by stopping at the first round $r$ with
\begin{equation}\label{eq:msdstop}
\sum_{r'\le r}o_{r'}L_{r'}\,C_0(\sigma/|h_{r'}|)\ge K\qquad\text{and}\qquad\sum_{r'\le r}o_{r'}L_{r'}\,C_1(\sigma/|h_{r'}|)\ge N,
\end{equation}
where $o_{r'}=\mathbf 1\{\text{round }r'\text{ is not erased}\}$.
Replacing these two conditions by those of Proposition~\ref{prop:outer} gives the asymptotic outer benchmark. Since $C_1(\sigma)<1$ for every finite $\sigma$, the level-$1$ condition can never hold at $n=N$; numerically we evaluate it in the form $\sum_{r'\le r}o_{r'}L_{r'}(1-C_1)\le\sum_{r'\le r}o_{r'}L_{r'}-N$, with the deficit $1-C_1$ computed directly to prevent rounding from producing a spurious first-round stop. Maximizing the MSD limit over $\chi$ for a fixed $K+N$ yields the throughput-optimal split $\chi^\star(\overline{\mathrm{SNR}})$, shown in Fig.~\ref{fig:theory}(b). In the QC construction of Section~\ref{sec:construction}, $\chi=1-(2Z-\kappa)/(w_rZ)\approx1-2/w_r$, and therefore the split is tuned by the row weight $w_r$ alone.

\section{Capacity Results and a Provable Advantage}\label{sec:capacity}

The split price of Section~\ref{sec:theorems} is not intrinsic to rateless lattices: single-level $p$-ary rateless lattices achieve the accumulated capacity of uniform PAM, within $\frac12\log\frac{\pi e}{6}\approx0.254$ bits per dimension of the Gaussian IR capacity, a Gaussian labeling closes this gap (Section~\ref{sec:shaped}), and multilevel rateless lattices achieve capacity under joint decoding (Section~\ref{sec:mlcap}). Section~\ref{sec:vsrep} compares with repetition and layering. Throughout, ``for every noise sequence'' is meant in the sense of Remark~\ref{rem:quant}: the error bounds are averages over a random code and dither, for each noise sequence fixed independently of them.

\subsection{$p$-ary Rateless Lattices Achieve the Accumulated Capacity}
Let $p$ be a prime and $k_p\ge1$. Let $\mathbf{a}_1,\mathbf{a}_2,\dots$ be i.i.d.\ uniform on $\mathbb{F}_p^{k_p}$, let $\mathbf{A}_n\in\mathbb{F}_p^{n\times k_p}$ have rows $\mathbf{a}_1^{\mathsf T},\dots,\mathbf{a}_n^{\mathsf T}$, and put $\mathcal{C}^{(p)}_n=\{\mathbf{A}_n\mathbf{u}:\mathbf{u}\in\mathbb{F}_p^{k_p}\}, \Lambda^{(p)}_n=\mathcal{C}^{(p)}_n+p\mathbb{Z}^n$, where $\mathbb F_p$ is identified with $\{0,\dots,p-1\}$. Every new dimension is a fresh random linear functional of the message, as in a dense $p$-ary fountain code.

\begin{lemma}\label{lem:pary}
$(\Lambda^{(p)}_n)_{n\ge1}$ satisfies $p\mathbb Z^n\subseteq\Lambda^{(p)}_n$ and $\pi_n(\Lambda^{(p)}_{n'})=\Lambda^{(p)}_n$ for all $n'\ge n$. On the event that $\mathbf{A}_{n_0}$ has rank $k_p$, which has probability at least $1-p^{k_p-n_0}/(p-1)$, the sequence $(\Lambda^{(p)}_n)_{n\ge n_0}$ is a $p$-ary rateless lattice with codebooks of size $p^{k_p}$, $\mathrm{vol}(\Lambda^{(p)}_n)=p^{\,n-k_p}$, and an encoder with $O(k_p)$ operations per symbol.
\end{lemma}

\begin{IEEEproof}
$\Lambda^{(p)}_n$ is a union of cosets of $p\mathbb Z^n$, and it is closed under addition because $\mathcal C^{(p)}_n$ is linear; hence it is a lattice containing $p\mathbb Z^n$. The first $n$ rows of $\mathbf A_{n'}$ are the rows of $\mathbf A_n$, and hence $\pi_n(\mathcal C^{(p)}_{n'})=\mathcal C^{(p)}_n$ and hence $\pi_n(\Lambda^{(p)}_{n'})=\Lambda^{(p)}_n$. If $\mathbf A_{n_0}$ has rank $k_p$, then every $\mathbf A_n$ with $n\ge n_0$ also has rank $k_p$, because appending rows cannot decrease the rank. Then $\mathbf u\mapsto\mathbf A_n\mathbf u$ is injective, the codebook $\Lambda^{(p)}_n\cap\{0,\dots,p-1\}^n=\mathcal C^{(p)}_n$ has $p^{k_p}$ points, and $\mathrm{vol}(\Lambda^{(p)}_n)=[\mathbb Z^n:\Lambda^{(p)}_n]=p^n/p^{k_p}$. For the rank bound, $\mathbf A_{n_0}$ is rank-deficient iff $\mathbf A_{n_0}\mathbf u=\mathbf 0$ for some $\mathbf u\neq\mathbf 0$, and it suffices to check one $\mathbf u$ per line through the origin. There are $(p^{k_p}-1)/(p-1)$ lines and $\Pr\{\mathbf A_{n_0}\mathbf u=\mathbf 0\}=p^{-n_0}$ for each $\mathbf u\neq\mathbf 0$. The union bound gives the claim. Each symbol $\mathbf a_j^{\mathsf T}\mathbf u$ costs $k_p$ multiply-adds.
\end{IEEEproof}

The full-rank event $\mathcal F_{n_0}=\{\mathrm{rank}\,\mathbf A_{n_0}=k_p\}$ matters whenever messages, rather than codewords, are recovered: on its complement two messages share a codeword, which happens necessarily if $n<k_p$. The error bounds below are for the unconditioned ensemble; conditionally on $\mathcal F_{n_0}$ we use $\mathbb E[P_{\mathrm e}\,|\,\mathcal F_{n_0}]\le\mathbb E[P_{\mathrm e}]/\Pr(\mathcal F_{n_0})\le2\,\mathbb E[P_{\mathrm e}]$ for $n_0\ge k_p+1$, since then $\Pr(\mathcal F_{n_0})\ge1-p^{-1}/(p-1)\ge\frac12$.

\emph{Channels.} In the power-constrained case, symbol $x\in\{0,\dots,p-1\}$ is sent as $\Delta_p(x-\frac{p-1}2)$, where $\Delta_p^2=12P/(p^2-1)$, such that uniform $p$-PAM has power $P$. The receiver sees $y_j=\Delta_p(x_j-\frac{p-1}2)+\sigma_jz_j$, with $\sigma_j=\infty$ allowed for erasures, and $\mathrm{SNR}_j=P/\sigma_j^2$. Let $I_p(\sigma)$ and $E_0^{(p)}(\rho,\sigma)$ be the mutual information and the Gallager function ($p$-ary analogue of~\eqref{eq:E0}, with weights $1/p$) of this channel with uniform input. In the unconstrained case, the receiver reduces modulo $p$: $W^{\bmod}_\sigma(y|x)=\sum_{l\in\mathbb Z}\varphi_\sigma(y-x-lp)$ for $y\in[0,p)$, with uniform-input mutual information $C^{\bmod}_p(\sigma)$ and Gallager function $E_0^{\bmod}$. A public uniform dither $\mathbf D\in\mathbb F_p^n$ is added before transmission: the codeword of $\mathbf u$ is $\mathbf x_{\mathbf u}=(\mathbf A_n\mathbf u+\mathbf D)\bmod p$.

\begin{theorem}[Rateless lattices achieve the accumulated capacity]\label{thm:capacity}
For every $n$ and every noise sequence $\sigma_1,\dots,\sigma_n\in(0,\infty]$, the ensemble-average ML error probability of the dithered $p$-ary rateless lattice code at prefix length $n$ satisfies
\begin{equation}\label{eq:capbound}
\bar P_{\mathrm e}(n)\le\min_{\rho\in[0,1]}2^{-\left(\sum_{j=1}^nE_0^{(p)}(\rho,\sigma_j)-\rho k_p\log p\right)}.
\end{equation}
In the unconstrained setting ($\mathbf x\in\mathbf x_{\mathbf u}+p\mathbb Z^n$), the receiver first decodes $\mathbf u$ by ML for the modulo-$p$ channel; the probability that $\mathbf u$ is decoded wrongly, which upper-bounds the probability of a wrong residue $\mathbf x\bmod p$ (the two events coincide on the full-rank event of Lemma~\ref{lem:pary}), is at most the right-hand side of~\eqref{eq:capbound} with $E_0^{\bmod}$ in place of $E_0^{(p)}$, and erasures are allowed. If, in addition, $\sigma_j<\infty$ for every $j\le n$, the receiver recovers the lattice point itself by rounding, and its lattice-point error probability is at most this quantity plus $\sum_{j=1}^n2Q\big(\frac{p}{2\sigma_j}\big)$. (With an erased coordinate $j$, the points $\mathbf x$ and $\mathbf x+p\mathbf e_j$ of the same coset induce identical observations, and no decoder recovers the lattice point with error below $\frac12$.) Under the conditions of Corollary~\ref{cor:universal} on $(\sigma_j)$ and $n$, there is $c>0$ such that $\bar P_{\mathrm e}(n)\le2^{-ck_p}$ as soon as
\begin{equation}\label{eq:capcond}
\sum_{j=1}^nI_p(\sigma_j)\ge(1+\delta)\,k_p\log p\qquad\Big(\text{resp. }\sum_{j=1}^nC^{\bmod}_p(\sigma_j)\ge(1+\delta)\,k_p\log p\Big).
\end{equation}
Conversely, any code with $p^{k_p}$ equiprobable messages, possibly drawn together with public randomness $\mathsf S$ (code choice, dither) that is independent of the message and the noise and known to the decoder, whose symbols are uniform on the $p$-PAM alphabet and that is decoded from $(\mathbf Y,\mathsf S)$ with error at most $\varepsilon$ satisfies $(1-\varepsilon)k_p\log p-1\le\sum_jI_p(\sigma_j)$.
\end{theorem}

\begin{IEEEproof}
The proof rests on the pairwise independence of the codewords. Fix $\mathbf u\neq\mathbf u'$. For each $j$, $x_j(\mathbf u')-x_j(\mathbf u)=\mathbf a_j^{\mathsf T}(\mathbf u'-\mathbf u)\bmod p$ is uniform on $\mathbb F_p$, because $\mathbf a_j$ is uniform and $\mathbf u'-\mathbf u\neq\mathbf0$. These differences are independent over $j$, because the rows are independent. The transmitted symbol $x_j(\mathbf u)$ is uniform and independent of $(\mathbf a_1,\mathbf a_2,\dots)$, because $D_j$ is. Hence, for each $j$, $(x_j(\mathbf u),x_j(\mathbf u'))$ is a pair of independent uniform symbols, and the pairs are independent over~$j$. Now let $\alpha=1/(1+\rho)$ and $\mathsf g_j(y)=\frac1p\sum_{x}W_j(y|x)^\alpha$, where $W_j$ is the channel of coordinate $j$. Conditioned on the transmitted word $\mathbf x(\mathbf u)$, each competitor $\mathbf x(\mathbf u')$ is i.i.d.\ uniform and independent of $\mathbf x(\mathbf u)$, and hence $\mathbb E[W^n(\mathbf y|\mathbf x(\mathbf u'))^\alpha\,|\,\mathbf x(\mathbf u)]=\prod_j\mathsf g_j(y_j)$. Inserting this into~\eqref{eq:galavg} with $p^{k_p}-1\le p^{k_p}$ competitors gives
\[
\bar P_{\mathrm e}\le p^{\rho k_p}\,\mathbb E\Big[\prod_j\int W_j(y|x_j(\mathbf u))^\alpha\,\mathsf g_j(y)^\rho\,dy\Big]=2^{\rho k_p\log p}\prod_j\int\mathsf g_j(y)^{1+\rho}dy,
\]
where the equality averages over the i.i.d.\ uniform symbols $x_j(\mathbf u)$. Since $\int\mathsf g_j^{1+\rho}=2^{-E_0^{(p)}(\rho,\sigma_j)}$, this is~\eqref{eq:capbound}. Only pairwise independence was used, as in~\cite[Thm.~5.6.2]{Gallager1968}, and the channels may vary with $j$. In the unconstrained case, the received $y_j\bmod p$ depends on $\mathbf x$ only through $\mathbf x\bmod p=\mathbf x_{\mathbf u}$ and has the law $W^{\bmod}_{\sigma_j}(\cdot|x_{\mathbf u,j})$, and therefore the argument above applies verbatim. If $\mathbf u$ is decoded correctly, the receiver outputs $\hat{\mathbf x}=\mathbf x_{\mathbf u}+p\lfloor(\mathbf y-\mathbf x_{\mathbf u})/p\rceil$, which equals the transmitted point unless $|\sigma_jz_j|\ge p/2$ for some $j$. The union bound over $j$ gives the extra term.

Condition~\eqref{eq:capcond} follows as in Corollary~\ref{cor:universal}: for fixed $p$, $E_0^{(p)}(0,\sigma)=0$, $\partial_\rho E_0^{(p)}(0,\sigma)=I_p(\sigma)$, and $\partial^2_\rho E_0^{(p)}$ is bounded below uniformly on $[0,1]\times[\sigma_{\min},\sigma_{\max}]$, exactly as in the derivation of~\eqref{eq:E0quad}. The remainder of the proof of Corollary~\ref{cor:universal} then applies with $K$ replaced by $k_p\log p$. The modulo-$p$ case is identical. Finally, for the converse, $\mathsf S$ is independent of $\mathbf U$, and Fano's inequality gives $(1-\varepsilon)k_p\log p-1\le I(\mathbf U;\mathbf Y,\mathsf S)=I(\mathbf U;\mathbf Y|\mathsf S)\le I(\mathbf X;\mathbf Y|\mathsf S)$. Given $\mathsf S$, the channel is memoryless with noise independent of $\mathsf S$, and $I(\mathbf X;\mathbf Y|\mathsf S)\le\sum_jI(X_j;Y_j|\mathsf S)$ (as in the proof of Theorem~\ref{thm:converse}). Since mutual information is concave in the input law for a fixed channel, $I(X_j;Y_j|\mathsf S)\le I(X_j;Y_j)=I_p(\sigma_j)$, because $X_j$ is uniform. (The conditioning matters: for $X=(U+D)\bmod p$ with independent uniform $U$ and $D$, $I(U;X)=0$ but $I(U;X,D)=\log p$.)
\end{IEEEproof}

Theorem~\ref{thm:capacity} has two corollaries that are relevant to the scenarios of this paper. The first one concerns the unconstrained channel. There, the integer input incurs a small \emph{grid loss}: by the decomposition $Y=\lfloor Y\rfloor+(Y\bmod1)$, in which $\lfloor Y\rfloor$ is uniform on $\mathbb Z_p$ and independent of $Y\bmod 1=Z\bmod1$,
\begin{equation}\label{eq:cmod}
C^{\bmod}_p(\sigma)=\log p-h(\sigma Z\bmod p)+h(\sigma Z\bmod 1)\ \ge\ \log p-\tfrac12\log(2\pi e\sigma^2)-\epsilon_{\mathrm g}(\sigma),
\end{equation}
where $\epsilon_{\mathrm g}(\sigma)=-h(\sigma Z\bmod1)\ge0$. Here we used $h(\sigma Z\bmod p)\le h(\sigma Z)=\frac12\log(2\pi e\sigma^2)$: reducing modulo $p$ is a piecewise translation, and $h(\sigma Z)=h(\sigma Z\bmod p)+H(\lfloor\sigma Z/p\rfloor\,|\,\sigma Z\bmod p)$. By the Poisson summation formula, the density of $\sigma Z\bmod 1$ is $1+2\sum_{m\ge1}e^{-2\pi^2m^2\sigma^2}\cos(2\pi my)\le1+\eta_\sigma$ with $\eta_\sigma=2\sum_{m\ge1}e^{-2\pi^2m^2\sigma^2}$, and hence $\epsilon_{\mathrm g}(\sigma)\le\log(1+\eta_\sigma)$. For $\sigma\ge1$ this is below $10^{-8}$ bits.

\begin{corollary}[Rateless Poltyrev inequality]\label{cor:poltyrev}
Let $\sigma_1,\dots,\sigma_n<\infty$ satisfy the conditions of Corollary~\ref{cor:universal}, let $\gamma>0$ be a scale, and let $V_{\mathrm{nom}}(n)=\gamma^np^{\,n-k_p}$ be the \emph{nominal volume}, a deterministic quantity that equals $\mathrm{vol}(\gamma\Lambda^{(p)}_n)$ on the full-rank event $\mathcal F_n$ of Lemma~\ref{lem:pary}. At every prefix $n$ that satisfies the \emph{accumulated Poltyrev condition}
\begin{equation}\label{eq:accpol}
\frac{V_{\mathrm{nom}}(n)^{2/n}}{2\pi e\,\big(\prod_{j=1}^n\sigma_j^2\big)^{1/n}}\ \ge\ p^{2\delta k_p/n}\cdot2^{\frac2n\sum_j\epsilon_{\mathrm g}(\sigma_j/\gamma)},
\end{equation}
the scaled dithered rateless lattice $\gamma\Lambda^{(p)}_n$ has, averaged over the ensemble and the dither, a message (and hence residue) error probability at most $2^{-ck_p}$ and a lattice-point error probability at most $2^{-ck_p}+\sum_j2Q(\frac{\gamma p}{2\sigma_j})$, with $c>0$ depending on $(p,\gamma,\delta,\sigma_{\min},\sigma_{\max},\beta)$. Conditionally on $\mathcal F_{n_0}$ with $k_p+1\le n_0\le n$, the same holds with $2^{-ck_p}$ replaced by $2\cdot2^{-ck_p}$, and the rounding term is unchanged. For constant $\sigma$ and on the full-rank event, the left-hand side of~\eqref{eq:accpol} is the VNR of $\gamma\Lambda^{(p)}_n$; for varying $\sigma_j$, the relevant noise level is the \emph{geometric mean} of the noise variances.
\end{corollary}

\begin{IEEEproof}
Scaling the lattice by $\gamma$ is equivalent to keeping $\Lambda^{(p)}_n$ and replacing $\sigma_j$ by $\sigma_j/\gamma$. By~\eqref{eq:cmod},
\[
\sum_jC^{\bmod}_p(\sigma_j/\gamma)\ge n\log p-\tfrac n2\log\frac{2\pi e\,\bar\sigma^2}{\gamma^2}-\sum_j\epsilon_{\mathrm g}(\sigma_j/\gamma),\qquad\bar\sigma^2=\Big(\prod_j\sigma_j^2\Big)^{1/n}.
\]
Since $V_{\mathrm{nom}}(n)^{2/n}=\gamma^2p^{2(n-k_p)/n}$, the right-hand side equals $k_p\log p+\frac n2\log\frac{V_{\mathrm{nom}}(n)^{2/n}}{2\pi e\bar\sigma^2}-\sum_j\epsilon_{\mathrm g}(\sigma_j/\gamma)$, a deterministic quantity. Under~\eqref{eq:accpol} it is at least $(1+\delta)k_p\log p$, which is the unconstrained condition in~\eqref{eq:capcond}. Theorem~\ref{thm:capacity} gives the unconditioned bounds. For the conditional statement, $\mathbb E[P_{\mathrm e}\,|\,\mathcal F_{n_0}]\le\mathbb E[P_{\mathrm e}]/\Pr(\mathcal F_{n_0})\le2\,\mathbb E[P_{\mathrm e}]$ for the message term (Lemma~\ref{lem:pary}), whereas the rounding term bounds the probability that some $|\sigma_jz_j|\ge\gamma p/2$, which holds for every fixed code.
\end{IEEEproof}

Corollary~\ref{cor:poltyrev} is a finite-parameter inequality: the ensemble is reliable at every prefix whose geometric-mean VNR exceeds $1$ by the slack factor in~\eqref{eq:accpol}, up to the rounding term. An asymptotic rateless Poltyrev theorem requires a \emph{joint} limit, which we do not carry out. (i) The rounding term satisfies $\sum_j2Q(\frac{\gamma p}{2\sigma_j})\le2nQ(\frac{\gamma p}{2\sigma_{\max}})$, which vanishes as $n\to\infty$ if, e.g., $\frac{\gamma p}{2\sigma_{\max}}\ge\sqrt{2(1+\tau)\ln n}$ for some $\tau>0$; $\gamma p/\sigma_{\max}\to\infty$ alone does not suffice (with $\frac{\gamma p}{2\sigma_{\max}}=\sqrt{\ln n}$ the union bound diverges like $\sqrt{n/\ln n}$, and for equal noise variances the coarse-lattice error $1-(1-2Q(\frac{\gamma p}{2\sigma}))^n$ tends to one, cf.\ Remark~\ref{rem:residue}). (ii) The slack factor tends to one only if $\delta k_p\log p/n\to0$ and the grid loss $\epsilon_{\mathrm g}(\sigma_j/\gamma)$ vanishes, i.e., $\gamma/\sigma_{\min}\to0$. (iii) Conditions~(i) and~(ii) force $p\to\infty$ and $\gamma/\sigma\to0$, whereas the constant $c$ of Theorem~\ref{thm:capacity} was obtained for fixed $p$ on a fixed compact noise range; a uniform exponent along a sequence $(p_K,\gamma_K,\delta_K,n_K)$, together with control of the rank deficiency (Lemma~\ref{lem:pary}) and the selection of a single code (Remark~\ref{rem:quant}), remains to be proved. We list it among the open problems.

\begin{corollary}[IR capacity within the cubic shaping gap]\label{cor:shaping}
For every $\sigma\in(0,\infty]$,
\begin{equation}\label{eq:Iplower}
I_p(\sigma)\ \ge\ \tfrac12\log\Big(1+\tfrac{6}{\pi e}\mathrm{SNR}\Big)-\tfrac12\log\Big(1+\tfrac{\mathrm{SNR}}{p^2-1}\Big).
\end{equation}
Consequently, under the conditions of Corollary~\ref{cor:universal}, the $p$-ary rateless lattice delivers $k_p\log p$ bits with error probability at most $2^{-ck_p}$ at the first $n$ such that
\begin{equation}\label{eq:shapinggap}
\sum_{j=1}^n\Big[\tfrac12\log\Big(1+\tfrac{6}{\pi e}\mathrm{SNR}_j\Big)-\tfrac12\log\Big(1+\tfrac{\mathrm{SNR}_j}{p^2-1}\Big)\Big]\ \ge\ (1+\delta)\,k_p\log p .
\end{equation}
The first term is at most $\frac12\log\frac{\pi e}{6}\approx0.254$ bits per dimension below the Gaussian IR limit $\frac12\log(1+\mathrm{SNR}_j)$, and the second term vanishes as $p\to\infty$. This holds without channel knowledge at the transmitter and for any fading or erasure pattern. The bound~\eqref{eq:Iplower} holds for any integer $p\ge2$, in particular for $p=2^a$.
\end{corollary}

\begin{IEEEproof}
Let $X$ be uniform $p$-PAM with spacing $\Delta_p$, let $V$ be uniform on $[-\frac12,\frac12)$, and let $X,V,Z$ be independent. The output $Y'=Y+\Delta_pV$ is a degraded version of $Y=X+\sigma Z$. Therefore, $I_p(\sigma)=I(X;Y)\ge I(X;Y')=h(X+\Delta_pV+\sigma Z)-h(\Delta_pV+\sigma Z)$. The variable $X+\Delta_pV$ is uniform on an interval of length $p\Delta_p$, and hence $2^{2h(X+\Delta_pV)}=p^2\Delta_p^2\ge12P$. By the entropy power inequality, $2^{2h(X+\Delta_pV+\sigma Z)}\ge12P+2\pi e\sigma^2$. The Gaussian maximum-entropy bound gives $2^{2h(\Delta_pV+\sigma Z)}\le2\pi e(\sigma^2+\Delta_p^2/12)$. Hence
\[
I_p(\sigma)\ge\frac12\log\frac{12P+2\pi e\sigma^2}{2\pi e(\sigma^2+\Delta_p^2/12)}=\frac12\log\frac{1+\frac6{\pi e}\mathrm{SNR}}{1+\frac{\mathrm{SNR}}{p^2-1}},
\]
using $\Delta_p^2/(12\sigma^2)=\mathrm{SNR}/(p^2-1)$. This is~\eqref{eq:Iplower}; for an erasure both sides are $0$. Primality of $p$ was not used. Condition~\eqref{eq:shapinggap} then implies~\eqref{eq:capcond}. Finally, $\log(1+\gamma)-\log(1+\frac{6}{\pi e}\gamma)\le\log\frac{\pi e}{6}$ for all $\gamma\ge0$, because $\frac{1+\gamma}{1+c'\gamma}\le\frac1{c'}$ for $c'=\frac6{\pi e}<1$.
\end{IEEEproof}

\subsection{Closing the Shaping Gap: Gaussian-Labelled Rateless Lattices}\label{sec:shaped}
The $0.254$-bit gap of Corollary~\ref{cor:shaping} is the loss of the uniform input and can be removed while keeping the rateless lattice as the coding layer, by the many-to-one mapping of~\cite[Sec.~6.2]{Gallager1968}, also used for probabilistic shaping~\cite{BSS2015}: a fixed scalar labeling maps the uniform $p$-ary symbols of the rateless lattice codeword to points of a fine one-dimensional grid with a nearly Gaussian distribution.

\begin{definition}[Gaussian-labelled rateless lattice code]\label{def:shaped}
Let $p$ be a prime, $\Delta>0$, and let $g:\mathbb Z_p\to\Delta\mathbb Z$ be a \emph{labeling} with $\frac1p\sum_{x\in\mathbb Z_p}g(x)^2\le P$. With the dithered $p$-ary rateless lattice of Section~\ref{sec:capacity}, the codeword of $\mathbf u\in\mathbb F_p^{k_p}$ is $x_j=g\big((\mathbf a_j^{\mathsf T}\mathbf u+D_j)\bmod p\big)$, $j=1,2,\dots$. Let $X=g(U)$ with $U$ uniform on $\mathbb Z_p$, let $P_X$ be its law, and define the \emph{entropy deficit} of $g$ as $\epsilon_g=\tfrac12\log(2\pi eP)-H(X)-\log\Delta$.
\end{definition}

The index sequence $\big((\mathbf a_j^{\mathsf T}\mathbf u+D_j)\bmod p\big)_j$ is a dithered codeword of the rateless lattice $\Lambda^{(p)}$, and every prefix of the transmitted sequence lies in the scaled integer lattice $\Delta\mathbb Z^n$. The transmitted codebook is the image of a lattice codebook under a coordinatewise map, and is not itself a lattice; the decoder recovers the lattice codeword. The dither makes every transmitted symbol distributed as $X$, and therefore the average power is $\mathbb E[X^2]\le P$.

\begin{lemma}[Entropy-power bound for labeled inputs]\label{lem:epi}
Let $X$ take values in $\Delta\mathbb Z$ with $\mathbb E[X^2]\le P$. For every $\sigma\in(0,\infty]$, with $\mathrm{SNR}=P/\sigma^2$,
\begin{equation}\label{eq:epilab}
I(X;X+\sigma Z)\ \ge\ \tfrac12\log(1+\mathrm{SNR})-[\epsilon_g]^+-\tfrac12\log\Big(1+\mathrm{SNR}\,\frac{\Delta^2}{12P}\Big),
\end{equation}
where $\epsilon_g$ is the entropy deficit of Definition~\ref{def:shaped} with the law of $X$.
\end{lemma}

\begin{IEEEproof}
For $\sigma=\infty$ both sides are at most $0$ and the left side is $0$. Let $\sigma<\infty$, let $V$ be uniform on $[-\frac12,\frac12)$, and let $X,V,Z$ be independent. As in the proof of Corollary~\ref{cor:shaping}, $Y'=X+\sigma Z+\Delta V$ is degraded with respect to $Y=X+\sigma Z$, and hence $I(X;Y)\ge h(X+\Delta V+\sigma Z)-h(\Delta V+\sigma Z)$. Because $X\in\Delta\mathbb Z$, the variable $X+\Delta V$ has the density $P_X(x)/\Delta$ on the cell $[x-\frac\Delta2,x+\frac\Delta2)$, hence $h(X+\Delta V)=H(X)+\log\Delta=\frac12\log(2\pi eP)-\epsilon_g$. The entropy power inequality gives $2^{2h(X+\Delta V+\sigma Z)}\ge2\pi eP\,2^{-2\epsilon_g}+2\pi e\sigma^2$, and the Gaussian maximum-entropy bound gives $2^{2h(\Delta V+\sigma Z)}\le2\pi e(\sigma^2+\Delta^2/12)$. Therefore
\[
I(X;Y)\ge\tfrac12\log\frac{P2^{-2\epsilon_g}+\sigma^2}{\sigma^2+\Delta^2/12}=\tfrac12\log\big(1+2^{-2\epsilon_g}\mathrm{SNR}\big)-\tfrac12\log\Big(1+\mathrm{SNR}\frac{\Delta^2}{12P}\Big).
\]
If $\epsilon_g\le0$, the first term is at least $\frac12\log(1+\mathrm{SNR})$. If $\epsilon_g>0$, then $\log(1+c\gamma)\ge\log c+\log(1+\gamma)$ for $c=2^{-2\epsilon_g}\le1$ and $\gamma\ge0$ gives the first term at least $\frac12\log(1+\mathrm{SNR})-\epsilon_g$.
\end{IEEEproof}

\begin{lemma}[Nearly Gaussian labelings exist]\label{lem:labels}
For every $\epsilon>0$ and $\mathrm{SNR}_{\max}<\infty$ there are a prime $p$, a step $\Delta$ and a labeling $g$ as in Definition~\ref{def:shaped} such that $[\epsilon_g]^++\frac12\log(1+\mathrm{SNR}_{\max}\Delta^2/(12P))\le\epsilon$.
\end{lemma}

\begin{IEEEproof}
Take $P=1$ by scaling. Fix $m\in\mathbb N$, put $\Delta=1/m$, fix an integer $K_0$, let $T=(K_0+\frac12)\Delta$, and let $s\in(0,1)$. Let $f$ be the density of a $\mathcal N(0,s^2)$ variable conditioned on $[-T,T)$, and let $\pi(k)=\int_{(k-1/2)\Delta}^{(k+1/2)\Delta}f$ for $|k|\le K_0$; the cells partition $[-T,T)$. We first show that this binned law has nearly Gaussian entropy and power, and then realize it with a uniform index. If $X_\pi$ has law $\pi$ on $\Delta\mathbb Z$, then $X_\pi+\Delta V$ has the density $\bar f$ that is constant on each cell and equal to the cell average of $f$. Since $t\mapsto-t\log t$ is concave, Jensen's inequality on each cell gives $h(\bar f)\ge h(f)$. Hence $H(\pi)+\log\Delta=h(\bar f)\ge h(f)$, and $h(f)\to\frac12\log(2\pi es^2)$ as $T/s\to\infty$. Moreover, binning a variable $G_T\sim f$ gives a coupling with $|X_\pi-G_T|\le\Delta/2$, and hence $\sqrt{\mathbb E[X_\pi^2]}\le s+\Delta/2$.

To realize $\pi$ with a uniform index, let $n_k=\lfloor p\,\pi(k)\rfloor$ for $k\neq0$ and $n_0=p-\sum_{k\neq0}n_k$, and let $g$ map $n_k$ elements of $\mathbb Z_p$ to $k\Delta$. Then $P_X=(n_k/p)_k$ satisfies $\|P_X-\pi\|_1\le2(2K_0+1)/p$. By the continuity of entropy~\cite[Thm.~17.3.3]{CoverThomas2006}, $|H(X)-H(\pi)|\le-\|P_X-\pi\|_1\log(\|P_X-\pi\|_1/(2K_0+1))$ once $\|P_X-\pi\|_1\le\frac12$, and $|\mathbb E[X^2]-\mathbb E[X_\pi^2]|\le\|P_X-\pi\|_1T^2$.

Finally, choose $s=1-\Delta$, $K_0=\lceil m\log m\rceil$ and $p\ge m^4$ prime. As $m\to\infty$: $\Delta\to0$, $T/s\to\infty$, $\|P_X-\pi\|_1T^2\to0$ and the entropy correction tends to $0$. The power constraint holds exactly, not only up to $o(1)$: every nonzero point $k\Delta$ receives $n_k/p=\lfloor p\pi(k)\rfloor/p\le\pi(k)$ and all leftover mass is put at $0$, and therefore
\[
\mathbb E[X^2]=\sum_{k\neq0}\frac{\lfloor p\pi(k)\rfloor}{p}(k\Delta)^2\le\sum_k\pi(k)(k\Delta)^2=\mathbb E[X_\pi^2]\le(s+\Delta/2)^2=(1-\Delta/2)^2<1 .
\]
Moreover, $\epsilon_g\le\frac12\log\frac1{s^2}+o(1)\to0$ and $\frac12\log(1+\mathrm{SNR}_{\max}\Delta^2/12)\to0$.
\end{IEEEproof}

Lemma~\ref{lem:labels} is not only an asymptotic statement. Evaluating the construction of its proof with $K_0=\lceil4.5m\rceil$, for example, $p=65521$ and $\Delta=\sqrt P/32$ give $\epsilon_g=0.009$ bits, and the total loss in~\eqref{eq:epilab} is at most $0.015$ bits at $20$~dB.

\begin{theorem}[Gaussian-labelled rateless lattices achieve the Gaussian IR capacity]\label{thm:shaped}
For every $n$ and every noise sequence, the ensemble-average ML error probability of the code of Definition~\ref{def:shaped} at prefix length $n$ satisfies
\begin{equation}\label{eq:shapedbound}
\bar P_{\mathrm e}(n)\le\min_{\rho\in[0,1]}2^{-\left(\sum_{j=1}^nE_0^{g}(\rho,\sigma_j)-\rho k_p\log p\right)},\qquad
E_0^{g}(\rho,\sigma)=-\log\int\Big(\sum_xP_X(x)\varphi_\sigma(y-x)^{\frac1{1+\rho}}\Big)^{1+\rho}dy .
\end{equation}
Consequently, fix $\epsilon>0$, $\delta>0$ and $\mathrm{SNR}_{\max}$, and choose $(p,\Delta,g)$ by Lemma~\ref{lem:labels}. Under the conditions of Corollary~\ref{cor:universal} on $(\sigma_j)$ and $n$, with $\mathrm{SNR}_j\le\mathrm{SNR}_{\max}$ for the received coordinates, there is $c>0$ such that $\bar P_{\mathrm e}(n)\le2^{-ck_p}$ whenever
\begin{equation}\label{eq:shapedcond}
\sum_{j=1}^n\tfrac12\log(1+\mathrm{SNR}_j)\ \ge\ (1+\delta)\,k_p\log p+n_{\mathrm{rx}}\,\epsilon,
\end{equation}
where $n_{\mathrm{rx}}$ is the number of non-erased coordinates among the first $n$. Conversely, every code with $\mathbb E[X_j^2]\le P$ for every $j$, possibly drawn with public randomness known to the receiver and independent of the message and the noise, that delivers $b$ bits with error probability $\varepsilon$ satisfies $(1-\varepsilon)b-1\le\sum_j\frac12\log(1+\mathrm{SNR}_j)$.
\end{theorem}

\begin{IEEEproof}
The proof of Theorem~\ref{thm:capacity} carries over because the labeling acts symbol by symbol. That proof shows that, for $\mathbf u'\neq\mathbf u$, the index symbols of the two codewords are independent and uniform on $\mathbb Z_p$, independently over $j$. Applying the fixed map $g$ preserves this independence, and therefore the transmitted symbol and the competitor symbol are independent and both distributed as $X$. The Gallager argument of that proof then applies with $\mathsf g_j(y)=\sum_xP_X(x)\varphi_{\sigma_j}(y-x)^\alpha$: conditioned on the transmitted word, the expected $\alpha$-th power of a competitor's likelihood is $\prod_j\mathsf g_j(y_j)$, and averaging $\varphi_{\sigma_j}(y-X_j)^\alpha$ over $X_j\sim P_X$ gives $\int\mathsf g_j^{1+\rho}=2^{-E_0^g(\rho,\sigma_j)}$. The decoder estimates $\mathbf u$; it never needs to invert $g$. To prove~\eqref{eq:shapedcond}, note that for the finitely supported input $P_X$, $E_0^g(0,\sigma)=0$, $\partial_\rho E_0^g(0,\sigma)=I(X;X+\sigma Z)$~\cite[Thm.~5.6.3]{Gallager1968}, and $\partial^2_\rho E_0^g$ is bounded below uniformly on $[0,1]\times[\sigma_{\min},\sigma_{\max}]$, as in the derivation of~\eqref{eq:E0quad}. The remainder of the proof of Corollary~\ref{cor:universal} shows that $\sum_jI(X;X+\sigma_jZ)\ge(1+\delta)k_p\log p$ implies $\bar P_{\mathrm e}(n)\le2^{-ck_p}$. By Lemmas~\ref{lem:epi} and~\ref{lem:labels}, every received coordinate has $I(X;X+\sigma_jZ)\ge\frac12\log(1+\mathrm{SNR}_j)-\epsilon$, and erased coordinates contribute $0$ to both sides; hence~\eqref{eq:shapedcond} suffices. Finally, for the converse, conditioning on the public randomness $\mathsf S$ as in the proof of Theorem~\ref{thm:capacity} gives $(1-\varepsilon)b-1\le\sum_jI(X_j;Y_j|\mathsf S)$. By the Gaussian maximum-entropy bound and the concavity of the logarithm, $I(X_j;Y_j|\mathsf S)\le\mathbb E_{\mathsf S}\big[\frac12\log(1+\mathbb E[X_j^2|\mathsf S]/\sigma_j^2)\big]\le\frac12\log(1+\mathbb E[X_j^2]/\sigma_j^2)\le\frac12\log(1+\mathrm{SNR}_j)$.
\end{IEEEproof}

Theorem~\ref{thm:shaped} is the rateless-lattice counterpart of the Gaussian fountain-capacity result of~\cite{STV2007}. The labeling is fixed for an SNR range; for coordinates with $\mathrm{SNR}_j>\mathrm{SNR}_{\max}$, which occur under Rayleigh fading, monotonicity of $I(X;X+\sigma Z)$ in the SNR still gives $I\ge\frac12\log(1+\mathrm{SNR}_{\max})-\epsilon$; the guarantee therefore degrades gracefully but is no longer within $\epsilon$ of $\frac12\log(1+\mathrm{SNR}_j)$ (cf.\ Remark~\ref{rem:rayleigh}). Because the labeling is many-to-one, however, the transmitted vectors are not lattice points. We now show that an \emph{injective} labeling, i.e., geometric rather than probabilistic shaping, removes this restriction.

\begin{definition}[Lattice-consistent labeling]\label{def:latlab}
A labeling $g:\mathbb Z_p\to\Delta\mathbb Z$ is \emph{lattice-consistent} if $g(c)\in\Delta(c+p\mathbb Z)$ for every $c\in\mathbb Z_p$. Equivalently, $g(c)=\Delta(c+p\,w(c))$ for an integer function $w$. A lattice-consistent labeling is injective, and $X=g(U)$ is uniform on the $p$ points $g(\mathbb Z_p)$.
\end{definition}

\begin{proposition}[Geometrically shaped codes are lattice codes]\label{prop:latcode}
Let $g$ be lattice-consistent, and let $\mathcal R_g=\bigcup_{c\in\mathbb Z_p}[g(c)-\frac\Delta2,g(c)+\frac\Delta2)$. For every prefix length $n$, the codebook of Definition~\ref{def:shaped} equals $\Delta(\Lambda^{(p)}_n+\mathbf D_{[1:n]})\cap\mathcal R_g^n$. If, in addition, $n\ge n_0$ and $\mathbf A_{n_0}$ has rank $k_p$ (the event $\mathcal F_{n_0}$ of Lemma~\ref{lem:pary}), the map from messages to codewords is a bijection onto this set; without full rank, distinct messages can share a codeword, which happens necessarily for $n<k_p$. In particular, every transmitted prefix is a point of the translate $\Delta(\Lambda^{(p)}_n+\mathbf D_{[1:n]})$ of the scaled rateless lattice.
\end{proposition}

\begin{IEEEproof}
Let $c_j=(\mathbf a_j^{\mathsf T}\mathbf u+D_j)\bmod p$ and $x_j=g(c_j)=\Delta(c_j+p\,w(c_j))$. Then $\mathbf x/\Delta-\mathbf D\equiv\mathbf A_n\mathbf u\pmod p$, and therefore $\mathbf x/\Delta-\mathbf D\in\mathcal C^{(p)}_n+p\mathbb Z^n=\Lambda^{(p)}_n$; clearly $\mathbf x\in\mathcal R_g^n$. Conversely, let $\mathbf z=\Delta(\boldsymbol\lambda+\mathbf D)$ with $\boldsymbol\lambda\in\Lambda^{(p)}_n$ lie in $\mathcal R_g^n$. Each interval of $\mathcal R_g$ has length $\Delta$ and contains exactly one point of $\Delta\mathbb Z$, and hence $z_j=g(c_j)$ for a unique $c_j$, and $c_j\equiv\lambda_j+D_j\pmod p$ by lattice consistency. Since $\boldsymbol\lambda\bmod p\in\mathcal C^{(p)}_n$, the vector $\mathbf c-\mathbf D$ equals $\mathbf A_n\mathbf u$ for some $\mathbf u$, i.e., $\mathbf z$ is the codeword of $\mathbf u$. On $\mathcal F_{n_0}$ and for $n\ge n_0$, injectivity of $g$ and of $\mathbf u\mapsto\mathbf A_n\mathbf u$ gives the bijection. Error bounds conditioned on $\mathcal F_{n_0}$ follow from the unconditioned ones by $\mathbb E[P_{\mathrm e}|\mathcal F_{n_0}]\le\mathbb E[P_{\mathrm e}]/\Pr(\mathcal F_{n_0})$, as noted after Lemma~\ref{lem:pary}.
\end{IEEEproof}

The shaping region $\mathcal R_g^n$ is a product of unions of small intervals rather than a Voronoi region or a ball, and it is the same for every $n$, which makes the shaping compatible with projection; a Voronoi-shaped or lattice-Gaussian codebook~\cite{23silva} in a horizon dimension $n'$ does not project onto a codebook of the same kind. For the Gaussian-quantile constellation below, equiprobable signaling approaches the Gaussian capacity~\cite{SvT1993}; we prove a version that is uniform in the SNR and holds on the lattice grid.

\begin{theorem}[Geometrically shaped rateless lattice codes achieve the Gaussian IR capacity]\label{thm:geoshape}
For a lattice-consistent labeling $g$, the bound~\eqref{eq:shapedbound} holds with $P_X$ uniform on $g(\mathbb Z_p)$, and the code is reliable under the conditions of Corollary~\ref{cor:universal} as soon as $\sum_jI(X;X+\sigma_jZ)\ge(1+\delta)k_p\log p$. Let $g_p$ be the lattice-consistent labeling that maps $c$ to the point of $\Delta_p(c+p\mathbb Z)$ nearest to the centroid of the $c$-th of $p$ equiprobable cells of $\mathcal N(0,(1-\eta)P)$, with $\Delta_p=p^{-3}\sqrt P$ and $\eta\in(0,1)$. Then, for every $\mathrm{SNR}_{\max}<\infty$,
\begin{equation}\label{eq:geogap}
\limsup_{p\to\infty}\ \sup_{0\le\mathrm{SNR}\le\mathrm{SNR}_{\max}}\Big[\tfrac12\log(1+\mathrm{SNR})-I\big(g_p(U);g_p(U)+\sigma Z\big)\Big]\ \le\ \tfrac12\log\frac1{1-\eta}.
\end{equation}
Consequently, for every $\epsilon>0$ there are $p$ and a lattice-consistent $g$ such that the rateless lattice codes of Proposition~\ref{prop:latcode} satisfy~\eqref{eq:shapedcond}, i.e., lattice codes come within $\epsilon$ bits per received dimension of the Gaussian incremental-redundancy capacity for every noise sequence with $\mathrm{SNR}_j\le\mathrm{SNR}_{\max}$.
\end{theorem}

\begin{IEEEproof}
The first statement is the special case of Theorem~\ref{thm:shaped} in which $g$ is injective. For the second, let $G\sim\mathcal N(0,(1-\eta)P)$, let $\bar X_p=\mathbb E[G\,|\,\text{cell of }G]$ be the centroid of the equiprobable cell containing $G$, and let $X_p=g_p(U)$, which we couple with $G$ such that $|X_p-\bar X_p|\le p\Delta_p/2=\sqrt P/(2p^2)$. We first show that $X_p\to G$ in $L^2$. For every $T>0$, the cells inside $[-T,T]$ have width at most $1/(p\,\varphi_{\sqrt{(1-\eta)P}}(T))$, and the cells meeting the complement contribute at most $4\,\mathbb E[G^2\mathbf 1\{|G|>T-1\}]$ to $\mathbb E[(G-\bar X_p)^2]$ for large $p$. Letting $p\to\infty$ and then $T\to\infty$ gives $\mathbb E[(G-\bar X_p)^2]\to0$, and the snapping error vanishes as well. In particular $\mathbb E[X_p^2]\to(1-\eta)P$, and $\mathbb E[X_p^2]\le P$ for all large $p$.

Fix $\sigma\in(0,\infty)$. The $L^2$ convergence implies the weak convergence of the joint law of $(X_p,X_p+\sigma Z)$ to that of $(G,G+\sigma Z)$, and of the product of its marginals to the product of the limiting marginals. Since mutual information is the relative entropy between these two laws, and relative entropy is jointly lower semicontinuous under weak convergence~\cite{Posner1975}, we obtain $\liminf_pI(X_p;X_p+\sigma Z)\ge\frac12\log(1+(1-\eta)\mathrm{SNR})$. To make this uniform, let $f_p(s)$ and $f(s)=\frac12\log(1+(1-\eta)s)$ denote the mutual information and its limit at $\mathrm{SNR}=s$. Both are nondecreasing in $s$ (a larger SNR gives a less degraded channel), and $f$ is uniformly continuous on $[0,\mathrm{SNR}_{\max}]$. Given $\epsilon'>0$, pick $0=s_0<\dots<s_m=\mathrm{SNR}_{\max}$ with $f(s_{i+1})-f(s_i)<\epsilon'$ and then $p_0$ such that $f_p(s_i)\ge f(s_i)-\epsilon'$ for all $i$ and $p\ge p_0$. For $s\in[s_i,s_{i+1}]$, $f_p(s)\ge f_p(s_i)\ge f(s_i)-\epsilon'\ge f(s)-2\epsilon'$. Finally, $\frac12\log(1+s)-\frac12\log(1+(1-\eta)s)\le\frac12\log\frac1{1-\eta}$, which gives~\eqref{eq:geogap}. The last claim follows by choosing $\eta$ and then $p$, exactly as in the proof of Theorem~\ref{thm:shaped}.
\end{IEEEproof}

\begin{table}[!t]
\centering
\footnotesize
\caption{Gap (bits/dim) of the lattice-consistent constellations $g_p$ of Theorem~\ref{thm:geoshape} to the Gaussian capacity ($P=1$).}
\label{tab:geo}
\begin{tabular}{@{}rccc@{}}
\toprule
$p$ & $10$~dB & $20$~dB & $30$~dB\\
\midrule
$127$ & $0.0012$ & $0.0141$ & $0.0666$\\
$509$ & $0.0002$ & $0.0028$ & $0.0138$\\
$1021$ & $0.0001$ & $0.0013$ & $0.0064$\\
$4093$ & $0.0000$ & $0.0003$ & $0.0014$\\
\bottomrule
\end{tabular}
\end{table}

Geometric shaping is also far more efficient. Table~\ref{tab:geo} lists the gap $\frac12\log(1+\mathrm{SNR})-I(X;X+\sigma Z)$ for the labelings $g_p$ (with $\eta$ chosen as small as the power constraint allows). Already $p=127$ is within $0.015$ bits of capacity at $20$~dB, and $p=1021$ is within $0.007$ bits at $30$~dB, whereas the probabilistic labeling needs $p=65521$ for $0.015$ bits at $20$~dB.

\subsection{Multilevel Rateless Lattices Achieve Capacity}\label{sec:mlcap}
The $p$-ary family needs decoders over $\mathbb F_p$. We now show that \emph{multilevel} rateless lattices, whose levels are binary, also achieve capacity. What they require is joint rather than stage-by-stage decoding, together with a balanced split.

Let $a\ge2$ and $q=2^a$. For a symbol $x\in\mathbb Z_q$, its $l$-th \emph{level} is the $l$-th bit $b_l$ of its binary expansion $x=\sum_{l<a}2^lb_l$. Fix $l\in\{0,\dots,a-1\}$. The lower levels $b_0,\dots,b_{l-1}$ determine the \emph{context} $x\bmod2^l$, i.e., the coset $\mathcal O=x+2^l\mathbb Z_q$, which contains $2^{a-l}$ points. Let $W_{\ge l}(\cdot;\sigma)$ be the channel whose input is the context together with the levels $l,\dots,a-1$, whose output is $Y=x+\sigma Z$ together with the context, and let
\begin{align}
C_{\ge l}(\sigma)&=I(B_l,\dots,B_{a-1};Y\,|\,B_0,\dots,B_{l-1}),\notag\\
E_0(\rho,W_{\ge l}(\cdot;\sigma))&=-\log\Big[2^{-l}\!\!\sum_{\mathcal O\in\mathbb Z_q/2^l\mathbb Z_q}\int\Big(2^{-(a-l)}\sum_{x\in\mathcal O}W_\sigma(y|x)^{\frac1{1+\rho}}\Big)^{1+\rho}dy\Big],\label{eq:E0ml}
\end{align}
both computed for uniform $X$ on $\mathbb Z_q$ (i.e., uniform independent levels). Thus $E_0(\rho,W_{\ge l})$ is the Gallager function of the channel from the upper levels to $Y$ with uniform input on the context coset, combined over the uniformly distributed context. As $\rho\to0$, $E_0(\rho,W_{\ge l})=\rho\,C_{\ge l}+O(\rho^2)$, because the derivative at $0$ of $-\log\sum_{\mathcal O}2^{-l}2^{-E_{\mathcal O}(\rho)}$ is the average of the context-wise mutual informations. By the chain rule, $C_{\ge l}=\sum_{m\ge l}C_m$, where $C_m=I(B_m;Y|B_0,\dots,B_{m-1})$ are the level capacities of uniform $2^a$-PAM~\cite{WFH1999}. For $a=2$, $C_{\ge0}=C_{\mathrm X}$, $C_{\ge1}=C_1$ and $W_{\ge1}$ is $W_1$ of Section~\ref{sec:theorems}.

\begin{definition}[$\mathbb Z_{2^a}$-linear rateless lattice ensemble]\label{def:zq}
Fix $1\le K\le N$. The precode $\mathcal C_{\mathrm{pre}}\subseteq\mathbb F_2^N$ is the image of a generator matrix $\mathbf A_{\mathrm{pre}}$ drawn uniformly from the $N\times K$ binary matrices of rank $K$, as in Definition~\ref{def:ens}. The \emph{message module} is
\[
\mathcal U=\big\{(\mathbf m_0+2\boldsymbol\mu)\bmod q:\ \mathbf m_0\in\mathcal C_{\mathrm{pre}},\ \boldsymbol\mu\in\mathbb Z_{2^{a-1}}^N\big\}=\big\{\boldsymbol\xi\in\mathbb Z_q^N:\ \boldsymbol\xi\bmod2\in\mathcal C_{\mathrm{pre}}\big\},
\]
which has $2^{K+(a-1)N}$ elements. Its level-$0$ message is $\mathbf u_0\in\mathbb F_2^K$ with $\mathbf m_0=\mathbf A_{\mathrm{pre}}\mathbf u_0$, and its level-$l$ message ($1\le l\le a-1$) is the $(l-1)$-th bit plane of $\boldsymbol\mu$, such that the level sizes are $k_0=K$ and $k_l=N$ for $l\ge1$. The LT coefficient vectors $\mathbf s_1,\mathbf s_2,\dots$ are i.i.d.\ uniform on $\mathbb Z_q^N$. The codeword of $\boldsymbol\xi\in\mathcal U$ is $\big(\boldsymbol\xi,(-\mathbf s_t^{\mathsf T}\boldsymbol\xi)_{t\ge1}\big)\bmod q$, and a public uniform dither $\mathbf D$ on $\mathbb Z_q^n$ is added. The prefix codebook is the set $\mathcal S_n\subseteq\mathbb Z_q^n$ of undithered prefix codewords, and $\Lambda_n=\{\mathbf x\in\mathbb Z^n:\mathbf x\bmod q\in\mathcal S_n\}$.
\end{definition}

$\mathcal U$ is a $\mathbb Z_q$-submodule of $\mathbb Z_q^N$, because reduction modulo~$2$ is a ring homomorphism and $\mathcal C_{\mathrm{pre}}$ is linear. Hence $\mathcal S_n$ is $\mathbb Z_q$-linear, $\Lambda_n$ is a lattice containing $q\mathbb Z^n$, and, exactly as in Theorem~\ref{thm:structure}, $(\Lambda_n)_{n\ge N}$ is a $q$-ary rateless lattice with $|\mathcal S_n|=2^{K+(a-1)N}$ and $\mathrm{vol}(\Lambda_n)=2^{an-K-(a-1)N}$. For $a=2$ and $\mathbf s_t\in\{0,1\}^N$ the same construction is the Raptor lattice of Definition~\ref{def:raptorlattice}, but this choice of coefficients is \emph{not} the ensemble of Definition~\ref{def:zq}, and Theorem~\ref{thm:mlcap} below does not cover it (see the discussion after its proof). Definition~\ref{def:zq} is an existence ensemble, not an implementation: a uniform row $\mathbf s_t\in\mathbb Z_q^N$ has about $N(1-1/q)$ nonzero entries; each parity symbol therefore costs $\Theta(N)$ ring operations and a prefix of length $n$ costs $\Theta(N(n-N))$ beyond the precode, which is quadratic when $n-N=\Theta(N)$. A sparse ensemble with ring coefficients would keep the cost per symbol bounded, but it would require its own coding theorem.

The admissible level sizes of Definition~\ref{def:zq} are restricted: $k_0=K\le N$ and $k_1=\dots=k_{a-1}=N$. The theorem below therefore describes rates only in this parameter domain; for $a=2$ it is the sector $K\le N$, and arbitrary independent level dimensions $(k_0,\dots,k_{a-1})$ would need a different message module and a new counting argument.

We call a code of this form \emph{nondegenerate up to length $n$} if every coordinate functional $\mathbf x\mapsto x_j$, $j\le n$, viewed modulo~$2$ as a functional of $\mathbf m_0$, does not vanish on $\mathcal C_{\mathrm{pre}}$. This is the condition of Theorem~\ref{thm:structure}(e); it holds for the systematic coordinates whenever no coordinate of $\mathcal C_{\mathrm{pre}}$ is identically zero, and for the coordinate of $\mathbf s_t$ whenever $(\mathbf s_t\bmod2)\notin\mathcal C_{\mathrm{pre}}^\perp$. In the ensemble, a systematic coordinate $i$ vanishes on $\mathcal C_{\mathrm{pre}}$ iff $\mathbf e_i\in\mathcal C_{\mathrm{pre}}^\perp$, which has probability $(2^{N-K}-1)/(2^N-1)$ because $\mathcal C_{\mathrm{pre}}^\perp$ is a uniformly random $(N-K)$-dimensional subspace, and a parity row is degenerate with probability $2^{-K}$. By the union bound, a code is nondegenerate up to length $n$ with probability at least $1-N\frac{2^{N-K}-1}{2^N-1}-(n-N)2^{-K}$ (e.g., exactly $\frac13$ for $N=2$, $K=1$, $n=N$). Nondegeneracy is a finite-horizon property, which is how the converse is used.

\begin{theorem}[Rateless capacity region of multilevel lattices]\label{thm:mlcap}
For every $n$ and every noise sequence, the ensemble of Definition~\ref{def:zq} under joint ML decoding satisfies
\begin{equation}\label{eq:mlbound}
\bar P_{\mathrm e}(n)\le\sum_{l=0}^{a-1}\min_{\rho\in[0,1]}2^{-\left(\sum_{j=1}^nE_0(\rho,W_{\ge l}(\cdot;\sigma_j))-\rho\sum_{m\ge l}k_m\right)},\qquad \sum_{m\ge l}k_m=\begin{cases}K+(a-1)N,&l=0,\\(a-l)N,&l\ge1.\end{cases}
\end{equation}
Conversely, for any code of this form that is nondegenerate up to length $n$, and any decoder (which may know $\mathbf D$) with error probability $\varepsilon$, for every $l$,
\begin{equation}\label{eq:mlconv}
(1-\varepsilon)\sum_{m\ge l}k_m-1\ \le\ \sum_{j=1}^nC_{\ge l}(\sigma_j).
\end{equation}
Hence, as $K,N\to\infty$ with fixed ratios $K/N\le1$ and $n/N$, and for noise sequences that satisfy the conditions of Corollary~\ref{cor:universal} (non-erased noise levels in a fixed compact interval $[\sigma_{\min},\sigma_{\max}]$), the rateless capacity region of this ensemble under ML decoding, within its admissible domain $k_0=K\le N=k_1=\dots=k_{a-1}$, is exactly $\{\sum_{m\ge l}k_m\le\sum_jC_{\ge l}(\sigma_j),\ l=0,\dots,a-1\}$. For $a=2$, this region is $\{K\le N,\ K+N\le\sum_jC_{\mathrm X}(\sigma_j),\ N\le\sum_jC_1(\sigma_j)\}$, and therefore the outer bound of Proposition~\ref{prop:outer} is achievable in the sector $K\le N$.
\end{theorem}

\begin{IEEEproof}
Let $\alpha=1/(1+\rho)$ and fix the transmitted $\boldsymbol\xi\in\mathcal U$. The transmitted word is $\mathbf X=(\mathbf x_{\boldsymbol\xi}+\mathbf D)\bmod q$. It is i.i.d.\ uniform on $\mathbb Z_q^n$ and independent of $(\mathbf A_{\mathrm{pre}},\mathbf s_1,\mathbf s_2,\dots)$, because $\mathbf D$ is. A competitor $\boldsymbol\xi'\neq\boldsymbol\xi$ produces $\mathbf X'=\mathbf X+\mathbf x_{\boldsymbol\Delta}$ with $\boldsymbol\Delta=\boldsymbol\xi'-\boldsymbol\xi$, where $\mathbf x_{\boldsymbol\Delta}=(\boldsymbol\Delta,(-\mathbf s_t^{\mathsf T}\boldsymbol\Delta)_t)$. We split the competitors by the $2$-adic valuation $l$ of $\boldsymbol\Delta$, i.e., the largest $l$ with $\boldsymbol\Delta\in2^l\mathbb Z_q^N$, and apply~\eqref{eq:galavg} to each class (union bound). For a context coset $\mathcal O$ of $2^l\mathbb Z_q$ let $\mathsf g^{(l)}_j(y|\mathcal O)=2^{-(a-l)}\sum_{x\in\mathcal O}W_{\sigma_j}(y|x)^\alpha$.

Consider first a class $l\ge1$, in which $\boldsymbol\Delta\in2^l\mathbb Z_q^N\setminus2^{l+1}\mathbb Z_q^N$. Every such $\boldsymbol\Delta$ is admissible, because $\boldsymbol\xi+\boldsymbol\Delta\equiv\boldsymbol\xi\pmod2$, and there are fewer than $2^{(a-l)N}$ of them. Write $\boldsymbol\Delta=2^l\boldsymbol\Delta'$, where $\boldsymbol\Delta'$ has at least one odd coordinate. Then $\mathbf s_t^{\mathsf T}\boldsymbol\Delta'$ is uniform on $\mathbb Z_q$ (condition on all coordinates of $\mathbf s_t$ except one that multiplies an odd entry), and hence $-\mathbf s_t^{\mathsf T}\boldsymbol\Delta$ is uniform on $2^l\mathbb Z_q$, independently over $t$. On a parity coordinate, the competitor symbol is therefore uniform on the context coset of $X_j$, and its expected $\alpha$-th power likelihood is $\mathsf g^{(l)}_j(y_j|X_j+2^l\mathbb Z_q)$. On the systematic coordinates we extend the sum over $\boldsymbol\Delta$ to all of $2^l\mathbb Z_q^N$, where it factorizes:
\[
\sum_{\boldsymbol\Delta\in2^l\mathbb Z_q^N}\prod_{i\le N}W_{\sigma_i}(y_i|X_i+\Delta_i)^\alpha=\prod_{i\le N}2^{a-l}\mathsf g^{(l)}_i(y_i|X_i+2^l\mathbb Z_q).
\]
Hence the expected sum over class $l$ is at most $2^{(a-l)N}\prod_{j\le n}\mathsf g^{(l)}_j(y_j|X_j+2^l\mathbb Z_q)$, and~\eqref{eq:galavg} gives
\[
P_{l}\le2^{\rho(a-l)N}\,\mathbb E_{\mathbf X}\Big[\prod_j\int W_{\sigma_j}(y|X_j)^\alpha\,\mathsf g^{(l)}_j(y|X_j+2^l\mathbb Z_q)^\rho dy\Big]=2^{\rho(a-l)N}\prod_j2^{-E_0(\rho,W_{\ge l}(\cdot;\sigma_j))}.
\]
The equality averages over the i.i.d.\ uniform $X_j$: averaging $W^\alpha$ over $X_j$ within its context $\mathcal O$ turns the integrand into $\mathsf g^{(l)}_j(y|\mathcal O)^{1+\rho}$, and the context is uniform, which gives~\eqref{eq:E0ml}.

Consider next the class $l=0$, in which $\mathbf e=\boldsymbol\Delta\bmod2=\mathbf A_{\mathrm{pre}}(\mathbf u_0'\oplus\mathbf u_0)\neq\mathbf 0$, and there are $2^K-1$ choices of $\mathbf u_0'$. As shown before~\eqref{eq:ewlaw}, $\mathbf e$ is uniform on $\mathbb F_2^N\setminus\{\mathbf 0\}$. Given $\mathbf e\neq\mathbf 0$, $\boldsymbol\Delta$ has an odd coordinate, and therefore each parity coordinate of the competitor is uniform on $\mathbb Z_q$, with expected $\alpha$-th power likelihood $\mathsf g^{(0)}_j(y_j)$. On the systematic coordinates, the competitors with a given $\mathbf u_0'$ range over all $\boldsymbol\xi'\in\mathbb Z_q^N$ with $\boldsymbol\xi'\bmod2=\mathbf m_0\oplus\mathbf e$. Averaging over $\mathbf e$ uniform on $\mathbb F_2^N\setminus\{\mathbf0\}$ is at most $\frac{2^N}{2^N-1}$ times the average over $\mathbf e$ uniform on $\mathbb F_2^N$, and the latter gives
\[
2^{-N}\sum_{\boldsymbol\xi'\in\mathbb Z_q^N}\prod_{i\le N}W_{\sigma_i}(y_i|\xi'_i+D_i)^\alpha=2^{-N}\prod_{i\le N}2^a \mathsf g^{(0)}_i(y_i)=2^{(a-1)N}\prod_{i\le N}\mathsf g^{(0)}_i(y_i).
\]
Since $(2^K-1)\frac{2^N}{2^N-1}\le2^K$ for $K\le N$, the expected sum over class $0$ is at most $2^{K+(a-1)N}\prod_j\mathsf g^{(0)}_j(y_j)$, and~\eqref{eq:galavg} gives the $l=0$ term of~\eqref{eq:mlbound} as for the classes $l\ge1$.

For the converse~\eqref{eq:mlconv}, fix $l$ and a decoder. Give the decoder, as genie information, the lower messages $\mathbf U_{<l}=(\mathbf u_0,\dots)$ of levels $0,\dots,l-1$, and let it know $\mathbf D$. Fano's inequality gives $(1-\varepsilon)\sum_{m\ge l}k_m-1\le I(\mathbf U_{\ge l};\mathbf Y|\mathbf U_{<l},\mathbf D)$. Write $\boldsymbol\xi=\boldsymbol\xi_{<l}+2^l\boldsymbol\eta$, where $\boldsymbol\xi_{<l}$ is determined by $\mathbf U_{<l}$ and $\boldsymbol\eta$ is uniform on $\mathbb Z_{2^{a-l}}^N$ and determined by $\mathbf U_{\ge l}$ (for $l=0$, $\boldsymbol\eta=\boldsymbol\xi$ is uniform on $\mathcal U$). Each coordinate is $X_j=L_j(\boldsymbol\xi)+D_j$ with a $\mathbb Z_q$-linear functional $L_j$. Given $(\mathbf U_{<l},\mathbf D)$, the context $X_j\bmod2^l$ is fixed, and $X_j=\mathrm{const}+2^lL_j(\boldsymbol\eta)$. By nondegeneracy, $L_j$ has an odd coefficient, and hence $L_j(\boldsymbol\eta)\bmod2^{a-l}$ is uniform and $X_j$ is uniform on its context coset; for $l=0$, nondegeneracy also makes $X_j\bmod2$ uniform. Moreover, the context $X_j\bmod 2^l$ is uniform over the cosets, because $D_j$ is. Since the channel is memoryless, $I(\mathbf U_{\ge l};\mathbf Y|\mathbf U_{<l},\mathbf D)\le\sum_jI(X_j;Y_j|\mathbf U_{<l},\mathbf D)=\sum_jC_{\ge l}(\sigma_j)$, where the inequality is the chain of inequalities in the proof of Theorem~\ref{thm:converse}, and the equality holds because, given the conditioning, $X_j$ is uniform on a uniformly distributed context coset. The two bounds together give the region: if $\sum_{m\ge l}k_m\le(1-\delta)\sum_jC_{\ge l}(\sigma_j)$ for all $l$, the quadratic lower bound~\eqref{eq:E0quad}, applied to each $W_{\ge l}$ on this compact noise range (its derivation applies verbatim to the finite-input channels $W_{\ge l}$), makes every term of~\eqref{eq:mlbound} exponentially small. If some constraint is violated by a factor $1+\delta$, \eqref{eq:mlconv} bounds $\varepsilon$ away from zero. For $a=2$, $C_{\ge0}=C_{\mathrm X}$ and $C_{\ge1}=C_1$.
\end{IEEEproof}

\begin{figure}[!t]
\centering
\begin{tikzpicture}[x=5.2cm,y=5.2cm,>=latex,font=\footnotesize]
\def\Cx{1.102}\def\Cone{0.861}\def\Czero{0.241}\def\Chalf{0.551}
\draw[->] (0,0)--(1.25,0) node[right]{$K/n$};
\draw[->] (0,0)--(0,1.05) node[above]{$N/n$};
\fill[pattern=north east lines,pattern color=black!25] (0,0)--(\Chalf,\Chalf)--(\Cx,0)--cycle;
\fill[blue!12] (0,0)--(\Czero,\Czero)--(\Czero,\Cone)--(0,\Cone)--cycle;
\fill[red!10] (\Czero,\Czero)--(\Czero,\Cone)--(\Chalf,\Chalf)--cycle;
\draw[thick,blue!70!black] (\Czero,\Czero)--(\Czero,\Cone)--(0,\Cone);
\draw[thick,red!70!black] (0,\Cone)--(\Czero,\Cone)--(\Chalf,\Chalf);
\draw[thin,black!50,dashed] (\Chalf,\Chalf)--(\Cx,0);
\node[font=\scriptsize,black!60,align=center,fill=white,inner sep=1pt] at (0.66,0.13) {$K>N$: outside the\\ensembles (Defs.~\ref{def:ens}, \ref{def:zq})};
\draw[dashed] (0,0)--(0.95,0.95) node[above]{$K=N$};
\node[circle,fill=black,inner sep=1.3pt,label={[font=\scriptsize]above right:{MSD corner $(\bar C_0,\bar C_1)$}}] at (\Czero,\Cone) {};
\node[circle,fill=red!70!black,inner sep=1.3pt,label={[font=\scriptsize]right:{$K/n=N/n=\bar C_{\mathrm X}/2$ (joint ML)}}] at (\Chalf,\Chalf) {};
\node[font=\scriptsize] at (-0.06,\Cone) {$\bar C_1$};
\node[font=\scriptsize] at (\Czero,-0.04) {$\bar C_0$};
\node[font=\scriptsize] at (\Cx,-0.04) {$\bar C_{\mathrm X}$};
\node[font=\scriptsize,blue!70!black] at (0.12,0.45) {MSD box};
\node[font=\scriptsize,red!70!black,align=center] at (1.2,0.72) {gained by\\joint decoding};
\draw[->,red!70!black,thin] (1.05,0.72)--(0.35,0.64);
\end{tikzpicture}
\caption{Rateless capacity regions of a two-level lattice at prefix length $n$, with $\bar C_i=\frac1n\sum_{j\le n}C_i(\sigma_j)$, in the admissible sector $K\le N$ (hatched: $K>N$). Blue: multistage decoding (Theorems~\ref{thm:achiev}--\ref{thm:converse}); blue and red: joint ML decoding of the dense ensemble (Theorem~\ref{thm:mlcap}). Values: AWGN at $6$~dB ($C_0=0.241$, $C_1=0.861$, $C_{\mathrm X}=1.102$ bits).}
\label{fig:region}
\end{figure}
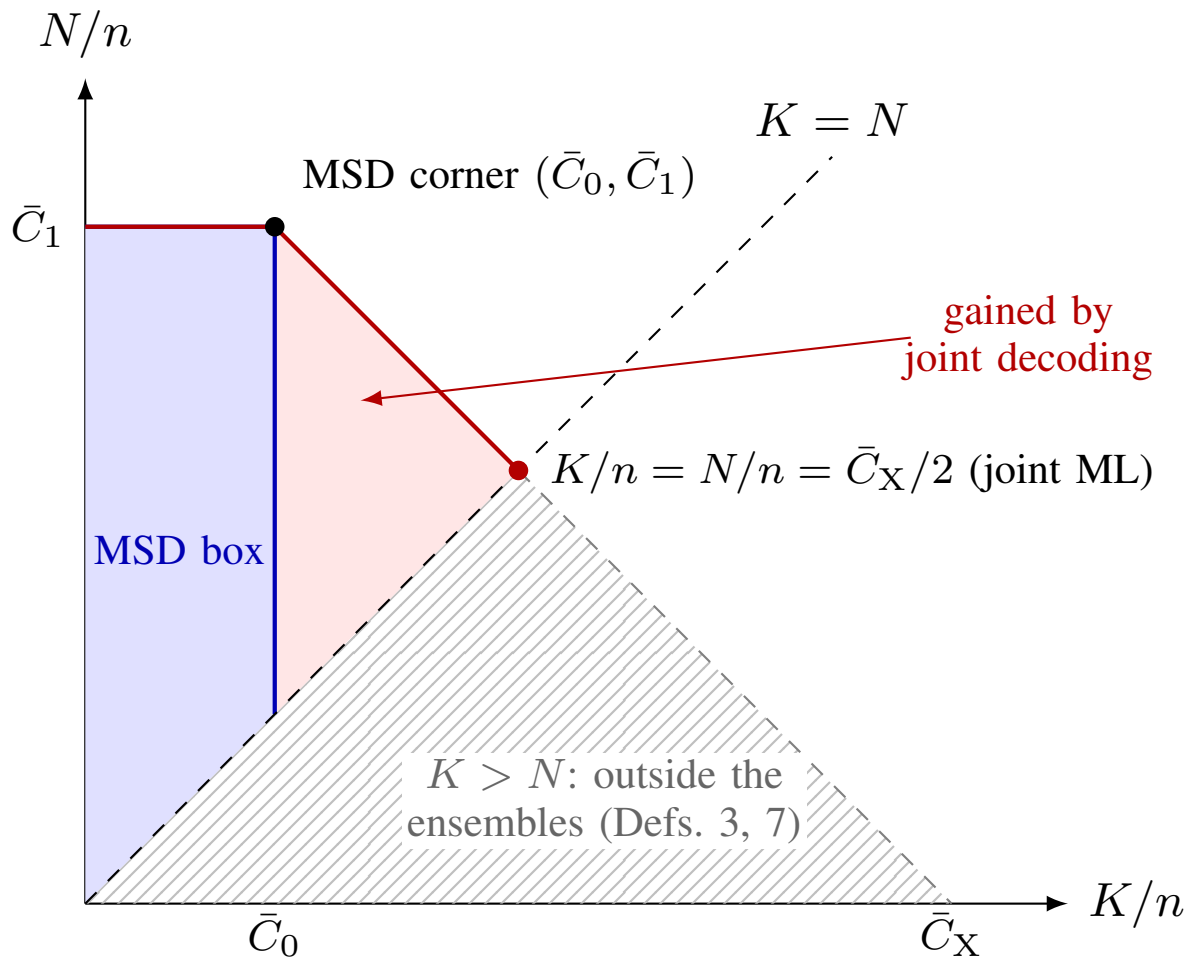
\emph{Scope of Theorem~\ref{thm:mlcap}.} The achievability proof uses one fact that depends on the coefficient law: if $\boldsymbol\Delta=2^l\boldsymbol\Delta'$ and $\boldsymbol\Delta'$ has an odd coordinate, then $\mathbf s_t^{\mathsf T}\boldsymbol\Delta'$ is uniform on $\mathbb Z_q$. This fails for binary coefficients (with $N=1$, $q=4$, $\Delta'=1$, a binary $s_t$ gives only $0$ or $1$), and the public dither, which randomizes symbols but not codeword differences, does not restore it. Theorem~\ref{thm:mlcap} therefore does not cover the sparse binary Raptor lattice of Sections~\ref{sec:construction}--\ref{sec:numerics}, the converse~\eqref{eq:mlconv} applies to codes of the stated form only, and the joint-BP gains of Section~\ref{sec:numjoint} are evidence about a code outside the ensemble, not a confirmation of the theorem.

\begin{lemma}\label{lem:monotone}
For $a=2$, $C_0(\sigma)\le C_1(\sigma)$ for all $\sigma$. For $a\in\{3,4,5\}$ the level capacities satisfy $C_0\le C_1\le\dots\le C_{a-1}$ at every point of a grid of SNRs between $-10$ and $40$~dB (numerical evaluation; this is evidence, not a proof, for $a\ge3$).
\end{lemma}

\begin{IEEEproof}
Let $f(d/\sigma)$ denote the capacity of binary antipodal signalling with distance $d$ in Gaussian noise of standard deviation $\sigma$; it depends only on $d/\sigma$. It is nondecreasing in $d/\sigma$, because a channel with noise $\sigma'>\sigma$ is a degraded version of the channel with noise $\sigma$. Since $B_0$ and $B_1$ are independent, $C_0=I(B_0;Y)\le I(B_0;Y,B_1)=I(B_0;Y|B_1)$. Given $B_1=b_1$, the bit $B_0$ selects one of the two points $\{2b_1,2b_1+1\}$ at distance $1$, and hence $I(B_0;Y|B_1)=f(1/\sigma)$. Given $B_0$, $B_1$ selects one of two points at distance $2$, and hence $C_1=f(2/\sigma)\ge f(1/\sigma)$.
\end{IEEEproof}

\begin{corollary}[Universal capacity with binary levels]\label{cor:mlcap}
Take $K=N$, i.e., equal level sizes and $\mathcal U=\mathbb Z_q^N$. Suppose that the \emph{tail inequalities}
\begin{equation}\label{eq:tail}
C_{\ge l}(\sigma)\ \ge\ \tfrac{a-l}{a}\,C_{\mathrm X}^{(q)}(\sigma),\qquad l=0,\dots,a-1,
\end{equation}
hold for all $\sigma$, where $C_{\mathrm X}^{(q)}$ is the mutual information of uniform $q$-PAM. They hold for $a=2$ by Lemma~\ref{lem:monotone}, and for any $a$ they follow from the ordering $C_0\le\dots\le C_{a-1}$. Then $\sum_jC_{\ge l}(\sigma_j)\ge\frac{a-l}{a}\sum_jC_{\mathrm X}^{(q)}(\sigma_j)$ for every $l$, and all constraints of Theorem~\ref{thm:mlcap} are implied by $aN\le\sum_jC_{\mathrm X}^{(q)}(\sigma_j)$. The ensemble of Definition~\ref{def:zq} thus achieves the accumulated capacity of uniform $2^a$-PAM for every noise sequence; this is proved for $a=2$ and holds for $a\ge3$ under~\eqref{eq:tail}, which we have verified only numerically on a grid (Lemma~\ref{lem:monotone}). Under~\eqref{eq:tail} for all $a$, and scaled to power $P$, the accumulated rate is at least $\sum_j[\frac12\log(1+\frac6{\pi e}\mathrm{SNR}_j)-\frac12\log(1+\frac{\mathrm{SNR}_j}{4^a-1})]$, i.e., within $\frac12\log\frac{\pi e}6\approx0.254$ bits per dimension of the Gaussian IR capacity as $a\to\infty$, with binary component codes at every level.
\end{corollary}

\begin{IEEEproof}
Fix $j$. If $C_0\le\dots\le C_{a-1}$ at $\sigma_j$, the average of the largest $a-l$ level capacities is at least the average of all $a$ of them, and hence $C_{\ge l}(\sigma_j)=\sum_{m\ge l}C_m(\sigma_j)\ge\frac{a-l}{a}C^{(q)}_{\mathrm X}(\sigma_j)$, which is~\eqref{eq:tail}; for $a=2$, \eqref{eq:tail} with $l=1$ reads $C_1\ge\frac12(C_0+C_1)$, i.e., Lemma~\ref{lem:monotone}. Under~\eqref{eq:tail}, summing over $j$ and using $\sum_{m\ge l}k_m=(a-l)N\le\frac{a-l}{a}\sum_jC^{(q)}_{\mathrm X}(\sigma_j)$ gives all constraints. The last claim is~\eqref{eq:Iplower} with $p=2^a$, since $C^{(q)}_{\mathrm X}=I_{2^a}$ after scaling to power $P$.
\end{IEEEproof}

\begin{remark}[Where the split price comes from]\label{rem:mlvsmsd}
Theorems~\ref{thm:achiev}, \ref{thm:converse} and~\ref{thm:mlcap} together separate three effects (Fig.~\ref{fig:region}), within the admissible domain of Definition~\ref{def:zq} ($k_0=K\le N$, $k_1=\dots=k_{a-1}=N$). Under MSD (whose proofs extend level by level to any $a$) the achievable region is the box $\{k_m\le\sum_jC_m(\sigma_j)\ \forall m\}$, which a fixed split meets at its corner only at isolated SNRs (for $a=2$, at one or two SNRs, see the discussion after Corollary~\ref{cor:price}); under joint decoding it is the polymatroid $\{\sum_{m\ge l}k_m\le\sum_jC_{\ge l}(\sigma_j)\ \forall l\}$ intersected with this domain, which with equal level sizes contains the sum-rate point at every SNR for $a=2$, and for $a\ge3$ under the tail inequalities~\eqref{eq:tail}; and, for codes of the stated form, no decoder does better (converse~\eqref{eq:mlconv}).
The split price of Corollary~\ref{cor:price} is therefore exactly the gap between the box and the polymatroid. Decoders that pass soft information between the levels on the joint factor graph aim at the polymatroid. With equal level sizes ($K=N$) and level~$0$ as the bottleneck, MSD needs $n\ge N/C_0$, and therefore its efficiency is $aC_0/C^{(q)}_{\mathrm X}$, whereas joint decoding has efficiency $1$ for $a=2$, and for $a\ge3$ under~\eqref{eq:tail}.
\end{remark}

\subsection{Rateless Lattices Versus Repetition and Layering}\label{sec:vsrep}
All lattice families in the literature have a fixed dimension: LDLC, LDPC and LDA Construction-A/D$'$ lattices, polar lattices, and the families of~\cite{11kah,silva,Pietro2018,Polarm}. A transmitter that uses such a family and must add redundancy after a failed decoding attempt has three options: (i) \emph{repeat the codeword} (Chase combining, CC); (ii) \emph{send further coordinates of a longer codeword} (puncturing); or (iii) \emph{superpose fixed-rate codewords in layers}~\cite{ETW2012}, compared at the end of this subsection. Option~(ii) does not in general give a rateless lattice, since the projection of a lattice need not be a lattice: $\Lambda=\{(a+\sqrt2b,b):a,b\in\mathbb Z\}$ has $\pi_1(\Lambda)=\mathbb Z+\sqrt2\mathbb Z$, which is not discrete, although $\pi_1$ is injective on $\{(0,0),(\sqrt2,1)\}$. It does when the projected lattices are integral, contain $q\mathbb Z^n$ and have codebooks of constant size, and a finite puncturing schedule does not define an extension beyond its length.
The next theorem quantifies option~(i) for every fixed-dimension code.

\begin{theorem}[Rateless lattices versus repetition]\label{thm:advantage}
Consider block fading with $r$ rounds of $n_1$ real dimensions each, known gains at the receiver, and per-round SNRs $\gamma_1,\dots,\gamma_r$ at average power $P$. Let \emph{any} code of dimension $n_1$ and average power $P$ (in particular, any lattice code of any family) be repeated in every round, with arbitrary decoding of all $r$ observations. If it delivers $b$ bits with error probability $\varepsilon$, then
\begin{equation}\label{eq:ccconv}
(1-\varepsilon)b-1\ \le\ \tfrac{n_1}{2}\log\Big(1+\sum_{i=1}^r\gamma_i\Big).
\end{equation}
In contrast, by Corollary~\ref{cor:shaping}, a rateless lattice that sends $n_1$ new dimensions per round delivers, with vanishing error, any $b$ with
\begin{equation}\label{eq:irach}
(1+\delta)\,b\ \le\ \tfrac{n_1}{2}\sum_{i=1}^r\Big[\log\Big(1+\tfrac{6}{\pi e}\gamma_i\Big)-\log\Big(1+\tfrac{\gamma_i}{p^2-1}\Big)\Big].
\end{equation}
For equal SNRs $\gamma_i=\gamma$, the limits must be taken in order. For a fixed $p$, uniform $p$-PAM carries at most $\log p$ bits per dimension; the comparison is therefore meaningful only if the alphabet is large relative to the SNR: letting $p\to\infty$ first (or along any sequence with $\gamma/p^2\to0$), the ratio of the achievable payload in~\eqref{eq:irach} to the repetition bound~\eqref{eq:ccconv} tends to $r/(1+\delta)$ as $\gamma\to\infty$, and to $r$ as the coding slack $\delta\downarrow0$. At a fixed SNR, the same ratio grows like $r/\log r$ as $r\to\infty$. With the Gaussian-labelled code of Theorem~\ref{thm:shaped}, the factor $\frac6{\pi e}$ in~\eqref{eq:irach} can be replaced by $1$ at the cost of $\epsilon$ bits per received dimension.
\end{theorem}

\begin{IEEEproof}
Let coordinate $j$ of round $i$ be received as $y_{i,j}=h_ix_j+w_{i,j}$ with known gain $h_i$ and $w_{i,j}\sim\mathcal N(0,\sigma^2)$, and hence $\gamma_i=h_i^2P/\sigma^2$. The maximum-ratio-combined output $\tilde y_j=\sum_ih_iy_{i,j}/\sum_ih_i^2=x_j+\tilde w_j$, with $\tilde w_j\sim\mathcal N(0,\sigma^2/\sum_ih_i^2)$, is a sufficient statistic for $(y_{1,j},\dots,y_{r,j})$: the likelihood of $x_j$ depends on the observations only through $\tilde y_j$. Hence the $r$ rounds are equivalent to one Gaussian channel use per coordinate with SNR $\sum_i\gamma_i$. Fano's inequality gives $(1-\varepsilon)b-1\le I(\mathbf X;\tilde{\mathbf Y})\le\sum_{j}\frac12\log(1+P_j\sum_ih_i^2/\sigma^2)$, where $P_j=\mathbb E[X_j^2]$, by the Gaussian maximum-entropy bound. Concavity of the logarithm and $\frac1{n_1}\sum_jP_j\le P$ give~\eqref{eq:ccconv}. Equation~\eqref{eq:irach} is~\eqref{eq:shapinggap} for the $rn_1$ dimensions sent over the $r$ rounds, with $b=k_p\log p$. For the ratio, $r\log(1+c\gamma)/\log(1+r\gamma)\to r$ as $\gamma\to\infty$, for any $c>0$, while for fixed $\gamma$ it behaves like $r\log(1+c\gamma)/\log r$ as $r\to\infty$; the factor $1/(1+\delta)$ in~\eqref{eq:irach} remains in both limits until $\delta\downarrow0$. The two bounds compared are an achievability benchmark and a converse, not measured rates of a fixed finite code.
\end{IEEEproof}

At $\gamma=20$~dB and $r=3$ rounds, for example, the right-hand side of~\eqref{eq:irach} (with $p\to\infty$) exceeds that of~\eqref{eq:ccconv} by a factor of about $2.24$, i.e., $2.24/(1+\delta)$ for the payload.

\emph{Comparison with layered rateless codes.} Erez, Trott and Wornell~\cite{ETW2012} build rateless codes for the AWGN channel from $L$ layers of equal-rate base codes. Each redundancy block is a combination of the same $L$ base codewords with complex, block-dependent weights, and the receiver decodes the layers successively; the gain is unknown to the transmitter but constant over the blocks. Layering turns any good single-rate AWGN code, possibly a lattice code, into a rateless code that is (near-)perfect at the design points of a constant-gain channel. Rateless lattices provide a single algebraic object instead: an encoder with no design horizon, whose prefixes are lattice points, and whose guarantees (Corollary~\ref{cor:universal}, Theorems~\ref{thm:capacity} and~\ref{thm:shaped}) hold, as ensemble guarantees with a shared random seed (Remark~\ref{rem:quant}), for every noise, fading and erasure sequence without any channel knowledge at the transmitter. A layered block, a complex-weighted sum of lattice points, is in general not a lattice point; the lattice structure of every prefix matters for applications that rely on linearity, such as compute-and-forward~\cite{21kah} and lattice wiretap codes~\cite{19silva,LCOMcomp}.

\section{A Raptor Realization with Prefix-Consistent Girth}\label{sec:construction}

The ensembles of Section~\ref{sec:theorems} use LT vectors with i.i.d.\ entries, either dense or with logarithmic density (Theorem~\ref{thm:sparse}). We now describe the sparse realization with a constant degree that is used in Section~\ref{sec:numerics}, prove its graph-theoretic properties, and describe its multistage and joint decoders.

\subsection{Girth-$8$ QC-LDPC Precode}
Let $Z$ be a circulant size and $w_r$ the row weight. The precode parity-check matrix is the $2\times w_r$ array of $Z\times Z$ circulant permutation matrices
\begin{equation}\label{eq:qc}
\mathbf{H}_{\mathrm{pre}}=\begin{bmatrix}\mathbf{I}^{(0)}&\mathbf{I}^{(0)}&\cdots&\mathbf{I}^{(0)}\\ \mathbf{I}^{(\nu_1)}&\mathbf{I}^{(\nu_2)}&\cdots&\mathbf{I}^{(\nu_{w_r})}\end{bmatrix},
\end{equation}
where $\mathbf{I}^{(\nu)}$ is the $Z\times Z$ identity matrix with its columns cyclically shifted by $\nu$, such that column $t$ has its one in row $t+\nu\bmod Z$. This gives $N=w_rZ$, column weight $2$, and an all-one $2\times w_r$ base matrix, as in~\cite{a}.

\begin{lemma}\label{lem:qc}
The Tanner graph of~\eqref{eq:qc} has girth at least $8$ iff $\nu_1,\dots,\nu_{w_r}$ are distinct modulo $Z$. Moreover, $\mathrm{rank}_{\mathbb{F}_2}\mathbf{H}_{\mathrm{pre}}=2Z-\kappa$, where $\kappa$ is the number of cosets of the subgroup of $\mathbb{Z}_Z$ generated by $\{\nu_j-\nu_1\}$. In particular, $K=N-2Z+1$ whenever these differences generate $\mathbb{Z}_Z$.
\end{lemma}

\begin{IEEEproof}
Index the check nodes by $(1,t)$ and $(2,t)$, $t\in\mathbb Z_Z$, according to the block row. Every variable node has exactly one neighbour in each block row; the variable node of column $t$ in block column $j$ is adjacent to $(1,t)$ and $(2,t+\nu_j)$. A cycle therefore leaves each variable node through the other block row from the one it entered, and therefore the check nodes along the cycle alternate between the two block rows. Hence the cycle visits an even number of variable nodes and its length is a multiple of $4$; in particular, there are no $6$-cycles. A $4$-cycle is $(1,t)\to v\to(2,s)\to v'\to(1,t)$ with distinct variable nodes $v$ in block column $j$ and $v'$ in block column $k$. Since $v$ and $v'$ are both adjacent to $(1,t)$, they are columns $t$ of blocks $j$ and $k$, and $j\neq k$ because a block has one column per $t$. They are both adjacent to $(2,s)$ iff $s=t+\nu_j=t+\nu_k$, i.e., iff $\nu_j\equiv\nu_k\pmod Z$, the condition of~\cite{Fossorier2004}. Hence the girth is at least $8$ iff the $\nu_j$ are distinct.

For the rank, note that every column has weight $2$, and hence $\mathbf{H}_{\mathrm{pre}}$ is the vertex--edge incidence matrix of a graph on the $2Z$ check nodes. Over $\mathbb F_2$, the rank of such a matrix is the number of vertices minus the number of connected components: the rows of one component sum to zero, and any proper subset of them does not. Hence $\mathrm{rank}\,\mathbf H_{\mathrm{pre}}=2Z-(\text{number of components})$ and $K=N-2Z+\kappa$. Check $(1,t)$ is adjacent to $(2,t+\nu_j)$ for every $j$, and hence $(1,t)$ and $(1,t+\nu_j-\nu_k)$ are connected through $(2,t+\nu_j)$. Consequently, $(1,t)$ and $(1,t')$ are connected iff $t'-t$ lies in the subgroup generated by the differences $\nu_j-\nu_1$, and every $(2,s)$ is connected to some $(1,t)$. The components are therefore in bijection with the cosets of this subgroup, and there are $\kappa$ of them.
\end{IEEEproof}

\begin{remark}[Bounded cycles of the QC precode]\label{rem:qccycles}
Distinct shifts exclude $4$- and $6$-cycles, but they do not make the precode locally tree-like. Collapse every (degree-$2$) variable node into an edge between its two checks; with top checks $T_t$ and bottom checks $B_t$, an edge of block column $j$ joins $T_t$ and $B_{t+\nu_j}$. For any three distinct shifts $a,b,c$ among the $\nu_j$ (i.e., $w_r\ge3$), the closed walk $T_0\to B_a\to T_{a-b}\to B_{a-b+c}\to T_{c-b}\to B_c\to T_0$ (indices modulo $Z$) uses edges of types $a,b,c,a,b,c$, visits three distinct top checks, three distinct bottom checks and six distinct variable nodes, and is therefore a Tanner cycle of length $12$; by translation it passes through every precode variable. These cycles persist as $Z\to\infty$; hence the QC ensemble does not have tree neighbourhoods of arbitrary fixed depth. The six variables of the cycle form a codeword; hence $d_{\min}(\mathcal C_{\mathrm{pre}})\le6$ for all parameters with $w_r\ge3$; LT checks do not remove these precode cycles. Lemma~\ref{lem:qc} is unaffected, but the density evolution of Section~\ref{sec:de} describes a different, locally tree-like surrogate ensemble.
\end{remark}

\subsection{Balanced LT Stage with Online $4$-Cycle Removal}
LT rows are generated in \emph{sub-blocks} of $\lfloor N/D\rfloor$ groups, following the design of~\cite{2019}. A sub-block draws a uniformly random permutation of $\{1,\dots,N\}$ and splits its first $D\lfloor N/D\rfloor$ entries into consecutive groups of size $D$. Within a sub-block every intermediate bit is used at most once, which balances the LT load. The raw construction contains many $4$-cycles, whose number grows quadratically with the number of LT rows ($1307$ at $n=995$ for $N=684$, $D=12$). We remove them \emph{online}, as follows.

\begin{algorithm}[t]
\caption{Online prefix-consistent LT generation with coverage completion}\label{alg:lt}
\begin{algorithmic}[1]
\State \textbf{Input:} $N$, degree $D$, precode rows $\mathcal{R}_{\mathrm{pre}}$; \textbf{state:} set $\mathcal{P}$ of covered pairs, initialized with all pairs $\{i,k\}$ contained in a row of $\mathcal{R}_{\mathrm{pre}}$
\For{each sub-block}
  \State draw a uniformly random permutation of $\{1,\dots,N\}$ and split its first $D\lfloor N/D\rfloor$ entries into $\lfloor N/D\rfloor$ groups of $D$ indices; $\mathcal{Q}\gets\emptyset$ (owed indices)
  \For{each group $\mathcal T$ of the sub-block, in order}
    \State $\mathcal{L}\gets(\mathcal{Q},\mathcal{T})$ in this order; $\mathcal{Q}\gets\emptyset$; $\mathcal{N}\gets\emptyset$
    \For{$i\in\mathcal{L}$ in list order}
       \If{$\{i,k\}\notin\mathcal{P}$ for all $k\in\mathcal{N}$} \State $\mathcal{N}\gets\mathcal{N}\cup\{i\}$ \Else \State append $i$ to $\mathcal{Q}$ \EndIf
    \EndFor
    \State $\mathcal{P}\gets\mathcal{P}\cup\{\{i,k\}:i\neq k\in\mathcal{N}\}$; \textbf{emit} the next row $\mathbf{s}_t$ with support $\mathcal N_t=\mathcal N$
  \EndFor
  \For{$i\in\mathcal Q$} \textbf{emit} the next row $\mathbf s_t$ with support $\mathcal N_t=\{i\}$ \EndFor
\EndFor
\end{algorithmic}
\end{algorithm}

\begin{lemma}[Prefix-consistent girth]\label{lem:girth}
Let $\mathbf{s}_1,\mathbf{s}_2,\dots$ be produced by Algorithm~\ref{alg:lt} and let $\mathbf{H}_{\mathrm{pre}}$ satisfy Lemma~\ref{lem:qc}. For every $n\ge N$:
\begin{enumerate}[label=(\roman*)]
\item the Tanner graphs of $\mathbf{H}_0^{(n)}$ and $\mathbf{H}_1^{(n)}$ in~\eqref{eq:H} have girth at least $6$;
\item $\mathbf{s}_1,\dots,\mathbf{s}_{n-N}$ do not depend on how many further rows are generated. Hence the lattices $\Lambda_n$ obtained by stopping Algorithm~\ref{alg:lt} at any length form a single rateless lattice in the sense of Definition~\ref{def:rateless};
\item every index proposed in a sub-block belongs to an LT row emitted before the next sub-block starts. In particular, if $D$ divides $N$, every intermediate index lies in an LT row of $\Lambda_n$ for all $n\ge N+N/D+\varsigma_1$, where $\varsigma_1$ is the number of degree-$1$ rows emitted at the end of the first sub-block.
\end{enumerate}
\end{lemma}

\begin{IEEEproof}
(i) A Tanner graph is bipartite, and therefore its girth is even, and it has a $4$-cycle iff two rows of the parity-check matrix share two columns. Each parity column $N+t$ appears in exactly one row, and therefore a shared pair consists of intermediate columns $i,k\le N$. Two precode rows share at most one column by Lemma~\ref{lem:qc}. Consider an LT row $t$ and any earlier row $r$ (precode or LT). Suppose $i,k\in\mathcal N_t\cap\mathcal N_r$ with $i\neq k$, and let $k$ be the one admitted to $\mathcal N_t$ later. When row $t$ is formed, the pair $\{i,k\}$ lies in $\mathcal P$, because it lies in row $r$ and $\mathcal P$ contains all pairs of earlier rows. At the moment $k$ is considered, $i$ is already in $\mathcal N_t$, and therefore the membership test of Algorithm~\ref{alg:lt} rejects $k$, a contradiction. Hence $|\mathcal{N}_t\cap\mathcal{N}_r|\le1$, and no two rows share two columns. The Tanner graph of $\mathbf{H}_1^{(n)}$ is a subgraph of that of $\mathbf{H}_0^{(n)}$. (ii) Row $t$ depends only on the random permutations of the sub-blocks generated up to that point, on $\mathcal{P}$, which is determined by $\mathcal{R}_{\mathrm{pre}}$ and $\mathbf{s}_1,\dots,\mathbf{s}_{t-1}$, and on the owed list, which is determined by the same quantities. The claim then follows from Theorem~\ref{thm:structure}. (iii) A proposed index is either admitted to the row being formed or appended to $\mathcal Q$, and an index of $\mathcal Q$ is proposed again at the next group of the same sub-block. At the end of the sub-block each remaining index of $\mathcal Q$ forms a row of its own. A degree-$1$ row contains no pair, and therefore it neither creates a $4$-cycle nor changes $\mathcal P$, which keeps~(i) and~(ii) valid. If $D$ divides $N$, the first sub-block proposes every index, which gives the bound.
\end{IEEEproof}

An offline removal, which deletes ones from a parity-check matrix that is fixed in advance, produces a code that depends on the final length. Algorithm~\ref{alg:lt} instead never revisits an emitted row, which is exactly what rateless operation requires. Rejecting an index changes only the degree of one LT parity, and therefore~\eqref{eq:H} and~\eqref{eq:enc} are unaffected. Without the owed list, an index rejected by the $4$-cycle test is not proposed again until a later sub-block happens to admit it, and the next proposition shows that such an index produces a vector of squared norm $4$ in $\Lambda_n$.

\emph{Finite pair budget.} The pair set $\mathcal P$ only grows and has at most $\binom N2$ elements, and every emitted row of degree $d\ge2$ consumes $\binom d2$ pairs that are not yet in $\mathcal P$. Hence, for every horizon,
\begin{equation}\label{eq:pairs}
\sum_t\binom{|\mathcal N_t|}{2}\ \le\ \binom N2-|\mathcal P_{\mathrm{pre}}|,
\end{equation}
where $\mathcal P_{\mathrm{pre}}$ is the initial pair set. For a fixed $N$, only finitely many rows can have degree at least $2$, and an infinite stream generated by Algorithm~\ref{alg:lt} eventually consists of degree-$1$ rows. It is still a rateless lattice (Lemma~\ref{lem:girth}(ii)), but it does not keep a nontrivial stationary LT degree profile for an unbounded horizon, and the degree statistics and the fixed-$\Omega$ density evolution below refer to a finite operating horizon. In this paper the horizon is $n_{\max}=4104=6N$. Up to it, the precode uses $7524$ of the $\binom{684}2=233{,}586$ pairs, and the LT rows use $21{,}463$ (profile~R) and $13{,}439$ (profile~M), i.e., $12.4\%$ and $9.0\%$ of all pairs together with the precode; pair exhaustion is far away. The emitted degrees lie in $\{1,\dots,7\}$ (profile~R, mean $3.99$) and $\{1,\dots,8\}$ (profile~M, mean $3.09$).

\begin{proposition}[Short vectors and LT coverage]\label{prop:short}
For $i\le N$ let $d_i(n)=|\{t\le n-N:\ i\in\mathcal N_t\}|$ be the number of LT checks of $\Lambda_n$ that contain $i$, and let $N_0(n)=|\{i\le N:\ d_i(n)=0\}|$. Then
\begin{enumerate}[label=(\roman*)]
\item $4\mathbf e_j\in\Lambda_n$ for every $j\le n$, and $\mathbf w_i=2\mathbf e_i+2\sum_{t\le n-N:\,i\in\mathcal N_t}\mathbf e_{N+t}\in\Lambda_n$ for every $i\le N$, with $\|\mathbf w_i\|^2=4(1+d_i(n))$. Hence $d_{\min}^2(\Lambda_n)\le\min\{16,\,4(1+\min_i d_i(n))\}$.
\item On the channel $\mathbf y=\boldsymbol\lambda+\mathbf z$ with $\boldsymbol\lambda\in\Lambda_n$ and $\mathbf z\sim\mathcal N(\mathbf 0,\sigma^2\mathbf I_n)$, the error probability of nearest-point (lattice) decoding satisfies
\begin{equation}\label{eq:floor}
P_e(\Lambda_n,\sigma^2)\ \ge\ \max\Big\{1-\big(1-2Q(1/\sigma)\big)^{N_0(n)},\ \ 1-\big(1-2Q(2/\sigma)\big)^{n}\Big\}.
\end{equation}
\end{enumerate}
\end{proposition}

\begin{IEEEproof}
(i) Adding $2\mathbf e_i$ to $\mathbf x\in\Lambda_n$ leaves $\mathbf x_{[1:N]}\bmod 2$ unchanged, and therefore the precode congruences in~\eqref{eq:lattice} still hold. It adds $2$ to $\mathbf s_t^{\mathsf T}\mathbf x_{[1:N]}$ for exactly the $d_i(n)$ rows with $i\in\mathcal N_t$, and the term $2\mathbf e_{N+t}$ restores the congruence modulo $4$ of each of them. The vector $4\mathbf e_j$ changes no congruence. Both vectors are nonzero, which gives the bound on $d_{\min}$. (ii) Let $i$ satisfy $d_i(n)=0$; then $\mathbf w_i=2\mathbf e_i$, and suppose $|z_i|>1$. The lattice point $\boldsymbol\lambda+2\,\mathrm{sgn}(z_i)\mathbf e_i$ is strictly closer to $\mathbf y$ than $\boldsymbol\lambda$, because $\|\mathbf z-2\,\mathrm{sgn}(z_i)\mathbf e_i\|^2-\|\mathbf z\|^2=4-4|z_i|<0$, and the decoder errs. The events $\{|z_i|>1\}$ for the $N_0(n)$ uncovered indices are independent and have probability $2Q(1/\sigma)$ each, which gives the first term of~\eqref{eq:floor}. The same argument with $4\mathbf e_j$ and the events $\{|z_j|>2\}$, $j\le n$, gives the second term.
\end{IEEEproof}

The first term of~\eqref{eq:floor} is a floor of order $2N_0(n)Q(1/\sigma)$ that depends only on the LT graph, while the second is intrinsic to any two-level Construction-D$'$ lattice, which contains $4\mathbb Z^n$. By Lemma~\ref{lem:girth}(iii), the owed list makes $N_0(n)=0$ for every $n\ge N+N/D+\varsigma_1$. For the mother code of Section~\ref{sec:numerics} ($N=684$, $D=4$, profile~R) we have $\varsigma_1=0$, and full coverage is reached at $n=N+171$, at the end of the first sub-block. Without the owed list, $N_0(n)>0$ up to $n=N+491$, and for instance $N_0(995)=10$. The owed list raises the mean LT degree from $3.71$ to $3.99$ and the maximum degree from $4$ to $7$, and it adds three degree-$1$ rows among the first $1716$.

\subsection{Encoding, Multistage BP Decoding and Complexity}
\emph{Encoding.} By~\eqref{eq:enc}, the transmitter computes $\mathbf{m}_0\in\mathcal{C}_{\mathrm{pre}}$ from $\mathbf{u}_0$ (QC systematic encoding, $O(N)$~\cite{a}), forms $\boldsymbol{\xi}=\mathbf{m}_0+2\mathbf{m}_1$, and then emits $x_{N+t}=(-\sum_{i\in\mathcal{N}_t}\xi_i)\bmod 4$, which costs $|\mathcal{N}_t|$ additions modulo~$4$ per symbol ($3.99$ on average and at most $7$ for $D=4$). A prefix of length $n$ thus costs $O(N+\bar d(n-N))=O(n)$ operations with a stored LT graph, without matrix inversion. Generating the graph online also requires the pair set (about $2.9\times10^4$ entries at $n_{\max}$); this cost does not depend on the message and can be paid offline.

\emph{Decoding.} The receiver computes level-$0$ log-likelihood ratios (LLRs) from~\eqref{eq:W0},
\begin{equation}\label{eq:llr0}
L_{0,j}=\ln\frac{\varphi_{\sigma_j}(y_j)+\varphi_{\sigma_j}(y_j-2)}{\varphi_{\sigma_j}(y_j-1)+\varphi_{\sigma_j}(y_j-3)},
\end{equation}
with $L_{0,j}=0$ for erased coordinates. In the unconstrained case the sums run over the full periodic sets $2\mathbb{Z}$ and $2\mathbb{Z}+1$. It runs sum-product belief propagation (BP) on the Tanner graph of $\mathbf{H}_0^{(n)}$ and obtains $\hat{\mathbf{c}}_0$, then $\hat{\mathbf{v}}=\mathrm{Enc}(\hat{\mathbf{m}}_0,\mathbf{0})$. Level-$1$ LLRs are
\begin{equation}\label{eq:llr1}
L_{1,j}=\frac{(y_j-t_{1,j})^2-(y_j-t_{0,j})^2}{2\sigma_j^2},\qquad t_{0,j}=\hat v_j,\ t_{1,j}=(\hat v_j+2)\bmod 4,
\end{equation}
and BP on $\mathbf{H}_1^{(n)}$ yields $\hat{\mathbf{c}}_1$ and $\hat{\mathbf{x}}=(\hat{\mathbf{v}}+2\hat{\mathbf{c}}_1)\bmod 4$. For the unconstrained lattice, $\hat{\mathbf{x}}=\hat{\mathbf{v}}+2\hat{\mathbf{c}}_1+4\lfloor(\mathbf{y}-\hat{\mathbf{v}}-2\hat{\mathbf{c}}_1)/4\rceil$. A decoding attempt with at most $\mathfrak{I}$ iterations costs $O(\mathfrak{I}(|\mathcal{E}_0|+|\mathcal{E}_1|))$, where $|\mathcal{E}_0|=2N+\sum_t(|\mathcal{N}_t|+1)$ and $|\mathcal{E}_1|=\sum_t(|\mathcal{N}_t|+1)$ are the numbers of edges, i.e., $O(\mathfrak{I}Dn)$. Unlike~\cite{fan,silva}, no linear system has to be solved to re-encode between the stages. In an IR session with decoding attempts after rounds $r=1,\dots,r^\star$, the total cost is $O(\mathfrak{I}D\sum_{r\le r^\star}n_r)$. Warm-starting BP from the messages of the previous attempt reduces this further but is not used in our simulations.

\subsection{Joint BP Decoding Over $\mathbb Z_4$}\label{sec:joint}
Section~\ref{sec:mlcap} shows that the price of a fixed level split disappears under joint decoding. A practical joint decoder follows directly from Theorem~\ref{thm:structure}: the codebook $\mathcal S_n$ is the set of $\mathbf x\in\mathbb Z_4^n$ with
\[
\textstyle\sum_{i\in\mathcal N_t}x_i+x_{N+t}\equiv0\pmod 4\quad(t=1,\dots,n-N),\qquad \mathbf H_{\mathrm{pre}}(\mathbf x_{[1:N]}\bmod2)=\mathbf 0 ,
\]
and therefore it has a factor graph with $\mathbb Z_4$-valued variable nodes, one $\mathbb Z_4$ check per LT row and one binary check per precode row. This is the Tanner graph of Fig.~\ref{fig:tanner}, read over $\mathbb Z_4$ instead of level by level. The joint decoder runs sum-product BP on this graph. A variable node $j$ starts from the channel likelihoods $\varphi_{\sigma_j}(y_j-a)$, $a\in\mathbb Z_4$, and multiplies the incoming messages. An LT check of degree $d$ computes the law of the negated sum of the other $d-1$ symbols, i.e., a pointwise product of $d-1$ discrete Fourier transforms $\hat\mu(m)=\sum_a\mu(a)\,\mathrm i^{am}$ followed by one inverse transform (leave-one-out products by prefix and suffix products). A precode check acts on the low bits only: it converts each incoming message to the log-likelihood ratio $\ln\frac{\mu(0)+\mu(2)}{\mu(1)+\mu(3)}$, applies the tanh rule, and returns $\mu(a)\propto e^{\pm L/2}$ according to the parity of $a$.
After the last iteration the decoder decides $\hat{\boldsymbol\xi}$ symbol by symbol and re-encodes the parity symbols as $\hat x_{N+t}=-\sum_{i\in\mathcal N_t}\hat\xi_i\bmod4$. No level is decided before the other, and therefore the redundancy that level~$1$ does not need also helps level~$0$; this is the mechanism whose information-theoretic limit Theorem~\ref{thm:mlcap} establishes for the dense ring-coefficient ensemble under ML decoding (the binary-coefficient code decoded here is not covered by that theorem). An LT check of degree $d$ costs $O(4d)$ operations per iteration, and therefore an attempt costs $O(\mathfrak I n)$ operations, as for MSD, with a constant about four times larger.

\subsection{Density Evolution for the Joint Decoder}\label{sec:de}
We analyze the joint graph by density evolution (DE)~\cite{RU2008}. DE does not describe the QC realization itself: by Remark~\ref{rem:qccycles} its precode has unavoidable $12$-cycles, and Algorithm~\ref{alg:lt} changes the emitted degree law and correlates the rows. We therefore analyze a locally tree-like \emph{surrogate ensemble} with the same degree structure, and the agreement between DE and the simulations of the QC code (Section~\ref{sec:numjoint}) is an empirical observation, not a consequence of a concentration theorem. In the surrogate ensemble, $N\to\infty$ with the split $\chi=K/N$ and the \emph{LT overhead} $\theta=(n-N)/N$ fixed. Every LT check has $d$ systematic neighbours, with i.i.d.\ $d\sim\Omega$ (the nominal input profile of Algorithm~\ref{alg:lt}, not the emitted law) of mean $\bar d$, plus one parity leaf with its own channel message. The LT degree $D_{\mathrm v}$ of a systematic variable equals $\lfloor\theta\bar d\rfloor+1$ with probability $\theta\bar d-\lfloor\theta\bar d\rfloor$ and $\lfloor\theta\bar d\rfloor$ otherwise (the balanced assignment of Section~\ref{sec:construction}); we write $p_d=\Pr\{D_{\mathrm v}=d\}$ and $\lambda_d=dp_d/(\theta\bar d)$ for its node and LT-edge laws. Every systematic variable lies in two precode checks of degree $w_r$, independently of $D_{\mathrm v}$; hence the LT degree seen along a precode edge has law $p_d$. Finally, every computation graph is a tree up to any fixed depth.
A complication is that the $4$-PAM channel is not symmetric in the sense of~\cite{RU2008}: the outer points $0$ and $3$ are better protected than the inner ones, and the error probability of BP depends on the transmitted codeword. The public dither removes this dependence, as the next lemma shows.

\begin{lemma}[Rotated messages]\label{lem:rot}
Let the codeword $\mathbf x\in\mathcal S_n$ be sent as $X_j=(x_j+D_j)\bmod4$ with a uniform public dither $\mathbf D$, which the decoder removes by computing its channel messages for the codeword symbols. Rotate every $\mathbb Z_4$ message $\mathsf M$ into or out of variable node $j$ by the codeword symbol, $\tilde{\mathsf M}(a)=\mathsf M(a+x_j)$, and every binary precode message by $(-1)^{x_j\bmod2}$. Then the rotated messages obey the same update rules as the original ones, and the rotated channel messages $\tilde\ell_j(a)=\log\varphi_{\sigma_j}(Y_j-(X_j+a)\bmod4)$ are independent across $j$, with $X_j$ uniform on $\{0,1,2,3\}$ and independent of $\mathbf x$. Consequently, on a tree the law of the rotated messages after any number of iterations does not depend on the message, and the symbol error probability equals the probability that a rotated posterior does not have its maximum at $0$.
\end{lemma}

\begin{IEEEproof}
For an LT check with neighbours $j_1,\dots,j_d$ and constraint $\sum_ix_{j_i}\equiv0\pmod4$, the outgoing message to $j_d$ is $\mathsf M_{\mathrm{out}}(a)=\sum_{b_1+\dots+b_{d-1}\equiv-a}\prod_i\mathsf M_i(b_i)$, where $\mathsf M_i$ is the message from $j_i$. Substituting $b_i=b_i'+x_{j_i}$ and using $\sum_{i<d}x_{j_i}\equiv-x_{j_d}$ gives $\mathsf M_{\mathrm{out}}(a+x_{j_d})=\sum_{b_1'+\dots+b_{d-1}'\equiv-a}\prod_i\tilde{\mathsf M}_i(b_i')$, which is the same rule applied to the rotated messages. A precode check acts on the low bits, whose transmitted values have even parity; multiplying each incoming log-likelihood ratio by $(-1)^{x_j\bmod2}$ therefore commutes with the tanh rule, as for linear codes over binary-input channels. Variable nodes multiply messages pointwise, which commutes with a common rotation. Finally, the channel message of codeword symbol $j$ is $\ell_j(a)=\log\varphi_{\sigma_j}(Y_j-(a+D_j)\bmod4)$, and its rotation by $x_j$ is $\tilde\ell_j$ as stated; the transmitted symbols $X_j$ are i.i.d.\ uniform and independent of $\mathbf x$ because $\mathbf D$ is.
\end{IEEEproof}

Lemma~\ref{lem:rot} assumes a uniform public dither; the DE below therefore describes dithered transmission (the simulations of Section~\ref{sec:numerics} use the undithered mother code, see Section~\ref{sec:setup}). By Lemma~\ref{lem:rot}, DE tracks four families of rotated messages, which are updated in parallel in every iteration. A \emph{variable-to-LT} message combines a rotated channel message, $D_{\mathrm v}-1$ LT-check messages with $D_{\mathrm v}\sim\lambda$, and the two precode messages (as $\mathbb Z_4$ messages $\pm L/2$ according to parity). A \emph{variable-to-precode} message combines a channel message, $D_{\mathrm v}\sim p$ LT-check messages and the other precode message, reduced to the low-bit log-likelihood ratio. An \emph{LT-to-variable} message is the law of the negated sum of $d-1$ variable-to-LT messages and the parity-leaf message, with $d\sim d\,\Omega(d)/\bar d$, computed by DFT. A \emph{precode-to-variable} message applies the tanh rule to $w_r-1$ variable-to-precode messages.
All check-to-variable messages are initialized as uninformative, and the channel messages are resampled in every iteration with a uniform transmitted symbol. We evaluate the recursions by population dynamics with $M=2\times10^4$ samples per family. Every five iterations the decision statistic (channel message, $D_{\mathrm v}\sim p$ LT messages and two precode messages) is formed for $M$ fresh samples, and the estimated symbol error probability $\hat P_{\mathrm s}$ is the fraction of samples whose rotated posterior does not have its maximum at $0$. DE \emph{reaches the target} if $\hat P_{\mathrm s}<10^{-4}$ within $150$ iterations. The \emph{fixed-target DE operating point} $\theta^\star(\mathrm{SNR})$ is the smallest overhead in $[0,6]$ at which the target is reached, located by bisection with ten steps (resolution $6/2^{10}$), and the corresponding rateless throughput is
\begin{equation}\label{eq:derate}
R^\star(\mathrm{SNR})=\frac{K+N}{N(1+\theta^\star)}=\frac{1+\chi}{1+\theta^\star(\mathrm{SNR})}.
\end{equation}
The target $10^{-4}$ corresponds to at most one erroneous sample among $2\times10^4$. Even for independent samples, zero or one observed error gives one-sided $95\%$ upper confidence limits of $1.5\times10^{-4}$ and $2.4\times10^{-4}$; the target is therefore a nominal decision rule, not a guaranteed error level, and $\theta^\star$ and $R^\star$ are \emph{estimated} operating points. One bisection step ($6/2^{10}$) changes $R^\star$ by at most $0.011$ bits/dim. The sensitivity of $\theta^\star$ to the population size, the seed and the iteration limit is reported in Section~\ref{sec:numjoint}. The scripts that generate the graphs and evaluate DE are part of the supplementary material.

$R^\star$ is a finite-error design criterion and not a vanishing-error threshold. In fact, with a bounded mean LT degree and no precode on level~$1$, no decoder has vanishing error at a fixed overhead and noise level. Give a genie all message bits except the level-$1$ bit $m_{1,i}$. Flipping $m_{1,i}$ changes exactly $1+d_i$ transmitted symbols by $2$ modulo~$4$, i.e., by Euclidean distance $2$, where $d_i$ is the LT degree of $i$, and hence even the genie-aided optimal detector errs with probability $Q(\sqrt{1+d_i}/\sigma)$. At least half of the indices satisfy $d_i\le2\bar d_{\mathrm v}$, where $\bar d_{\mathrm v}=\theta\bar d$, and therefore the average level-$1$ bit error probability of every decoder, joint or multistage, is at least $\frac12Q(\sqrt{1+2\theta\bar d}/\sigma)>0$. For the operating points of Table~\ref{tab:de} this bound is far below the target (about $5\times10^{-8}$ at $8$~dB); it therefore does not affect $R^\star$, but it rules out a zero-error interpretation; a precode on level~$1$ removes the obstruction (Section~\ref{sec:conclusion}). Proposition~\ref{prop:outer} does not formally bound $R^\star$, but $R^\star$ never exceeds the outer benchmark $\min\{C_{\mathrm X},(1+\chi)C_1\}$ in our evaluations.

\section{Numerical Results}\label{sec:numerics}

\subsection{Setup and Methodology}\label{sec:setup}
All results come from end-to-end Monte Carlo simulation of the encoder~\eqref{eq:enc} and of the multistage BP decoder of Section~\ref{sec:construction} (sum-product, flooding schedule, at most $\mathfrak{I}=50$ iterations per level, syndrome-based stopping). A decoding attempt succeeds only if the decoded lattice point equals the transmitted one; this is the ideal error detection of Section~\ref{sec:theorems}, and a practical CRC would add its own overhead and undetected-error probability.

The simulations use the \emph{undithered} mother code: every session transmits $\mathrm{Enc}(\mathbf m_0,\mathbf m_1)$ of~\eqref{eq:enc} with fresh uniform messages, i.e., $\mathbf D=\mathbf 0$ and $\mathbf d=\mathbf 0$, and the decoders use the likelihoods~\eqref{eq:llr0}--\eqref{eq:llr1} (or the $\mathbb Z_4$ likelihoods of Section~\ref{sec:joint}) of the undithered symbols. The theorems are averages over random ensembles and a public dither, not statements about this finite code, and the DE of Section~\ref{sec:de} describes the dithered surrogate ensemble; the simulated values are empirical results for one fixed code.

\emph{Codes.} We use two precodes of the form~\eqref{eq:qc} with $N=684$, both satisfying Lemma~\ref{lem:qc}: design~A with $(Z,w_r)=(57,12)$, $K=571$ and $\chi=0.835$, and design~B with $(Z,w_r)=(171,4)$, $K=343$ and $\chi=0.501$. The payloads are $K+N=1255$ bits (A) and $1027$ bits (B). The LT stage follows Algorithm~\ref{alg:lt} up to a mother length $n_{\max}=4104$. We consider two LT degree profiles: profile~R, which is regular with $D=4$, and profile~M, which uses degrees $\{2,3,4,6\}$ with probabilities $\{0.4,0.3,0.2,0.1\}$ and assigns intermediate bits to rows by least current load. Unless stated otherwise, every curve uses \emph{one} mother lattice, and every prefix length is a prefix of it. In the convention of~\eqref{eq:qc}, the second block row uses the shifts $(\nu_j)=(56,33,9,15,7,36,44,50,1,21,40,11)$ (design~A) and $(43,85,9,92)$ (design~B). Profile~M draws the degree of each row of a sub-block from $\Omega$ and proposes the indices of smallest current LT load, with random tie-breaking, before the $4$-cycle test of Algorithm~\ref{alg:lt}. The supplementary material contains the generator code with all seeds, which reproduces the mother graphs, the decoders, and the per-session outcomes of all throughput experiments.

\emph{Statistics.} For FER curves we simulate until $50$ frame errors, or for at most $90$ minutes per point on a GPU implementation of the same decoder (single precision for $n\le1300$, where it agrees with the reference decoder, and double precision for $n\ge1800$), and show $95\%$ Wilson confidence intervals. Each throughput point averages $400$ independent IR sessions with fresh messages, noise and fading. Every session stops at the \emph{first} round after which decoding succeeds; the receiver attempts decoding after every received round (exact sequential stopping). (BP success need not be monotone in the received length, and no monotonicity is assumed.)

\emph{Benchmarks.} The information-theoretic curves are evaluated on the same channel realizations as the simulations, with the level capacities computed by quadrature (the deficits $1-C_1$ and $2-C_{\mathrm X}$ are evaluated directly, which keeps $C_1<1$ at every finite SNR). They are the asymptotic MSD benchmark of Theorems~\ref{thm:achiev}--\ref{thm:converse}, which stops at the first round at which the accumulated level capacities exceed $K$ and $N$; the asymptotic outer benchmark of Proposition~\ref{prop:outer}, which stops when $\sum_jC_{\mathrm X}\ge K+N$ and $\sum_jC_1\ge N$; the ensemble guarantee of Theorem~\ref{thm:achiev}, which stops when the right-hand side of~\eqref{eq:achiev} drops below $10^{-2}$; and the Gaussian IR limit $\sum_j\frac12\log(1+\mathrm{SNR}_j)\ge K+N$~\cite{CT2001}.
By Remark~\ref{rem:bench}, the first two are not finite-session upper bounds and the third holds for the ensemble, not for the simulated code. At the first round, $n_1=N$, level~$1$ is uncoded and $\sum_jC_1(\sigma_j)\ge N$ cannot hold, because $C_1<1$, while finite-length decoding of the uncoded level often succeeds at high SNR; the simulated throughput therefore exceeds the asymptotic benchmarks there. The maximum throughput is $(K+N)/n_1$.

\emph{Complexity.} With profile~R the Tanner graph of $\mathbf H_0^{(n)}$ has about $1368+4.99(n-N)$ edges (mean LT degree $3.99$; $3.09$ with profile~M), and one MSD attempt at $n=995$ takes about $1$~ms (just-in-time-compiled Python, two CPU cores).

\subsection{Price of the Split and Design Ablation}
\begin{figure}[!t]
\centering
\includegraphics[width=0.9\linewidth]{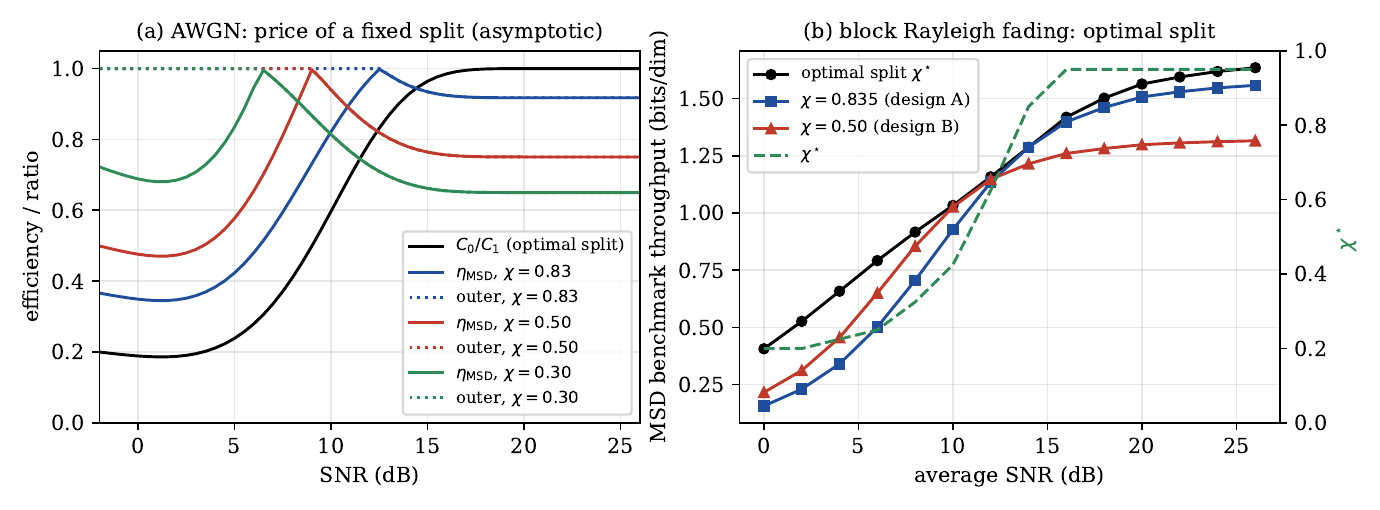}
\caption{(a) Asymptotic rateless efficiency on the AWGN channel under MSD (solid; Corollary~\ref{cor:price}) and outer bound (dotted; Proposition~\ref{prop:outer}), with the optimal split $C_0/C_1$. (b) Asymptotic MSD throughput under block Rayleigh fading for the optimal split and designs A and B, and the optimal split $\chi^\star$ (right axis).}
\label{fig:theory}
\end{figure}

Fig.~\ref{fig:theory}(a) evaluates Corollary~\ref{cor:price} and Proposition~\ref{prop:outer} for $4$-PAM. The ratio $C_0/C_1$, which is the optimal split, is not monotone (Section~\ref{sec:theorems}): it tends to $\frac14$ at low SNR, has a minimum $r_{\min}\approx0.186$ near $1.21$~dB, and rises to $0.99$ at $16$~dB and to $1$ as the SNR grows. The splits of designs A and B lie in $(\frac14,1)$; numerically, each is lossless at exactly one SNR, $12.5$~dB and $9.0$~dB, respectively. Design~A ($\chi=0.835$) loses up to $65\%$ of the capacity at low SNR under MSD, whereas the outer bound allows full efficiency there. Fig.~\ref{fig:theory}(b) shows the MSD-limit throughput under block Rayleigh fading, with rounds of $L_r=114$ dimensions, a first round of $N$ dimensions and $K+N=1255$. The optimal split increases from $\chi^\star\approx0.2$ at $0$~dB to the upper end $0.95$ of the search range above $16$~dB. Designs B and A are near-optimal around $10$ and $14$~dB, respectively, and their curves cross between $12$ and $14$~dB.

\emph{Design ablation.} Table~\ref{tab:ablation} reports, for profile~R with several degrees $D$, the gap between the level-$0$ BP threshold (noise level at which the level-$0$ FER equals $0.1$, located by bisection with $500$ frames per test) and the level-$0$ capacity limit $C_0^{-1}(K/n)$, at five prefix lengths of the same mother code. Algorithm~\ref{alg:lt} removes every $4$-cycle, and its benefit grows with $D$ and with $n$. For $D=12$, the gap at $n=3200$ falls from $3.25$~dB to $1.48$~dB. Without removal the threshold degrades steadily along the prefix family, which rateless operation cannot tolerate; for $D\le4$ the effect is within simulation accuracy ($\pm0.1$~dB). $D=4$ gives the best thresholds for $1300\le n\le2400$ ($0.59$--$0.68$~dB from the limit) and is used below.

\begin{table}[!t]
\centering
\footnotesize
\caption{Gap (dB) of the level-$0$ BP threshold to the level-$0$ capacity limit (design A, profile R with degree $D$), with and without online $4$-cycle removal; last column: $4$-cycles at $n=3200$.}
\label{tab:ablation}
\begin{tabular}{@{}cc ccccc r@{}}
\toprule
$D$ & removal & $n=995$ & $1300$ & $1800$ & $2400$ & $3200$ & $4$-cycles\\
 & & ($K/n=0.574$) & ($0.439$) & ($0.317$) & ($0.238$) & ($0.178$) & \\
\midrule
3 & on  & 1.38 & 1.04 & 0.83 & 0.71 & 0.56 & 0\\
3 & off & 1.41 & 1.10 & 0.80 & 0.71 & 0.59 & 341\\
4 & on  & 0.98 & 0.68 & 0.59 & 0.59 & 0.68 & 0\\
4 & off & 1.04 & 0.71 & 0.56 & 0.59 & 0.74 & 954\\
6 & on  & 0.71 & 0.80 & 1.10 & 1.28 & 1.54 & 0\\
6 & off & 0.71 & 0.83 & 1.10 & 1.41 & 1.70 & 4169\\
12 & on  & 1.16 & 1.61 & 1.80 & 1.70 & 1.48 & 0\\
12 & off & 1.13 & 1.70 & 2.32 & 2.83 & 3.25 & 63001\\
\bottomrule
\end{tabular}
\end{table}

\subsection{AWGN Channel: Error Rates and Rateless Efficiency}
\begin{figure}[!t]
\centering
\includegraphics[width=0.9\linewidth]{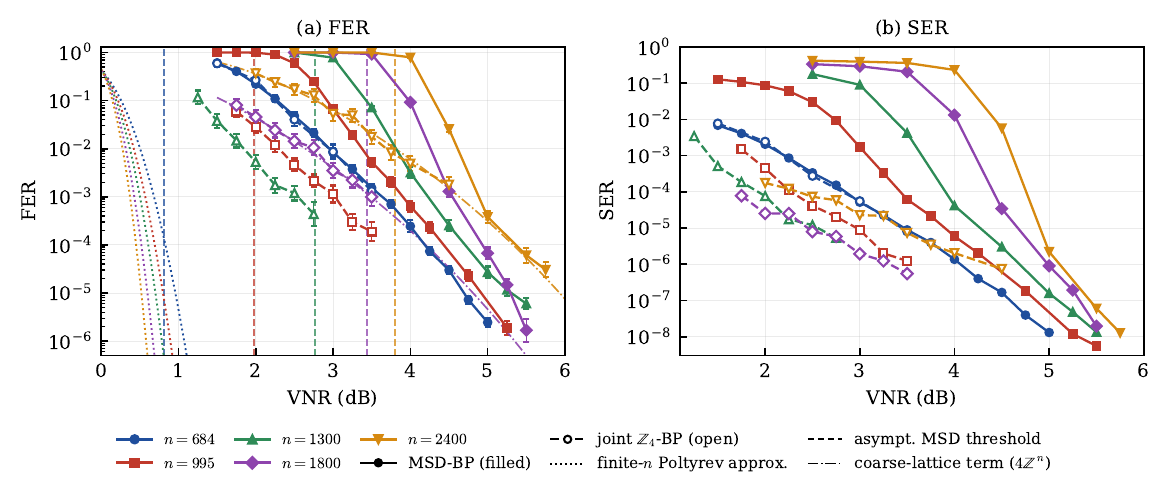}
\caption{FER ($95\%$ Wilson intervals) and SER of prefixes of one rateless Raptor lattice (design A, profile R) with MSD-BP (solid, filled markers) and joint $\mathbb Z_4$-BP (dashed, open markers) decoding. Dotted: finite-$n$ Poltyrev limit~\cite{IZF2013}; vertical dashed lines: asymptotic MSD residue threshold (level~$0$ only for $n=684$); dash-dotted: coarse-lattice term $1-(1-2Q(2/\sigma))^n$ of~\eqref{eq:floor} for $n=1800$ and $2400$. Joint-BP points rest on at least $30$ frame errors, except four with $11$ to $24$.}
\label{fig:awgnfer}
\end{figure}

Fig.~\ref{fig:awgnfer} shows FER and SER versus VNR on the unconstrained AWGN channel for prefixes $n\in\{684,995,1300,1800,2400\}$ of one mother lattice (design~A, profile~R). The dotted lines show the normal approximation of the finite-dimensional Poltyrev limit~\cite{IZF2013}, $\ln\mathrm{VNR}\approx\sqrt{2/n}\,Q^{-1}(\varepsilon)-\frac{\ln n}{n}$, for the same $n$. 

\emph{Error floors.} Points with FER below $10^{-3}$ rest on at least $50$ frame errors, except four points with $12$ to $34$ errors. A diagnosis of the error frames, guided by Proposition~\ref{prop:short}, identifies three mechanisms. (a) \emph{Uncovered coordinates.} Without the owed list, ten intermediate coordinates of $\Lambda_{995}$ lie in no LT check; the first term of~\eqref{eq:floor} ($5.8$, $3.1$ and $1.6\times10^{-5}$ at $5.25$, $5.5$ and $5.75$~dB) matches the measured FERs ($7.1$, $3.1$ and $1.6\times10^{-5}$), and $78$ of $87$ error frames are one uncovered symbol displaced by $\pm2$. The owed list lowers the FER at $5.25$~dB to $1.9\times10^{-6}$ and the waterfall FER at $n=995$--$1800$ by a factor of $2$ to $5$, at a cost of about $0.1$~dB at $n=2400$. (b) \emph{Low-weight precode words.} The column-weight-$2$ precode has words of weight $4$ and $6$ (Remark~\ref{rem:qccycles}). A precode word with support $\mathcal W$, $|\mathcal W|=w$, and signs $a_i\in\{\pm1\}$ extends to $\mathbf z\in\Lambda_n$ via $z_{N+t}\equiv-S_t\pmod4$, $S_t=\sum_{i\in\mathcal N_t\cap\mathcal W}a_i$, and the shortest extension has $\|\mathbf z\|^2=w+\#\{t:S_t\text{ odd}\}+4\,\#\{t:S_t\equiv2\pmod4\}$. At $n=N$ these vectors, of squared norm $4$, cause the floor of the base lattice. Without the owed list, most error frames at $n=1300$ and $1800$ near $5$~dB are of this type; with it, the FER keeps decreasing, to $6.0\times10^{-6}$ and $1.7\times10^{-6}$ at $5.5$~dB. (c) \emph{The sublattice $4\mathbb Z^n$.} The second term of~\eqref{eq:floor} is intrinsic to two-level Construction-D$'$ and grows with $n$ at fixed VNR, because $\mathrm{vol}(\Lambda_n)^{2/n}$ increases toward $16$ while $\|4\mathbf e_j\|=4$ stays fixed. At $n=2400$ it equals $5.4\times10^{-5}$ and $2.1\times10^{-5}$ at $5.5$ and $5.75$~dB against measured FERs of $6.1\times10^{-5}$ and $3.1\times10^{-5}$. It tends to one as $n\to\infty$ at fixed VNR (Remark~\ref{rem:residue}) and can be lowered only by a third level.

\emph{Thresholds along the family.} The VNR at which a prefix becomes reliable \emph{increases} with $n$, unlike for independently designed lattices. This is Corollary~\ref{cor:price} in Poltyrev normalization. The vertical dashed lines in Fig.~\ref{fig:awgnfer}(a) mark the asymptotic residue thresholds from $C_0^{-1}(K/n)$ and $C_1^{-1}(N/n)$; level~$0$ is the bottleneck, with MSD thresholds $1.98$, $2.77$, $3.44$ and $3.80$~dB at $n=995,1300,1800,2400$. At $n=684$ level~$1$ is uncoded and has no finite asymptotic threshold, and its line ($0.81$~dB) marks the level-$0$ threshold only. At FER $10^{-2}$ multistage decoding is within $2.1$~dB of it at $n=684$ and within $1.4$, $1.0$, $0.8$ and $0.8$~dB of the MSD thresholds at the other lengths; the gap \emph{shrinks} along the family, consistent with Table~\ref{tab:ablation}.

\emph{Joint decoding of the prefixes.} The dashed curves in Fig.~\ref{fig:awgnfer} use the joint $\mathbb Z_4$-BP decoder of Section~\ref{sec:joint} with the residue metric $\log\sum_k\varphi_\sigma(y-a-4k)$ and the same lattice-point decision. At $n=684$ both decoders coincide. For longer prefixes the gain of joint decoding at FER $10^{-2}$ is $1.1$~dB at $n=995$, $2.0$~dB at $n=1300$, $1.5$~dB at $n=1800$ and $0.9$~dB at $n=2400$, and the VNR that joint decoding needs no longer grows with $n$ from $995$ to $1300$ ($2.3$ and $1.9$~dB): the rise of the multistage thresholds is a property of stage-by-stage decoding, as Corollary~\ref{cor:price} and Theorem~\ref{thm:mlcap} predict. For $n\ge1800$ the joint curves coincide with the coarse-lattice term of~\eqref{eq:floor} (dash-dotted), e.g., $4.6\times10^{-2}$ against $4.3\times10^{-2}$ at $2$~dB for $n=1800$. The remaining errors are then, with few exceptions, single symbols displaced by $\pm4$, which no decoder can avoid with a fixed alphabet (Remark~\ref{rem:residue}). At $n=995$ and $1300$ the joint curves flatten instead because of short lattice vectors, discussed with Fig.~\ref{fig:cmpfixed} below.

\emph{Comparison with fixed-dimension lattices.} Panel~(a) of Fig.~\ref{fig:cmpfixed} isolates the code design: the fixed-dimension Raptor Construction-D$'$ lattice of~\cite{CCECE} with $n=995$ has the same $N=684$, QC precode and kind of LT stage as our prefix, and only its level-$1$ code, an ordinary high-rate code, differs. At $2$~dB its FER is $3.2\times10^{-4}$, between those of the LDLC lattice ($2.6\times10^{-4}$) and the multilevel LDPC lattice ($4.8\times10^{-4}$)~\cite{29silva,silva}, and it reaches FER $10^{-3}$ near $1.8$~dB, together with the LDLC, multilevel LDPC and polar~\cite{Polarm} lattices. Only the LDA lattice~\cite{Pietro2018} is clearly better.

Panel~(b) shows what the rateless level-$1$ extension costs in this comparison. Turning this lattice into a rateless one changes a single ingredient: level~$1$ must become the LT extension of rate $N/n$, which keeps every prefix a lattice. The extra level-$1$ redundancy enlarges $\mathrm{vol}(\Lambda_n)$ (Proposition~\ref{prop:vnr}) and moves the asymptotic MSD threshold from below $0.5$~dB (level-$1$ rate about $0.98$) to $1.98$~dB. The FER curve moves by about the same amount, $2.1$~dB at FER $10^{-3}$. Measured from its own threshold, the prefix is as efficient as the best fixed designs: it operates $1.4$~dB above that threshold at FER $10^{-2}$, while the LDLC and multilevel LDPC lattices operate $1.3$--$1.4$~dB above the Poltyrev limit. Part of this shift is due to stage-by-stage decoding: joint decoding, which is not bound by the MSD threshold, reduces the distance to the fixed-dimension lattice from $2.1$ to $1.2$~dB at FER $10^{-3}$. Below FER $10^{-3}$ it is limited by short lattice vectors: in six of eight error frames inspected at $2.75$~dB, the decoder output was a lattice point closer to $\mathbf y$ than the transmitted one, at squared distance $8$ to $10$, an error that ML decoding would also make. These vectors stem from the low-weight precode words and degree-$1$ coverage discussed above.

The shift is therefore the cost of the level-$1$ rate constraint and of the stage-by-stage decoder used in this experiment, which Corollary~\ref{cor:price} quantifies for MSD; it is not a lower bound for all rateless lattice architectures, since joint decoding recovers part of it and the equal-level and single-level ensembles of Section~\ref{sec:capacity} avoid the split loss. In exchange, one encoder serves every length. Literal repetition of a fixed-dimension codeword (Chase combining) loses a factor that is unbounded in the number of rounds (Theorem~\ref{thm:advantage}); puncturing and layering~\cite{ETW2012} are different options (Section~\ref{sec:vsrep}). Under a power constraint the split loss appears as the gap between the MSD and outer benchmarks, and Section~\ref{sec:numjoint} shows that the joint decoder recovers most of it there as well.

\begin{figure}[!t]
\centering
\includegraphics[width=0.95\linewidth]{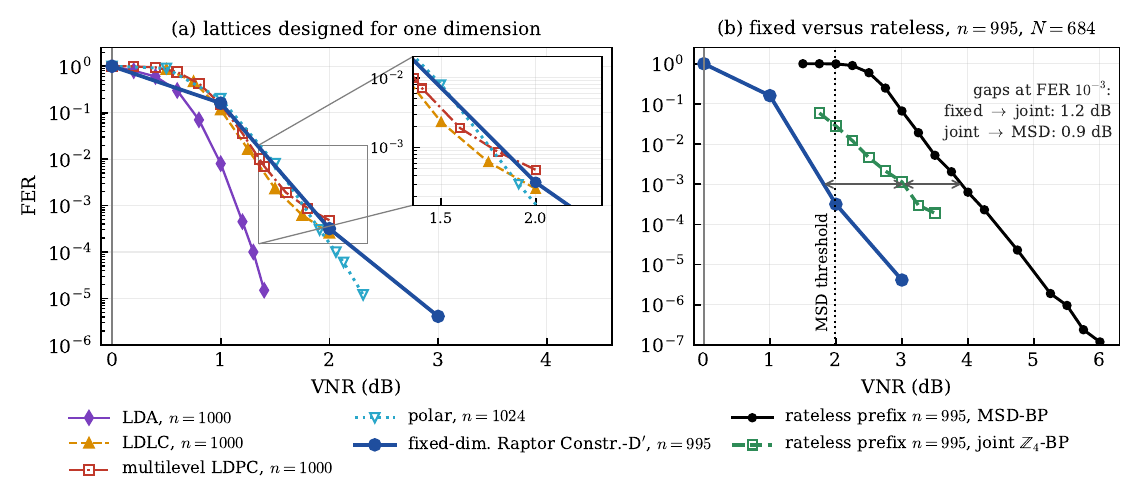}
\caption{(a) FER of lattices designed for one dimension near $1000$: LDA, LDLC, multilevel LDPC and polar lattices~\cite{Pietro2018,29silva,silva,Polarm}, and the fixed-dimension Raptor Construction-D$'$ lattice of~\cite{CCECE} ($n=995$, simulated in $1$-dB steps). (b) The same fixed-dimension Raptor lattice and the rateless prefix $n=995$ with MSD-BP and with joint $\mathbb Z_4$-BP decoding, both with $N=684$; arrows: VNR gaps at FER $10^{-3}$; the two last joint-BP points rest on $24$ and $19$ frame errors; dotted: asymptotic MSD threshold of the prefix.}
\label{fig:cmpfixed}
\end{figure}

\begin{figure}[!t]
\centering
\includegraphics[width=0.55\linewidth]{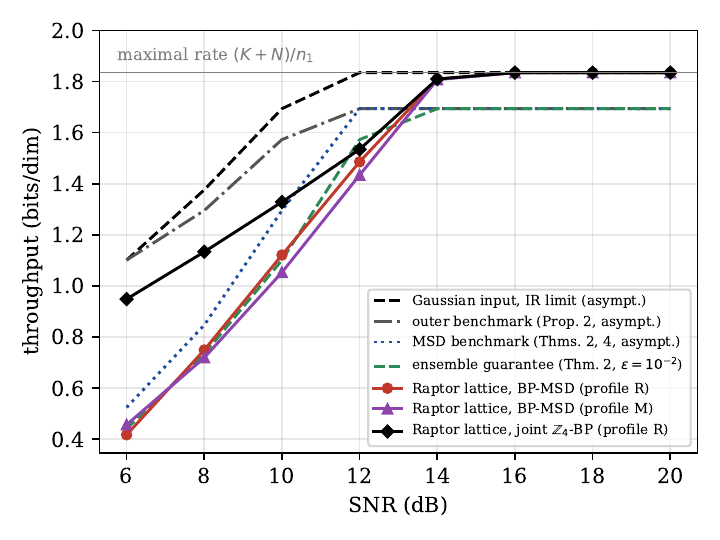}
\caption{Long-term throughput of IR with one rateless Raptor lattice on the $4$-PAM AWGN channel (design A, $n_1=684$, increments of $57$ dimensions, $400$ sessions per point): MSD-BP with profiles R and M, joint $\mathbb Z_4$-BP, the asymptotic benchmarks (Remark~\ref{rem:bench}) and the ensemble guarantee of Theorem~\ref{thm:achiev}.}
\label{fig:awgnrateless}
\end{figure}

Fig.~\ref{fig:awgnrateless} shows the long-term throughput of IR transmission over a fixed-SNR $4$-PAM AWGN channel. The first round has $n_1=N=684$ dimensions, followed by increments of $57$, and one mother lattice serves all SNRs. With profile~R the scheme attains $80$--$88\%$ of the asymptotic MSD benchmark between $6$ and $12$~dB (Table~\ref{tab:joint}). It reaches $1.810$ bits/dim at $14$~dB and, from $16$~dB on, the maximal rate $(K+N)/N=1.835$ bits/dim after the first round; from $14$~dB on it exceeds both asymptotic benchmarks (Remark~\ref{rem:bench}). The fixed code is above the ensemble guarantee of Theorem~\ref{thm:achiev} at $8$ and $10$~dB and from $14$~dB on, and $5$--$6\%$ below it at $6$~dB ($0.417$ versus $0.440$), where sessions use the longest prefixes, and at $12$~dB ($1.486$ versus $1.573$). The gap between the MSD and outer benchmarks at low SNR ($0.524$ versus $1.101$ bits/dim at $6$~dB) is the price of the fixed split (Corollary~\ref{cor:price}), and Section~\ref{sec:numjoint} shows that the joint decoder recovers most of it with the same code.

\subsection{Block Rayleigh Fading With and Without Erasures}
\begin{figure}[!t]
\centering
\includegraphics[width=0.9\linewidth]{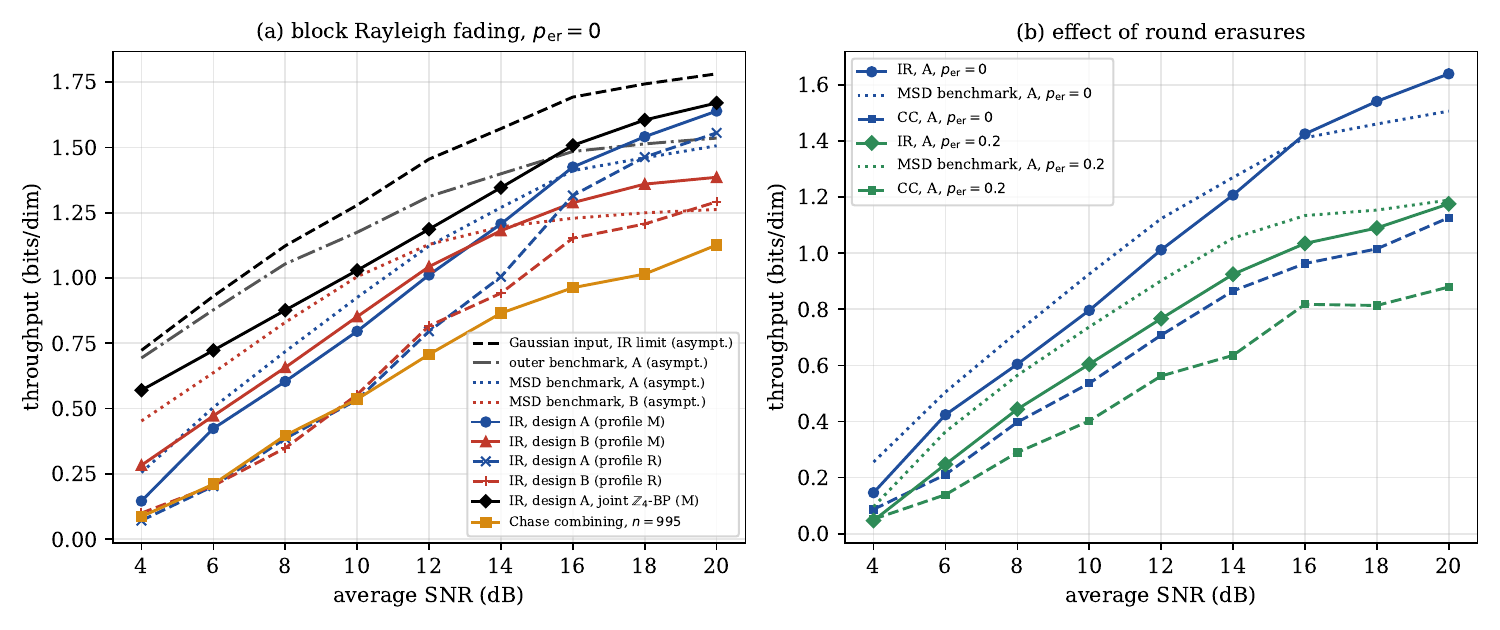}
\caption{Throughput under block Rayleigh fading ($n_1=684$, rounds of $114$ dimensions, budget $4104$, $400$ sessions per point). (a) Designs A and B with MSD-BP, design A with joint BP, Chase combining and asymptotic benchmarks. (b) Design A with round erasure probability $p_{\mathrm{er}}\in\{0,0.2\}$.}
\label{fig:fading}
\end{figure}

We now consider IR over block Rayleigh fading. The first round carries $n_1=684$ dimensions and every later round $114$ new dimensions, each round has an independent gain $|h_r|^2\sim\mathrm{Exp}(1)$, and the session budget is $n_{\max}=4104$ dimensions (31 rounds). The baseline is Chase combining (CC)~\cite{CT2001} of the $995$-prefix of the same mother lattice (same payload), with maximum-ratio combining and the same budget ($4\times995\le4104$). Fig.~\ref{fig:fading} and Table~\ref{tab:fading} summarize the results.

\emph{Degree profile matters under fading.} With the AWGN-optimized profile~R, IR suffers outages ($13.7\%$ at $10$~dB) even when the accumulated level capacities are several times $K$ and $N$. They occur when the first round, which carries all systematic symbols, is deeply faded: BP cannot start from LT checks that all have degree $4$, the fading analogue of the need for low-degree output symbols in LT codes~\cite{Luby2002,ES2006}. Profile~M fixes this: at $8$~dB the outage drops from $22.5\%$ to $0$ and the throughput rises from $0.386$ to $0.604$ bits/dim, at a cost of up to $0.07$ bits/dim on AWGN between $8$ and $12$~dB (Fig.~\ref{fig:awgnrateless}); at $6$~dB profile~M is even slightly better on AWGN.

\emph{IR versus CC.} With profile~M, IR outperforms CC at every SNR. Its throughput is $40$--$52\%$ higher between $8$ and $20$~dB, and it is twice as high at $6$~dB, where CC suffers a $42\%$ outage within the same budget. Designs~A and B attain $84$--$95\%$ and $79$--$99\%$ of their asymptotic MSD benchmarks between $8$ and $14$~dB and exceed them from $16$~dB on (Remark~\ref{rem:bench}).

\emph{Split.} As predicted by Fig.~\ref{fig:theory}(b), design~B outperforms design~A up to $12$~dB and design~A is better from $14$~dB on (Table~\ref{tab:fading}); the simulated crossover agrees with the theoretical one.

\emph{Erasures.} With round erasures of probability $p_{\mathrm{er}}=0.2$ (Fig.~\ref{fig:fading}(b)), an erased round simply contributes $\sigma_j=\infty$. Throughput decreases roughly in proportion to the received fraction and is $1$--$21\%$ below the asymptotic MSD benchmark for SNR $\ge8$~dB. CC degrades more, because an erased repetition wastes a whole $995$-dimensional block; the gain of IR over CC is $27$--$53\%$ for SNR between $8$ and $20$~dB.

\begin{table}[!t]
\centering
\footnotesize
\caption{Block-fading IR with profile M: throughput $\mathsf T$ (bits/dim), outage $P_{\mathrm{out}}$ and mean number of rounds $\bar r$, with Chase combining (CC) and the asymptotic MSD benchmark.}
\label{tab:fading}
\begin{tabular}{@{}c ccc c ccc c c@{}}
\toprule
 & \multicolumn{3}{c}{Design A, IR} & CC (A) & \multicolumn{3}{c}{Design B, IR} & MSD bench. & $\mathsf{T}_{\rm IR}/\mathsf{T}_{\rm CC}$\\
$\overline{\mathrm{SNR}}$ & $\mathsf T$ & $P_{\mathrm{out}}$ & $\bar r$ & $\mathsf T$ & $\mathsf T$ & $P_{\mathrm{out}}$ & $\bar r$ & A\,/\,B & (A)\\
\midrule
4 dB & 0.146 & 0.552 & 28.7 & 0.087 & 0.283 & 0.172 & 21.3 & 0.256\,/\,0.453 & 1.68\\
6 dB & 0.424 & 0.027 & 20.3 & 0.211 & 0.472 & 0.017 & 13.7 & 0.505\,/\,0.637 & 2.01\\
8 dB & 0.604 & 0.000 & 13.2 & 0.397 & 0.657 & 0.000 & 8.7 & 0.718\,/\,0.830 & 1.52\\
10 dB & 0.796 & 0.000 & 8.8 & 0.536 & 0.852 & 0.000 & 5.6 & 0.925\,/\,1.005 & 1.49\\
12 dB & 1.012 & 0.000 & 5.9 & 0.707 & 1.043 & 0.000 & 3.6 & 1.123\,/\,1.129 & 1.43\\
14 dB & 1.207 & 0.000 & 4.1 & 0.865 & 1.181 & 0.000 & 2.6 & 1.270\,/\,1.196 & 1.40\\
16 dB & 1.425 & 0.000 & 2.7 & 0.963 & 1.288 & 0.000 & 2.0 & 1.411\,/\,1.229 & 1.48\\
18 dB & 1.541 & 0.000 & 2.1 & 1.015 & 1.359 & 0.000 & 1.6 & 1.460\,/\,1.249 & 1.52\\
20 dB & 1.639 & 0.000 & 1.7 & 1.126 & 1.385 & 0.000 & 1.5 & 1.506\,/\,1.262 & 1.46\\
\bottomrule
\end{tabular}
\end{table}

\subsection{Joint $\mathbb Z_4$ Decoding Recovers the Split Loss}\label{sec:numjoint}
We repeat the AWGN and block-fading experiments of design~A with the joint $\mathbb Z_4$-BP decoder of Section~\ref{sec:joint} (at most $100$ iterations; same code, sessions and stopping rule). Figs.~\ref{fig:awgnrateless} and~\ref{fig:fading}(a) show the results, and Table~\ref{tab:joint} lists them.

On the AWGN channel the joint decoder attains $0.948$, $1.133$, $1.328$ and $1.535$ bits/dim at $6$, $8$, $10$ and $12$~dB. This is $86$, $87$, $84$ and $91\%$ of the asymptotic outer benchmark of Proposition~\ref{prop:outer}, and it \emph{exceeds the asymptotic MSD benchmark} at $6$ and $8$~dB by $81\%$ and $34\%$; the MSD-BP decoder attains $0.417$ and $0.749$ bits/dim there. At $14$~dB and above both decoders reach, or nearly reach, the maximal rate $(K+N)/N$ after the first round ($1.810$ bits/dim at $14$~dB). Under block Rayleigh fading with profile~M, joint decoding raises the throughput of design~A by a factor of $3.9$ at $4$~dB, by $71\%$ at $6$~dB, by $45\%$ at $8$~dB and by $29\%$ at $10$~dB, with no outage observed in the $400$ sessions per point (one-sided $95\%$ upper limit $0.75\%$; MSD-BP has a $55.2\%$ outage at $4$~dB). It exceeds the asymptotic MSD benchmark at every SNR, reaches $82$--$96\%$ of the outer benchmark between $4$ and $14$~dB, outperforms design~B with MSD at every SNR, and exceeds Chase combining by a factor of $1.9$--$6.6$ between $4$ and $10$~dB. This is consistent with Theorem~\ref{thm:mlcap} and Remark~\ref{rem:mlvsmsd} (which concern the dense ring-coefficient ensemble under ML decoding): the loss of Corollary~\ref{cor:price} is a property of stage-by-stage decoding, and a practical joint decoder removes most of it.

\emph{Density evolution.} Table~\ref{tab:de} compares the fixed-target DE operating points $R^\star$ of~\eqref{eq:derate} for the joint decoder with the smooth asymptotic MSD and outer rates and with the simulated throughputs. For profile~R, DE predicts $87$--$90\%$ of the asymptotic outer rate between $6$ and $12$~dB, well above the MSD rate, and the simulations attain $95$--$99\%$ of these values. Three seeds, a population of $5\times10^4$ instead of $2\times10^4$, and $300$ instead of $150$ iterations change $R^\star$ by less than $1\%$. Although the surrogate ensemble differs from the QC code (Remark~\ref{rem:qccycles}), DE thus agrees quantitatively with the joint decoder: most of the split loss is recovered, and $10$--$13\%$ of the outer rate remains. Profile~M loses $4$--$13\%$ on AWGN according to DE, consistent with the MSD-BP results.

\emph{Degree profile for the joint graph.} We used DE to search over $24$ LT degree distributions with degrees between $1$ and $12$, maximizing the average fraction of the asymptotic outer rate at $4$, $8$ and $12$~dB. The best, profile~J with $\Omega=\{3\!:\!0.2,\,4\!:\!0.5,\,6\!:\!0.2,\,10\!:\!0.1\}$, gains $2.5$--$3.2\%$ at $10$ and $12$~dB but loses $6\%$ at $4$~dB, and no distribution improved on profile~R by more than about $1.5\%$ on average; simulations confirm the ordering ($0.9$--$2.5\%$ above profile~R, Table~\ref{tab:de}). These profile gains of $1$--$3\%$ are small compared with the uncertainty of the DE operating points (Section~\ref{sec:de}): the $14$~dB entries of profiles M and J differ by about one bisection step, and the selected profile was not validated on held-out channel realizations. DE and the paired simulations therefore suggest, but do not establish, that for this precode the LT profile is not the main bottleneck and that closing the remaining gap requires a joint design of precode and LT stage.

\begin{table}[!t]
\centering
\footnotesize
\caption{AWGN throughput (bits/dim) of design A with joint BP: DE operating point $R^\star$ (surrogate ensemble) for profiles R, M and J, simulation, and the smooth asymptotic MSD rate $(1+\chi)\min\{C_0/\chi,C_1\}$ and outer rate $\min\{C_{\mathrm X},(1+\chi)C_1\}$.}
\label{tab:de}
\begin{tabular}{@{}c cc ccc cc@{}}
\toprule
 & & & \multicolumn{3}{c}{DE, joint BP} & \multicolumn{2}{c}{simulation, joint BP}\\
\cmidrule(lr){4-6}\cmidrule(l){7-8}
SNR & MSD rate & outer rate & R & M & J & R & J\\
\midrule
4 dB & 0.336 & 0.876 & 0.814 & 0.712 & 0.764 & -- & --\\
6 dB & 0.529 & 1.102 & 0.992 & 0.878 & 0.992 & 0.948 & 0.957\\
8 dB & 0.858 & 1.342 & 1.188 & 1.077 & 1.197 & 1.133 & 1.159\\
10 dB & 1.300 & 1.582 & 1.369 & 1.280 & 1.413 & 1.328 & 1.361\\
12 dB & 1.737 & 1.790 & 1.553 & 1.486 & 1.592 & 1.535 & 1.565\\
14 dB & 1.835 & 1.835 & 1.824 & 1.835 & 1.824 & 1.810 & 1.812\\
\bottomrule
\end{tabular}
\end{table}

\begin{table}[!t]
\centering
\footnotesize
\caption{Throughput (bits/dim) of design A with joint BP and with MSD-BP, and the round-quantized asymptotic benchmarks (AWGN: profile R; fading: profile M). No outage was observed with joint decoding ($400$ sessions per point).}
\label{tab:joint}
\begin{tabular}{@{}c cccc c cccc@{}}
\toprule
 & \multicolumn{4}{c}{AWGN} & & \multicolumn{4}{c}{Block Rayleigh fading}\\
\cmidrule{2-5}\cmidrule{7-10}
SNR & joint & MSD-BP & MSD b. & outer b. & & joint & MSD-BP & MSD b. & outer b.\\
\midrule
4 dB & -- & -- & -- & -- & & 0.570 & 0.146 & 0.256 & 0.694\\
6 dB & 0.948 & 0.417 & 0.524 & 1.101 & & 0.723 & 0.424 & 0.505 & 0.878\\
8 dB & 1.133 & 0.749 & 0.847 & 1.295 & & 0.877 & 0.604 & 0.718 & 1.054\\
10 dB & 1.328 & 1.121 & 1.295 & 1.573 & & 1.029 & 0.796 & 0.925 & 1.175\\
12 dB & 1.535 & 1.486 & 1.694 & 1.694 & & 1.187 & 1.012 & 1.123 & 1.312\\
14 dB & 1.810 & 1.810 & 1.694 & 1.694 & & 1.346 & 1.207 & 1.270 & 1.399\\
16 dB & 1.835 & 1.835 & 1.694 & 1.694 & & 1.508 & 1.425 & 1.411 & 1.484\\
18 dB & 1.835 & 1.835 & 1.694 & 1.694 & & 1.605 & 1.541 & 1.460 & 1.513\\
20 dB & 1.835 & 1.835 & 1.694 & 1.694 & & 1.670 & 1.639 & 1.506 & 1.535\\
\bottomrule
\end{tabular}
\end{table}

\section{Conclusion and Open Problems}\label{sec:conclusion}
We defined rateless lattices and showed that two-level Construction-D$'$ lattices built from nested Raptor codes form such a family. For transmitter-unknown noise and fading sequences independent of the code randomness, we proved a non-asymptotic ensemble random-coding bound and a converse for the multistage architecture, also for sparse LT stages of logarithmic degree. These results concern the residue modulo~$4$ (lattice-point recovery adds a coarse-lattice term) and isolate a loss specific to stage-by-stage decoding with a fixed split.

Under joint ML decoding this loss disappears: for a dense ensemble with ring coefficients we determined the rateless capacity region within its admissible level sizes, which with equal level sizes contains the accumulated capacity of uniform $2^a$-PAM for every noise sequence, for $a=2$ and, under a level-ordering condition that we verified only numerically, for $a\ge3$; single-level $p$-ary rateless lattices achieve the same and satisfy a finite-parameter rateless Poltyrev inequality. Geometric shaping with a lattice-consistent one-dimensional constellation keeps every transmitted prefix a lattice point and brings these codes within any $\epsilon>0$ of the Gaussian incremental-redundancy capacity, whereas repetition of a fixed-dimension lattice codeword loses a factor that grows with the number of rounds. In practice, a joint BP decoder over $\mathbb Z_4$ on the sparse realization, which the joint capacity theorem does not cover, recovers most of the split loss, both in throughput and in FER. Secrecy applications are developed in~\cite{LCOMcomp}.

Several questions remain open: (i) a deterministic universal construction, a joint-limit proof of the asymptotic rateless Poltyrev theorem, a joint-decoding theorem for sparse binary coefficients, and a proof of the level ordering for $a\ge3$; (ii) a constant LT degree, e.g., with a precode on level~$1$, and a joint design of precode and LT stage that closes the remaining gap and controls the error floors; (iii) an adaptive split and shaping in high dimension~\cite{EZ2004,23silva} compatible with projection; (iv) rateless compute-and-forward.

\bibliographystyle{IEEEtran}
\bibliography{refs}
\end{document}